\documentclass[a4paper,fleqn,usenatbib]{mnras}
\def\hi{\relax \ifmmode {\mbox H\,{\scshape i}}\else H\,{\scshape i}\fi}
\def\hii{\relax \ifmmode {\mbox H\,{\scshape ii}}\else H\,{\scshape ii}\fi}
\def\nii{\relax \ifmmode {\mbox N\,{\scshape ii}}\else N\,{\scshape ii}\fi}
\def\oi{\relax \ifmmode {\mbox O\,{\scshape i}}\else O\,{\scshape i}\fi}
\def\oii{\relax \ifmmode {\mbox O\,{\scshape ii}}\else O\,{\scshape ii}\fi}
\def\oiii{\relax \ifmmode {\mbox O\,{\scshape iii}}\else O\,{\scshape iii}\fi}
\def\sii{\relax \ifmmode {\mbox S\,{\scshape ii}}\else S\,{\scshape ii}\fi}
\def\siii{\relax \ifmmode {\mbox S\,{\scshape iii}}\else S\,{\scshape iii}\fi}
\def\ha{\relax \ifmmode {\mbox H}\alpha\else H$\alpha$\fi}                      
\def\nai{\relax \ifmmode {\mbox Na\,{\scshape i}}\else Na\,{\scshape i}\fi}
\def\mgi{\relax \ifmmode {\mbox Mg\,{\scshape i}}\else Mg\,{\scshape i}\fi}
\def\ki{\relax \ifmmode {\mbox K\,{\scshape i}}\else K\,{\scshape i}\fi}
\def\hb{\relax \ifmmode {\mbox H}\beta\else H$\beta$\fi}
\def\hei{\relax \ifmmode {\mbox He\,{\scshape i}}\else He\,{\scshape i}\fi}
\def\me{$^{-1}$}
\def\arcsec{\hbox{$^{\prime\prime}$}}

\def\msun{\relax \ifmmode {\mbox M}_{\odot}\else M$_{\odot}$\fi}                      
\usepackage[T1]{fontenc}
\usepackage{ae,aecompl}

\usepackage{graphicx}	% Including figure files
\usepackage{amsmath}	% Advanced maths commands
\usepackage{amssymb}	% Extra maths symbols
\usepackage{multicol}   % Multi-column entries in tables

\title[The multiphase wind in NGC~5394]{The multiphase starburst-driven galactic wind in NGC~5394}
\author[Mart\'\i n-Fern\'andez et al.]
{Pablo Mart\'\i n-Fern\'andez,$^{1}$\thanks{E-mail: pablomartin@ugr.es}
Jorge Jim\'enez-Vicente,$^{1,2}$
Almudena Zurita,$^{1,2}$\newauthor
Evencio Mediavilla,$^{3,4}$ and 
\'Africa Castillo-Morales$^{5}$\\
$^{1}$Dpto. de F\'\i sica y del Cosmos, Campus de Fuentenueva, Edificio Mecenas, Universidad de Granada, Granada, E-18071, Spain\\
$^{2}$Instituto Carlos I de F\'\i sica Te\'orica y Computacional, Facultad de Ciencias, Universidad de Granada, E-18071, Spain\\
$^{3}$Instituto de Astrof\'\i sica de Canarias, V\'\i a L\'actea, s/n, La Laguna, E-38200, Tenerife, Spain\\
$^{4}$Dpto. de Astrof\'\i sica,  Universidad de la Laguna, La Laguna, E-38200, Tenerife, Spain\\
$^{5}$Dpto. de Astrof\'\i sica y C.C. de la Atm\'osfera,  Universidad Complutense de Madrid, Madrid, E-28040, Spain
}

\date{Accepted XXX. Received YYY; in original form ZZZ}

\pubyear{2016}

\begin{document}
\label{firstpage}
\pagerange{\pageref{firstpage}--\pageref{lastpage}}
\maketitle

% Abstract of the paper
\begin{abstract}
We present a detailed study of the neutral and ionised gas phases in the galactic wind for the nearby starburst galaxy  NGC~5394 based on new integral field spectroscopy obtained with the INTEGRAL fibre system at the William Herschel Telescope. The neutral gas phase in the wind is detected via the interstellar \nai\ D doublet absorption. After a careful removal of the stellar contribution to these lines, a significant amount of neutral gas ($\sim10^7$~M$_\odot$) is detected in a central region of $\sim1.75$~kpc size. This neutral gas is blueshifted by $\sim165$~km~s\me\ with respect to the underlying galaxy. The mass outflow of neutral gas is comparable to the star formation rate of the host galaxy. Simultaneously, several emission lines (\ha, [\nii], [\sii]) are also analysed looking for the ionised warm phase counterpart of the wind. A careful kinematic decomposition of the line profiles reveals the presence of a secondary, broader, kinematic component. This component is found roughly in the same region where the \nai\ D absorption is detected. It presents higher  [\nii]/\ha\ and [\sii]/\ha\ line ratios than the narrow component at the same locations, indicative of contamination by shock ionization. This secondary component also presents blueshifted velocities, although smaller than those measured for the neutral gas, averaging to $\sim-30$~km~s\me. The mass and mass outflow rate of the wind is dominated by the neutral gas, of which a small fraction might be able to escape the gravitational potential of the host galaxy. The observations in this system can be readily understood within a bipolar gas flow scenario. 
\end{abstract}

% Select between one and six entries from the list of approved keywords.
% Don't make up new ones.

%%%%%%%%%%%%%%%%%%%%%%%%%%%%%%%%%%%%%%%%%%%%%%%%%%

%%%%%%%%%%%%%%%%% BODY OF PAPER %%%%%%%%%%%%%%%%%%

\section{Introduction}
Galactic winds (hereafter GWs) are large scale  outflows of gas from a galaxy.  These winds are produced by the energy input provided by nuclear activity and/or strong starbursts \citep[see ][for a comprehensive review]{veilleux05}. They are able to transport  interstellar material against the gravitational potential of the host galaxy, even reaching the intergalactic medium (IGM). As a consequence they have been proposed as a key ingredient in galaxy evolution.  Among their multiple effects are the enrichment of the IGM with heavy elements originated by massive stars in the parent galaxy \citep[e.g.][]{garnett,tremonti,dalcanton}. This gas removal can also effectively slow down the star formation rate and even quenching it \citep[e.g.][]{hopkins,cazzoli}. Nevertheless, in most cases the majority of the gas will not be able to reach the IGM, but will fall back to the host galaxy in a so called ``galactic fountain'' \citep[e.g.][]{shapiro,bregman}, redistributing heavy elements. % poner Krug2010?

GWs were more important in earlier stages of the Universe, when starbursts induced by mergers were more frequent. Most of the star formation in these epochs (z$\sim1-3$) is associated with luminous infrared galaxies (LIRGs, L$_{IR}\gtrsim10^{11}$~L$_{\odot}$), in which this phenomenon is ubiquitous \citep[e.g.][]{martin05,rupke05b}. At low redshift, LIRGs are less frequent, but more easily accessible observationally than their higher redshift counterparts, and these galaxies have indeed received most of the attention to study this phenomenon \citep[e.g.][]{2000ApJS..129..493H,rupke02,martin05}. Although GWs are not limited to LIRGS \citep[e.g.][]{schwartz,chen2010,krug2010}, galaxies with lower star formation rates have been less studied  and the properties of their associated winds are not so well known.

GWs are a complex phenomenon, in which warm ionised and neutral gas from the host galaxy are entrained by a hot flow, originated by an important energy supply from stellar winds and supernova explosions, or active galactic nuclei \citep{TenorioTagle98, veilleux05}. Their study therefore requires a multi-wavelength approach to trace the different involved gas phases. 

Absorption-line spectroscopy has proved to be an effective method to study GWs through the detection of  blueshifted absorbing material in front of the galaxy background. Since the starburst-driven winds usually move nearly perpendicular to the galaxy disc, this technique is favoured in face-on galaxies. Several absorption features in the ultraviolet and optical bands can be used to trace different gas phases \citep[e.g.][]{rupke02,martin05,heckman2015,wood}. At low-z, the \nai\ $\lambda\lambda$5890, 5896 doublet (\nai~D), are the most frequently used lines to trace neutral gas using ground-based observations. 

High blueshifted velocities in the \nai~D doublet have been  routinely detected in luminous and ultra-luminous infrared galaxies (ULIRGs, L$_{IR}\gtrsim10^{12}$~L$_{\odot}$)  \citep[e.g.][]{2000ApJS..129..493H,rupke05b,martin05,martin06,2013ApJ...768...75R,cazzoli}. These objects host the most powerful starbursts, with star formation rates (SFRs) of up to several hundreds solar masses per year. These kind of techniques have also been applied to non-LIRGs and dwarf starbursts \citep[see e.g.][]{roche,jjv2007,schwartz,chen2010}. 
The spatial projected extent of the \nai~D absorbers has typical sizes of a few kiloparsecs, % $\sim1$ to 12~kpc, 
and the velocities range from tens  to several hundreds of km s$^{-1}$. Despite a large scatter, the outflow velocities seem to increase with the galaxy SFRs \citep[cf.][]{martin05} for (U)LIRGs. However, for SFRs lower than $\sim10$~\msun~yr\me, this result is less clear. While this trend with SFR seems to hold for dwarf galaxies \citep{schwartz},   \cite{chen2010} do not find any significant trend for a large sample of galaxies with SFRs in the range of $\sim1$ to 13~\msun~yr\me. The measured mass outflow rates are comparable to their host galaxies global SFR, implying ``loading factors" ($\eta \equiv \dot M_{out} / SFR$) around unity \citep{rupke02,rupke05a,rupke05b,2013ApJ...768...75R,cazzoli}, and even higher \citep[reaching $\eta \sim 10$][]{2000ApJS..129..493H} in some extreme cases. Although some studies suggest that neutral outflows can remove large amounts of gas from the host galaxies \citep[e.g.][]{martin06,heckman2015},  a large fraction of the outflowing gas might not be able to reach the IGM, falling back afterwards to the host galaxy disc again \citep{jjv2007,cazzoli}, redistributing metals within the galaxy \citep{belfiore}.

Emission lines (tracing the warm gas phase) can supplement the information obtained from absorption-line observations. Emission-line imaging and  spectroscopy of winds are easier to obtain in those cases in which the wind is seen in front of a dark background, and it is not outshined by the underlying host galaxy \citep[e.g.][]{lehnert96}. The structure of the wind is frequently bipolar \citep{veilleux05}. information, Nevertheless, kinematic information in these cases  is very limited because the wind motions take place mainly perpendicular to the line-of-sight.
However, observations with high enough  spectral resolution may allow discrimination between the wind and the disc emission even for low inclination systems. This way, the kinematics of the outflowing gas can be obtained \citep[e.g.][]{shih10,soto,wood}. 
In this respect, integral field spectroscopy (IFS), joined to a careful decomposition of the line profiles, has proved to be a most valuable technique. It permits (provided high enough spatial resolution) to locally separate the low brightness wind emission, from that of the underlying galaxy, which could otherwise remain hidden due to the internal galaxy kinematics \citep[e.g.][]{sharp,westmoquette2012,2013ApJ...768...75R,ho2014,cazzoli,arribas2014}.

%The detection rate of ionised gas outflows is high
Although the measured velocities of the ionised component are usually smaller than those of the absorbing neutral gas \citep{2013ApJ...768...75R}, the maximum velocities of this ionised component seem to follow a similar relation with the SFR \citep{martin05,arribas2014}. The outflowing component systematically presents larger velocity dispersion and emission-line ratios indicative of shock excitation  \citep{monreal06,sharp,rich2011,soto,bellocchi,arribas2014}. When the kinematics of the neutral and ionised components are available for the same galaxy, their maximum velocities often correlate with each other \citep{2013ApJ...768...75R}, and most of the outflowing material seems to be contained in the neutral phase \citep{2013ApJ...768...75R}.
Nevertheless, \cite{arribas2014} find similar ionised mass loading factors than the average values available in the literature for the neutral gas in LIRGs, and even higher in ULIRGs. It is therefore still unclear the relative importance of the different gas phases in the winds and any possible dependence on the properties of their host galaxies. A better understanding of this issue (and other wind properties) would benefit from detailed simultaneous observations and analysis of both phases for the same targets.
This work aims at supplementing the double scarcity of detailed studies of GWs in objects with moderate star formation, but also of studies simultaneously  including  the different involved gas phases. In order to do so, we analyse new IFU observations of the nearby merger galaxy NGC~5394.

NGC~5394 is part of an  interacting pair \citep[NGC~5394/95, Arp~84,][]{arp66}, catalogued as strongly interacting with separate nuclei and strong tidal features by \cite{lanz}. The pair is at a distance of 48.73~Mpc \citep{catalan-torrecilla}.
NGC 5394 is classified as SB(s)b-pec  with a very bright inner-disk spiral arm at $\sim$0.9-3.0 kpc from the nucleus, and two long, open tidal arms with a tidal tail extending up to $\sim$20~kpc.  Its infrared luminosity  \citep[L$_{IR}$ = 10$^{10.75}$ L$_{\odot}$, ][]{catalan-torrecilla} and strong radio continuum emission from its nucleus \citep{becker} indicate  the presence of a strong starburst. According to \cite{2002AJ....123..702K}, the total gas mass (including 36\% of helium) is $5.5 \times 10^{9}$~M$_{\odot}$ with a molecular-to-atomic mass ratio of M(H$_{2}$)/M(\hi) = 2.5-2.7. The nuclear starburst is rich in molecular gas, containing almost 80\% of the total molecular mass in the galaxy, but only $\sim25$\% of the total galaxy atomic gas \citep{2002AJ....123..702K}.
 \cite{catalan-torrecilla} estimate a global SFR of $2.76\pm0.47\,M_{\odot}$~yr$^{-1}$ from corrected \ha\ emission. Its estimated stellar mass is  $4.79\times10^{10}\,M_{\odot}$ \citep{lanz}. 

The existence of an  outflow  in NGC 5394 was  suggested by \cite{1999AJ....118.1577K} from observed distortions in the velocity field and the asymmetry of the global \ha\ emission-line profile. The outflow has recently been confirmed by \cite{roche} through the presence of blueshifted interstellar \nai~D. Nevertheless, they find no signatures of the outflow in the emission lines.

This paper is structured as follows. In Section~\ref{sec:obs} we describe the observations and the analysis applied to the integral field spectroscopy. The results are then presented in  Section \ref{sec:Results} for the neutral and ionised phases. Finally, in Section~\ref{sec:Disc} we discuss the results, which can be interpreted in terms of a biconical geometry. Our conclusions  are summarised in Section~\ref{sec:Conclusions}.

\section{Observations and data analysis}
\label{sec:obs}
The observations were taken on the 6th of May 2014 at the Observatorio del Roque de los Muchachos with the fibre system INTEGRAL \citep{1998ASPC..152..149A} in combination with the WYFFOS spectrograph at the 4.2m William Herschel Telescope \citep{1994SPIE.2198...56B}. The weather conditions were good, with a seeing around 0.9\arcsec. The INTEGRAL fibre bundle SB2\footnote{The fibre bundle SB2 consists of 189 fibres of 0.9\arcsec\ which covers an approximate region of $16\arcsec\times12\arcsec$ on the sky. These fibres are surrounded by 30 sky fibres forming a 90\arcsec\ diameter ring (see http://www.iac.es/proyecto/integral/).} was used, in combination with the R1200R grating. The central wavelength and spectral resolution were 6300~\AA\ and 1.8~\AA\, ($R \approx 3500$) respectively. This configuration covers the spectral range 5420--7180~\AA, which contains the absorption features \nai\ $\lambda\lambda$5890, 5896 and the \ha, [\nii]$\lambda\lambda$6548, 6583 and [\sii]$\lambda\lambda$6716, 6731 emission lines (see Fig.~\ref{fig:spfit}a). The total integration time  (9000~s) was split in 5 exposures of 1800~s.

The data reduction has been performed using both the INTEGRAL package for {\sc IRAF}\footnote{IRAF is distributed by the National Optical Astronomy Observatory, which is operated by the Association of Universities for Research in Astronomy (AURA) under a cooperative agreement with the National Science Foundation.}, and some custom routines. It includes standard steps: bias subtraction, flat--fielding, spectra extraction, wavelength calibration, throughput and spectral response corrections and flux calibration. Typical wavelength calibration errors are $\sim0.07$~\AA\ and our  measured spectral resolution varies between $\sim1.8$ and $1.6$~\AA\ across the spectral range. For the flux calibration we used the spectrophotometric standard star BD+33 2642 \citep{oke}. We estimate that response calibration errors are typically of $\sim2$\%.

Fig.~\ref{fig:spfit} shows in black the integrated spectrum for the fibres with the highest signal-to-noise (S/N) in the reduced spectra . It clearly shows an emission-line-dominated spectrum. Absorption features are weak except for the  \nai~D doublet, which has mainly an interstellar origin (see below).% and a prominent . %This shows that the interestellar features dominate over the stellar component. 

\subsection{Subtraction of the stellar component}
\label{StellarFit}
\begin{figure}
   \centering
   \includegraphics[angle=0,width=8.5cm, clip=true]{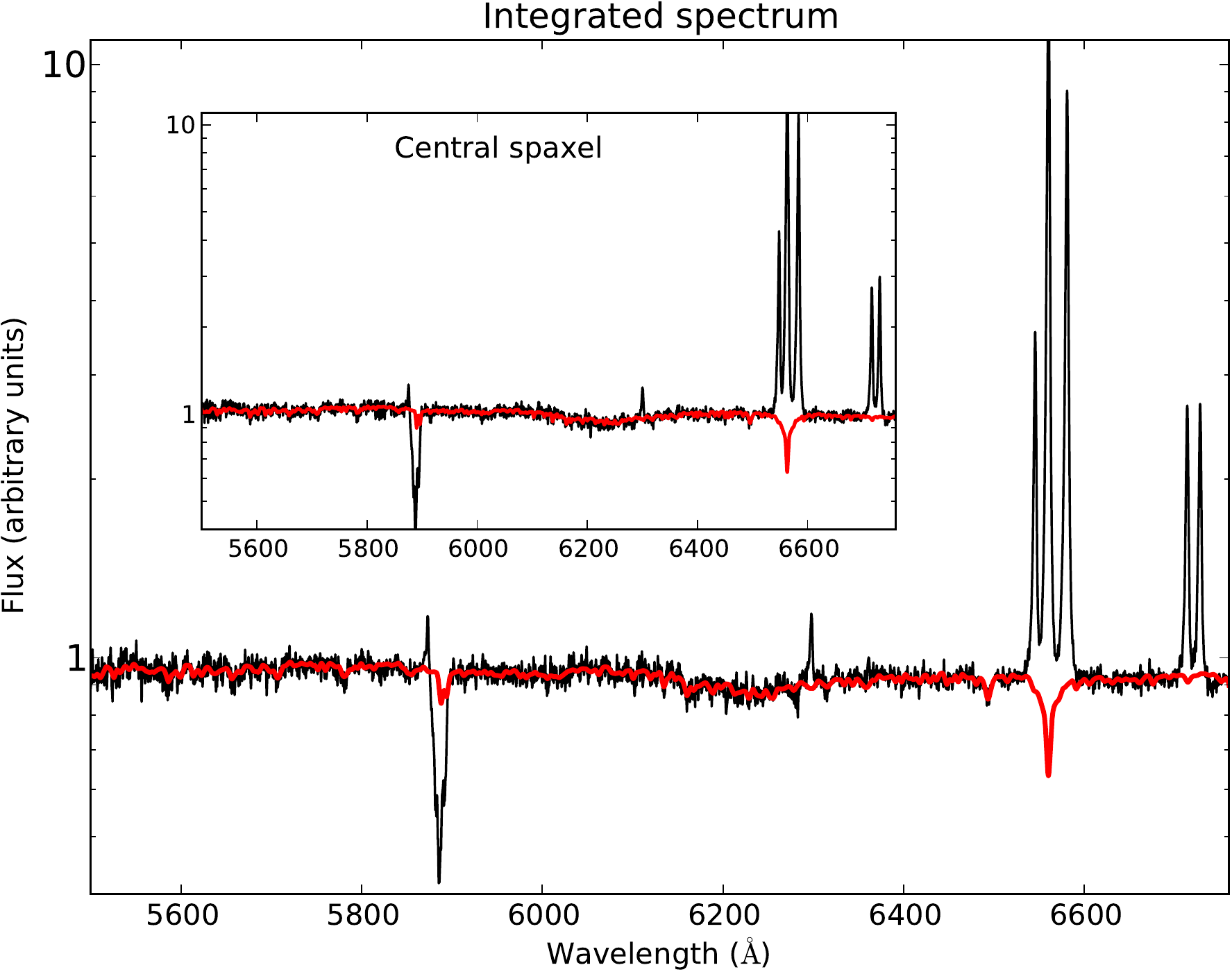}
   \caption{Integrated spectrum for the ten  innermost and highest S/N fibre spectra of NGC~5394. Overplotted in red is the fit to the stellar component (see Section~\ref{StellarFit} for details).  The inset panel shows the observed spectrum and corresponding stellar-component fit for the central (nuclear) fibre.}\label{fig:spfit}
\end{figure}

\begin{figure}
   \centering
   \includegraphics[angle=0,width=9cm, clip=true]{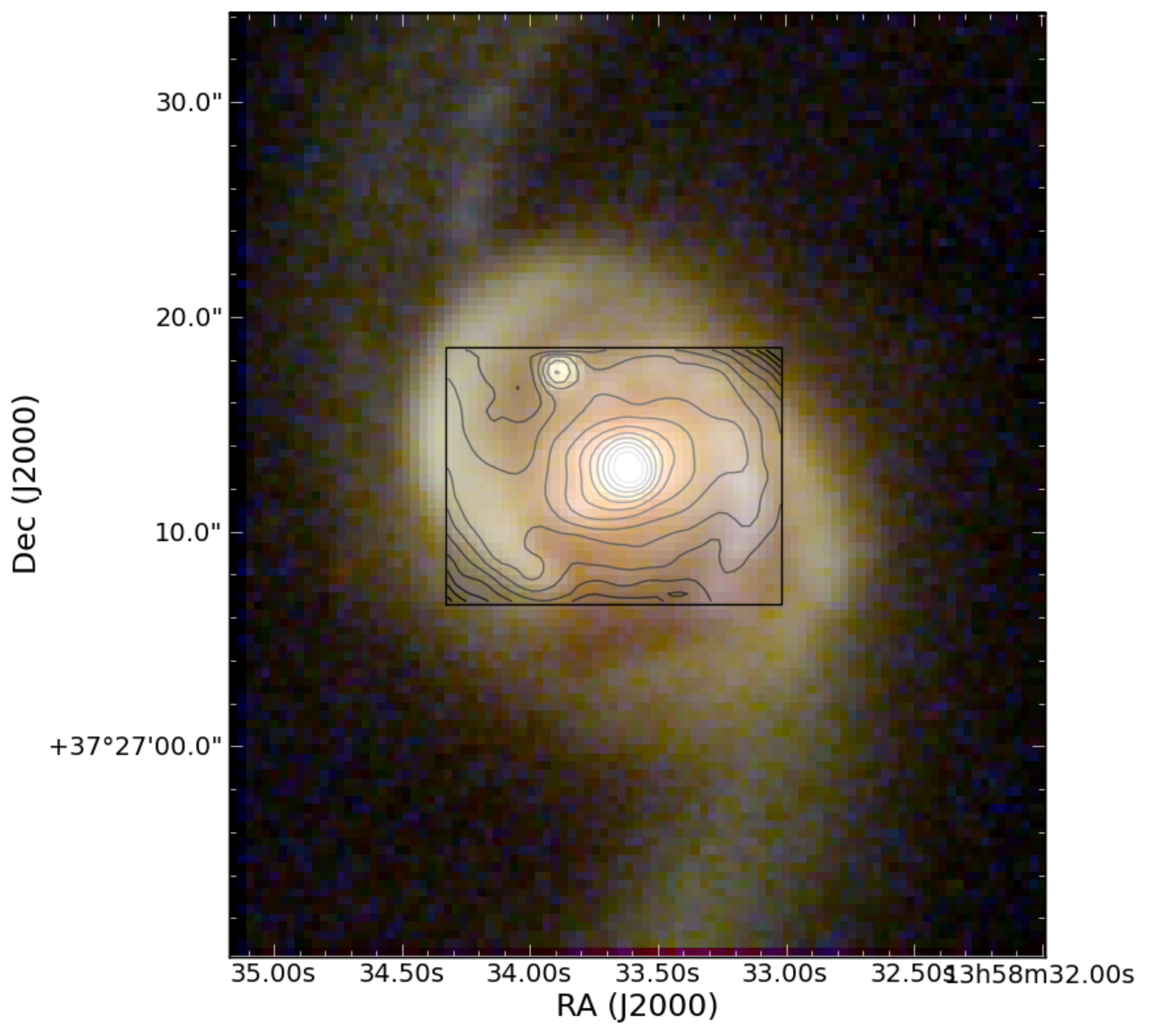}
   \caption{Colour image of NGC 5394 from the SDSS. The $r$-, $g$- and $u$-bands have been used for red, green and blue channels, respectively. Overplotted contours show the red-continuum emission from the INTEGRAL data presented in this paper.}\label{fig:sdsscont}
\end{figure}

Fig.~\ref{fig:sdsscont} shows a three-colour composite image of NGC~5394 made by assigning  Sloan Digital Sky Survey (SDSS)  $r$-, $g$- and $u$-band images to red, green and blue channels respectively. Our INTEGRAL red-continuum intensity map contours are overlaid in grey and match very well the inner morphology of the galaxy. Details on the continuum-map construction are given in Section~\ref{sec:kinem}.

Galactic winds can be directly observed in the local universe via the detection of relevant emission or absorption lines produced in the different wind gas phases. Interstellar medium (ISM) absorption lines in the optical range such as \nai~D have been used to trace the cold neutral gas phase of the outflowing gas in galaxies \citep[e.g.][]{2000ApJS..129..493H,rupke02,rupke05a,rupke05b,martin05,martin06,chen2010,jeong2013}. However, the \nai\ doublet is also a prominent absorption feature in the spectra of late-type stars. Therefore, the total observed \nai\ D  absorption in galaxies has contributions from both stars and clouds of cold gas. In order to study the ISM \nai\ D absorption alone, it is  necessary to remove the stellar contribution. % REMOVED from refereefrom the total observed  \nai\ absorption in our spectra.

In previous studies  the \nai\ stellar contribution has been estimated by scaling the equivalent width of the stellar \mgi\ triplet, $\lambda\lambda 5167, 5173, 5185$ \citep{2000ApJS..129..493H,rupke02,rupke05a},  by assuming a constant ratio between the equivalent width of both sets of lines for all galaxies.
These estimates can nowadays be improved due to the availability of synthesis stellar population  procedures that can be used to fit and subtract the  stellar contribution from a given observed galaxy spectrum \citep[e.g.][]{castillo07,jjv2007,jjv2010,lehnert11}. 
We have used the full-spectral fitting code STECKMAP  \citep{2006MNRAS.365...74O} in combination with the MILES library \citep{2006MNRAS.371..703S,2010MNRAS.404.1639V} to remove the stellar component. The procedure employed is as follows: first we combine the fibre spectra with the highest S/N (the innermost ten spectra) to ensure reliable fitting results. Then, in this combined spectrum we mask out the spectral windows containing strong emission lines, the \nai\ D doublet, and  bright sky line residuals. Afterwards we use STECKMAP to obtain an average age and metallicity for the stellar population that will be used for the whole field of view. 
Finally, the single stellar population MILES spectrum best-matching this age and metallicity was subtracted from each individual INTEGRAL fibre spectra after performing a line-of-sight velocity distribution (LOSVD) convolution with three free parameters in each fibre: velocity, velocity dispersion and amplitude. Our procedure then assumes a roughly uniform stellar population for the whole galaxy. While this is obviously an oversimplification, we remind that we are not interested in the stellar populations themselves, and our purpose is exclusively that of correcting the  \nai\ D absorption line (and eventually also the H$\alpha$ line) for stellar contribution. The above assumption should not therefore have a strong effect, and we are not using the estimated age and metallicity for any other purpose. 

\begin{figure*}
   \centering
   \includegraphics[angle=0,width=\textwidth]{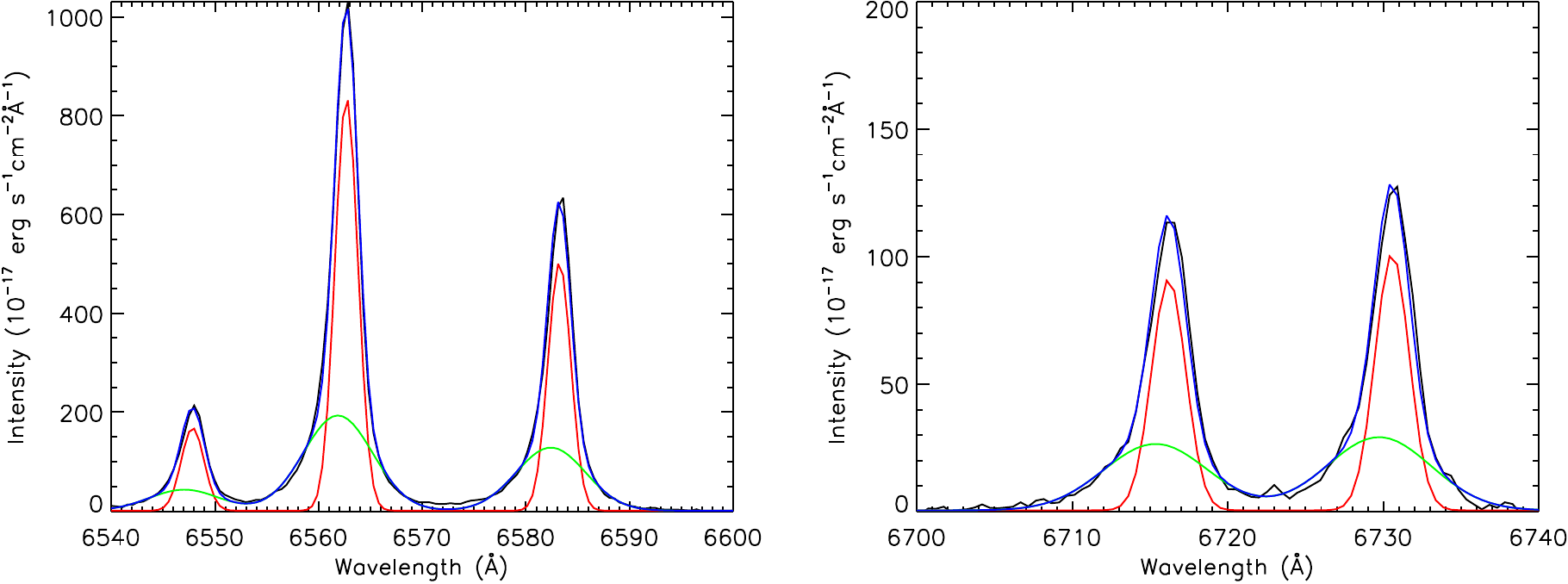}
   \caption{Example of the kinematic decomposition for our INTEGRAL data for the five brightest emission lines in our spectral range. The left panel shows the spectral range comprising [\nii]$\lambda\lambda$6548, 6583 and \ha, and the right panel shows the [\sii]$\lambda\lambda$6716, 6731 doublet range. The black, red and green solid lines show the observed spectrum, and the primary and secondary components, respectively. The blue line shows the fitted two-component spectrum. See Section~\ref{sec:kinem} for further details on the kinematic decomposition.}\label{fig:kine_decomp}
\end{figure*}

Fig.~\ref{fig:spfit} shows the integrated spectrum for fibres with the highest S/N (and also for the central fibre) with the stellar contribution to the spectra (obtained as described above) overplotted in red. It clearly shows  that the \nai~D absorption is mostly interstellar, with some (but small) stellar contribution. In the galaxy integrated spectrum for the 10 innermost fibres, only 25\% of the \nai~D absorption is stellar. This fraction is higher than the mean value of 10\% obtained by \citet{2013ApJ...768...75R} for a sample of ULIRGs. Hence, most of the \nai~D  absorption in the inner region of NGC~5394 is produced in the interstellar medium.

STECKMAP is very sensitive to the number of stellar features available in the spectrum, producing a more reliable fit if a high number of stellar features are available. However, the INTEGRAL spectral range covers only a few of them, and the most important one (\nai\ D) is masked due to the  interstellar absorption contribution. There are still some weak stellar lines, like the Ca+Fe blend at 6495~\AA\ and some Ca lines in the 6100~\AA\ region, which provide a reasonably good fit with the procedure described above.
Nevertheless, in order to test the validity of our methodology, we have applied the same procedure to available NGC~5394 spectroscopic data, covering a wider spectral range. Fortunately, NGC~5394 is part of the publicly available DR1 of the CALIFA survey \citep{2012A&A...538A...8S,2013A&A...549A..87H}. This survey contains PPAK integral field spectroscopy with the V500 setup, covering a spectral range of 3745--7300~\AA\ and resolution $R\sim850$ (cf. 5420-7180~\AA\ and $R\sim3500$ for our INTEGRAL data).
After applying the same procedure as for the INTEGRAL data (but now fitting with STECKMAP the spectrum for each spaxel of the CALIFA data),  we compare the final stellar-subtracted spectra for equivalent spaxels between both instruments. Differences between the \ha\ fluxes obtained with CALIFA and INTEGRAL data are within $\sim10$\%, and are lower than $\sim5$\% for \nai~D in all galaxy spaxels. This is expected as the average ages and metallicities obtained for CALIFA  are in good agreement with the value obtained for INTEGRAL (intermediate-age stellar population with solar metallicity). % Age: 1.0 Gyr; M/H=0.02
Therefore, we have checked that the lack of a wider spectral range in our data  is not biassing our measurements for the purposes of this paper.

\subsection{Kinematic analysis and map generation}
\label{sec:kinem}
In order to study the kinematics and the distribution of the ionised gas in NGC~5394,  the five brightest emission lines within our spectral range (\ha, [\nii]$\lambda\lambda$6548, 6583 and [\sii]$\lambda\lambda$6716,6730) are, in a first step, simultaneously fitted with a single  Gaussian component for each fibre in the stellar--subtracted spectrum. We impose the following constraints to the fits: the velocity and velocity dispersion ($\sigma$) are forced to be the same for all emission lines, where the later must be greater than the instrumental width ($\sigma_I$ = 40 km~s\me\ at $\lambda$=6000\AA). The [\nii]$\lambda$6548/[\nii]$\lambda$6583 line ratio is fixed to the theoretical value 1/3 \citep{osterbrock}, and the  [\sii]$\lambda$6716/[\sii]$\lambda$6731 line ratio is allowed to vary between 0.38 and 1.5.

However, in some spaxels this simple fit leaves large residuals, indicating that several kinematic components are needed to correctly fit the data. 
We therefore have repeated the above procedure but now fitting each emission line with two Gaussian components. These are fitted simultaneously, by imposing over each  component the same constraints explained above. In order to discriminate whether a single or a double kinematic component is needed for each spaxel, we use the reduced chi-square ($\chi^{2}_{red}$) statistics: if the $\chi^{2}_{red}$ for the fit with two Gaussian components improves at least 30\% over the $\chi^{2}_{red}$ for a single Gaussian fit, a double kinematic component is assumed. Otherwise, a single component is considered to describe the line profiles well enough. 

When two different kinematic components are required to properly fit a given spaxel spectrum, we identify a narrow and a broad component. The narrow component is similar to the single component fit in terms of intensity, velocity, and velocity dispersion. This narrow component is interpreted as the same kinematic component present in the spaxels with a single component, which traces the global ionised gas of the galaxy disc. However, the broad kinematic component has lower intensity and it is present in a smaller area of the galaxy, close to the galaxy centre (see below). 
In what follows we will refer to the narrow component as the {\em primary} kinematic component, while the broader one will be termed as the {\em secondary} component. The emission-line fit to only one kinematic component  will be refered to as the {\em single}-component fit. Fig.~\ref{fig:kine_decomp} shows an example of the result of our emission-line fitting.

\begin{figure}
   \centering
   \includegraphics[angle=0,width=0.47\textwidth]{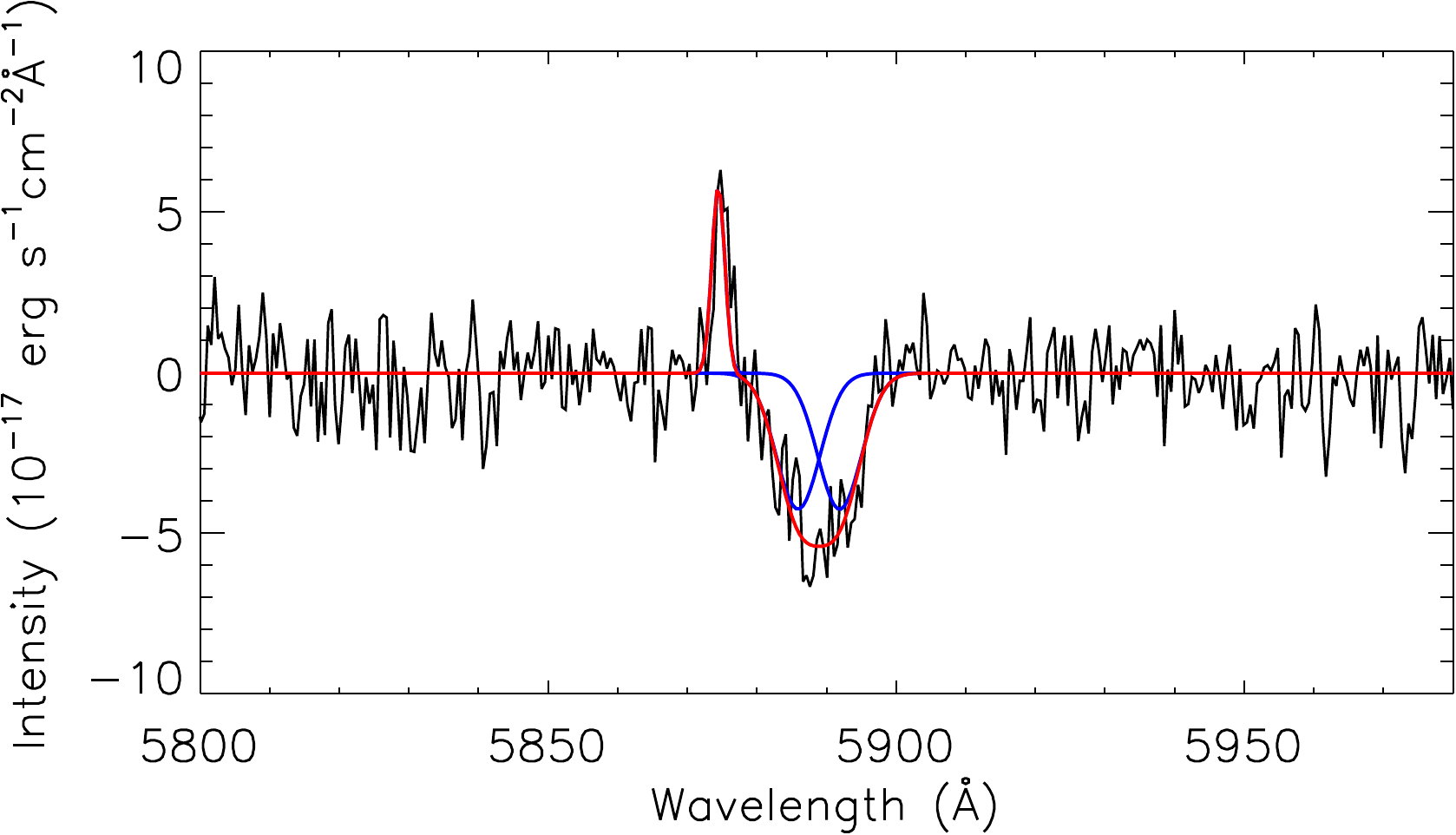}
   \caption{Example of the fit of the \nai\ D doublet for a fibre spectum including the
\hei\ line. The black and red lines show the (stellar subtracted) observed and fitted spectra respectively. 
Blue lines are the fit to the individual lines of the \nai\ D doublet.}\label{fig:nafit}
\end{figure}

We use the interstellar \nai\ D absorption to trace the neutral gas. This is done by fitting the \nai~D doublet with a single Gaussian component. As the S/N in absorption lines is considerably lower than in emission lines, in order to ensure reliable kinematic information on the interstellar neutral gas, the \nai~D doublet  is only fitted in fibre spectra where the \nai~D equivalent width (hereafter EW$_{Na I}$) is larger than 3~\AA. The  amplitude of the absorption feature is also required to be larger than twice the standard deviation of the continuum close to the doublet.
This is a very conservative approach, as the typical EW$_{Na I}$ for stellar spectra is well below this limit (the average EW$_{Na I}$ for our template stellar spectrum is 1.8~\AA) and, in addition, we are setting this limit on spectra from which we have pre\-vious\-ly removed the stellar component. Stellar contamination should therefore be minimal.

An additional difficulty in the \nai\ D fit is the fact that in some spaxels there is substantial \hei\ emission at 5876~\AA, which is blended with the \nai\ D doublet. In order to correct for this, we have adopted 
an approach similar to what was done for the emission lines: for each spaxel we perform two different \nai\ D fits, one including an extra Gaussian component to fit the \hei\ emission, and another one without it. The \hei\ fit is constrained to have the same velocity and velocity dispersion as the primary kinematic component of the emission lines.
The fit including the \hei\ emission line is only accepted as a best fit if the associated $\chi^{2}_{red}$  improves the value for the fit without it by at least 30\%. An example of a fit to
the \nai\ D doublet for a fibre spectrum including the \hei\ line is shown in Fig.~\ref{fig:nafit}.

We detect  interstellar \nai\ D roughly in the same region of the galaxy in which we detect a secondary kinematic component in the emission lines. The \hei\ is only fitted in the innermost four fibres of the galaxy data (close to the nucleus).

After the kinematic analysis for the ionised and neutral gas phases we end up with the following information for each spaxel: emission/absorption line fluxes of relevant spectral lines,  velocity and velocity dispersion\footnote{The velocity dispersion has been quadratically corrected for the instrumental width, but not for the thermal broadening, which is $\sim9.1$~km~s\me\ for \ha, for a temperature of $10^4$~K \citep{spitzer}.} for each kinematic component. 
From this information we map each parameter (flux, velocity, etc) at the fibre position to a regular grid. Final maps cover a field of view (FOV) of 15.7\arcsec $\times$11.3\arcsec (with 45$\times$32~pixels of  0.35\arcsec~pix\me). Our FOV is therefore smaller than that of the maps from CALIFA presented in \cite{roche} for the same galaxy albeit with a significantly better spatial and spectral resolution ($\sim0.9$\arcsec\ vs. $\sim3$\arcsec\ for the NGC~5394 CALIFA data, and $R\sim3500$ vs. $R\sim850$ for CALIFA). Our maps cover the innermost galaxy region, labeled by  \cite{roche} as regions 3, 6 and 7.

We have also derived a red-continuum map from our INTEGRAL data, by integrating the emission in the spectral range $6100\,$\AA$-6200\,$\AA. This map has been astrometrically ca\-li\-bra\-ted  by comparing the position of the nucleus and bright knots with the $r$-band image of the SDSS. The astrometric calibration was afterwards transferred to the rest of the maps. 
Fig.~\ref{fig:sdsscont} shows our continuum map intensity contours overlaid over a three-colour composite image from SDSS data.  It can be seen that the nuclear region of  NGC~5394 dominates the galaxy emission. It is in this region where the interstellar \nai~D and the secondary component of the emission lines are detected. Two spiral arms are clearly seen in the continuum image. A more prominent arm spreads towards the north and a weaker one towards the south.

\section{Results}
\label{sec:Results}
\subsection{Ionised gas phase}

\begin{figure}
   \centering
   \includegraphics[angle=0,width=8cm, clip=true]{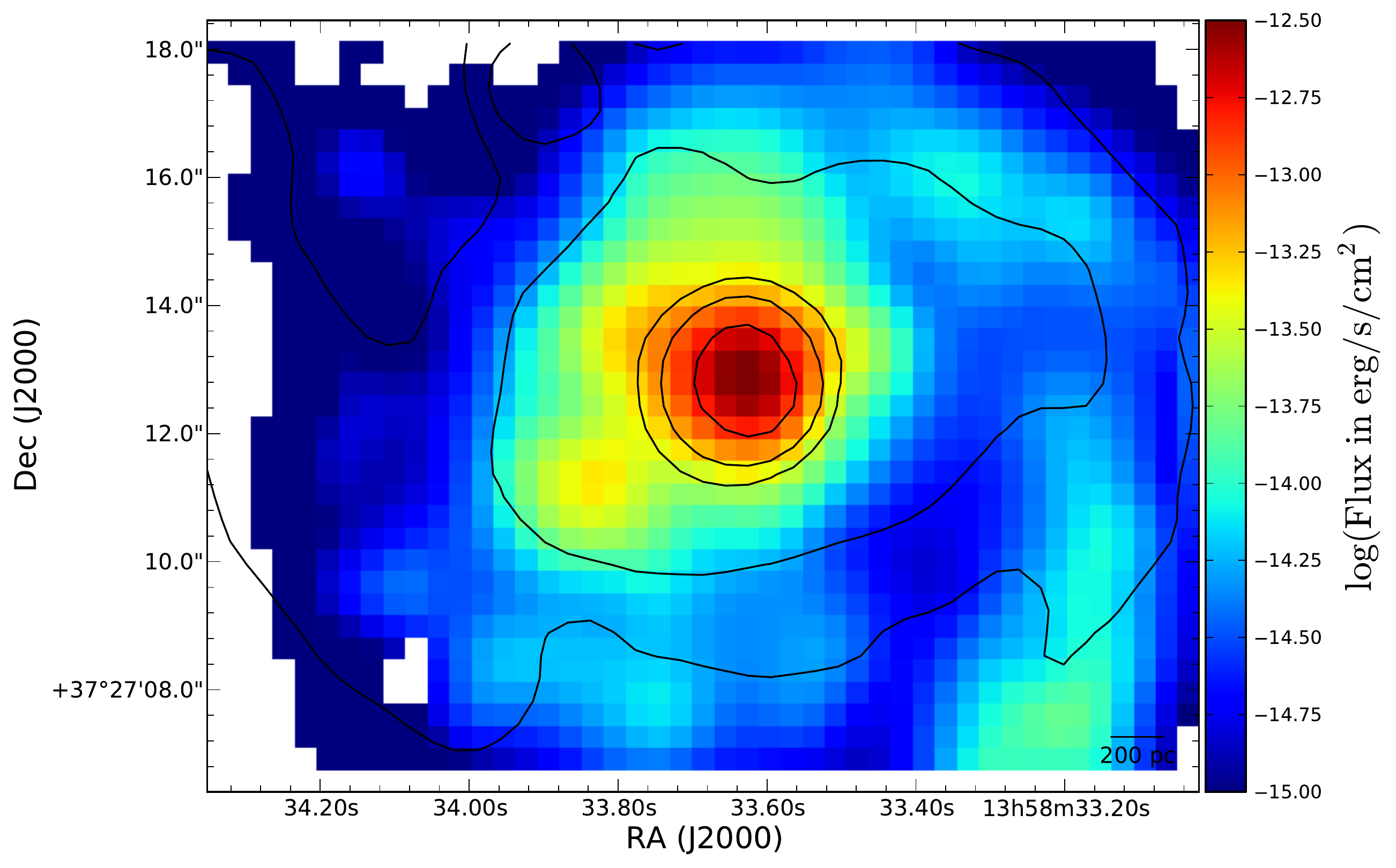}
   \includegraphics[angle=0,width=8cm, clip=true]{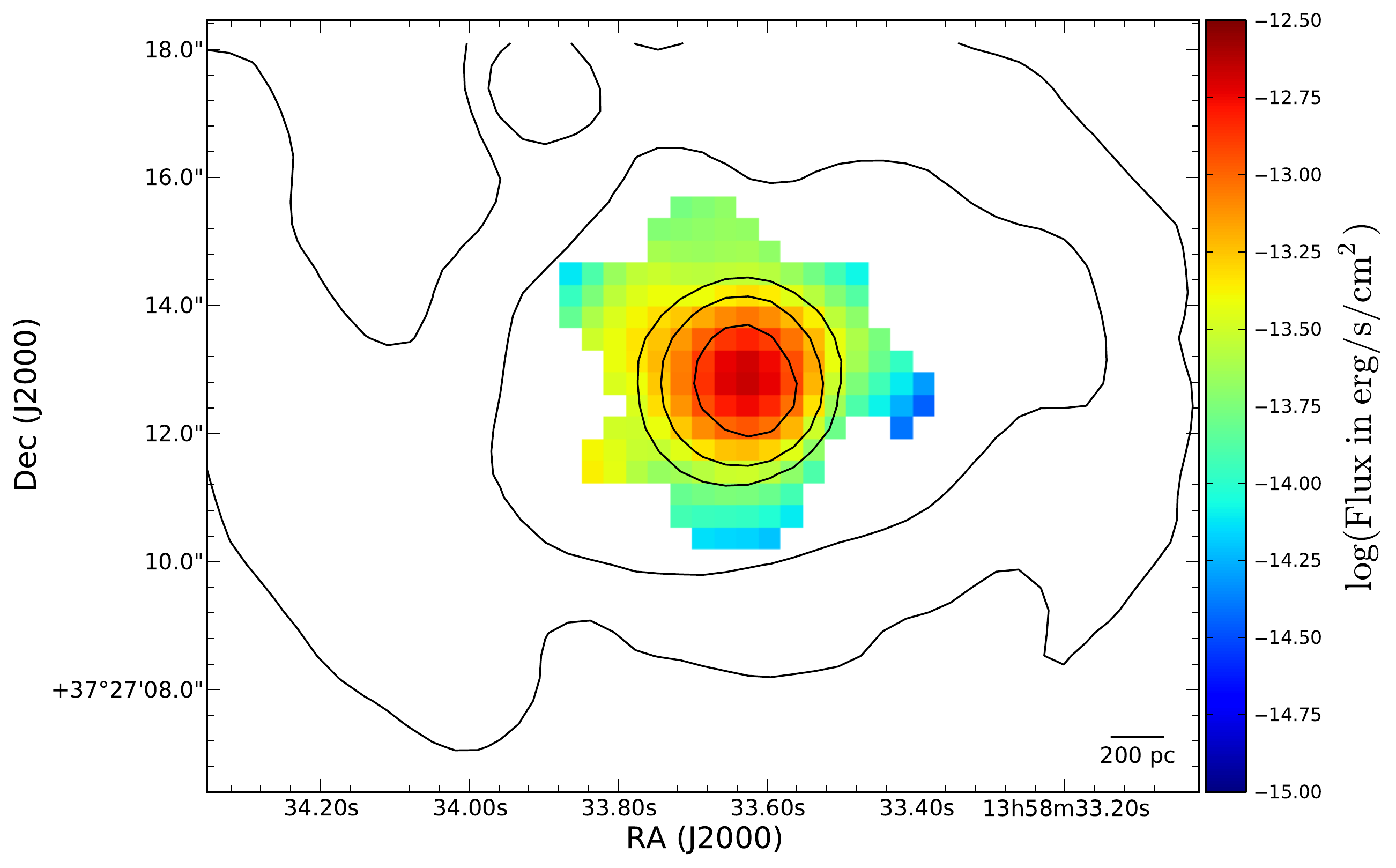}
   \includegraphics[angle=0,width=8cm, clip=true]{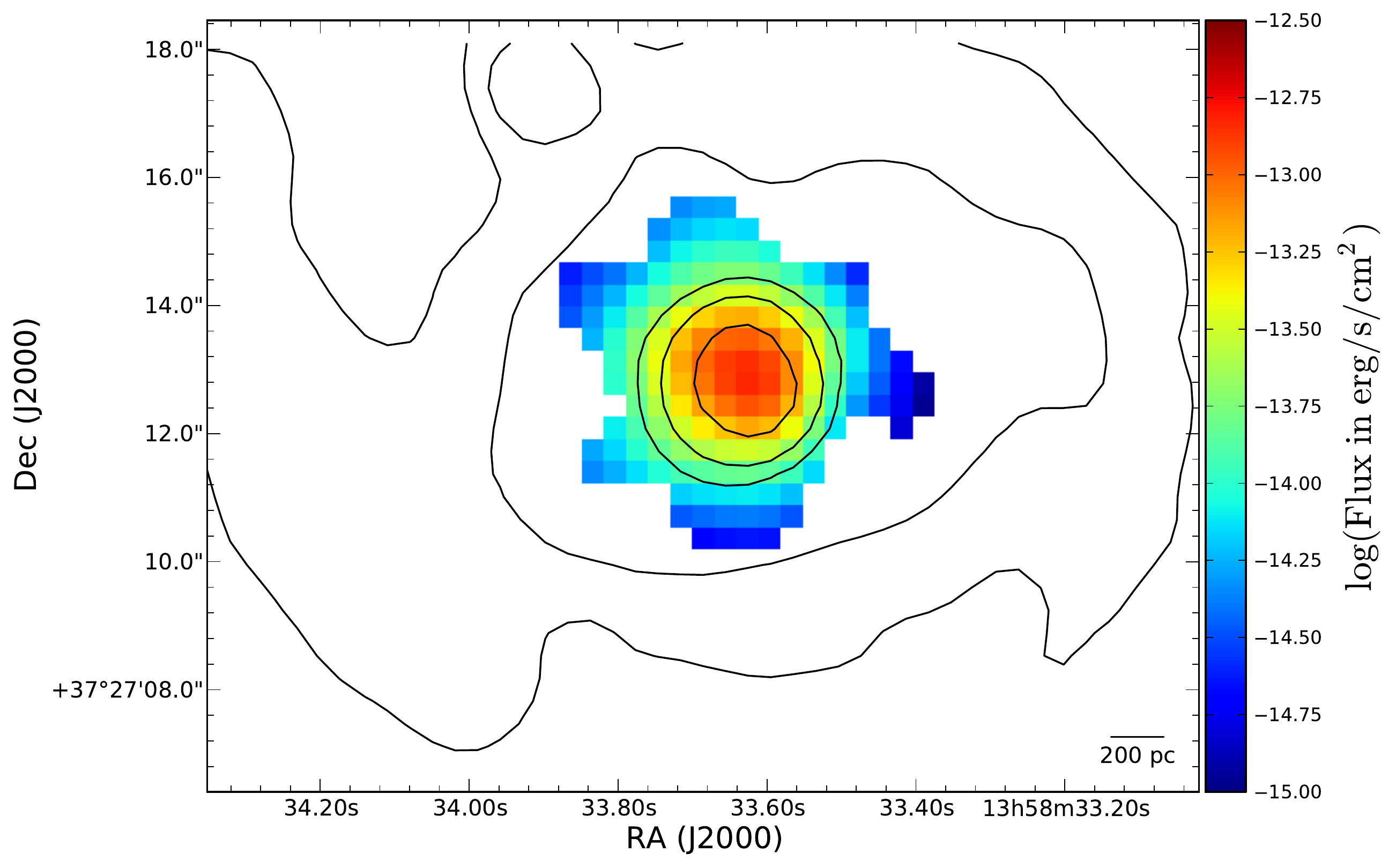}
   \caption{Observed H$\alpha$ emission for the single (top), primary (middle) and secondary (bottom) kinematic components. Overplotted contours show the red continuum emission.}\label{fig:flux_ha}
\end{figure}

\subsubsection{H$\alpha$ emission}
\label{ha_emission}
The H$\alpha$ emission map obtained from the single kinematic component fit is shown in the top panel of Fig.~\ref{fig:flux_ha}. Most of the \ha\ emission ($\sim98\%$ of the observed flux) within our FOV comes from the nuclear region, the inner $\sim2$\arcsec. The \ha\ intensity peak is spatially coincident with the photometric centre in the red-continuum map.  
This nuclear region is pear-like shaped in \ha, with the narrower end pointing NE. Some extended emission can be found towards the SE (with a secondary knot at $\sim4$\arcsec\ from the nucleus) and towards the N. The inner SW spiral arm shows substantial emission in \ha, with a bright knot located at $\sim7$\arcsec\ from the nucleus. Fainter knots are also present at  $\sim6$\arcsec\ from the nucleus (towards the NW and SE, roughly at the starting points of the inner SW and NE spiral arms, respectively). There is very little \ha\ emission coming from the NE inner arm within our FOV, in agreement with  \cite{kaufman}.

The high \ha\ emission at the nucleus is indicative of a starburst.
The total observed H$\alpha$ flux is $F(\ha)_{obs}=0.35\times 10^{-12}$~erg~s\me~cm$^{-2}$, which converts to $F(\ha)_{corr}=1.9\times 10^{-12}$~erg~s\me~cm$^{-2}$ after correction for internal dust extinction\footnote{The extinction has been estimated from the Balmer decrement using the \ha\ to \hb\ emission-line flux ratio, with fluxes measured over the CALIFA data. We have employed the \cite{seaton} reddening law with the \cite{howarth} parametrisation, assuming $R_V=3.1$. The average extinction in our FOV is $A(\ha)=1.7$~mag.}, by assigning to each INTEGRAL spaxel the estimated extinction at the closest CALIFA spaxel. 
This corrected  H$\alpha$ flux agrees within 10\% with the value obtained by \cite{roche} for the innermost region using CALIFA data, and corresponds to a total luminosity within our FOV of $L(\ha)_{corr}=5.7\times 10^{41}$~erg~s\me\ \citep[using a distance of 48.73~Mpc,][]{catalan-torrecilla}. SFR derived from this extinction-corrected H$\alpha$ luminosity is 2.95~\msun~yr\me\ \citep[from Eq.~4 of][]{catalan-torrecilla}. This value is in reasonable agreement with previous estimates for the innermost region \citep[3.3~\msun~yr\me,][assuming a  slightly larger distance of 52~Mpc]{roche}, and for the inner 36\arcsec \citep[2.76~\msun~yr\me,][]{catalan-torrecilla}, as most of the H$\alpha$ emission comes from the inner $\sim2-4$\arcsec.

The  middle and lower panels of Fig.~\ref{fig:flux_ha} show the H$\alpha$ intensity map corresponding to the primary and secondary kinematic components, respectively, with the same spatial and colour scales. The primary component resembles very closely the structure of the single component map, while the secondary one has more circular isophotes, and does not present the extended emission towards SE or N.  The amount of emission of the secondary component is similar to the primary one in the  inner 1.5-2\arcsec, and it is detected out to $\sim3$\arcsec\ from the nucleus (equivalent to $\sim0.8$~kpc) but the emission decays outwards much faster than the single/primary component (becoming ten times dimmer than the primary already at $\sim3$\arcsec\ from the centre).

It is interesting to notice that the integrated flux from the secondary broad component accounts for $\sim31$\% of the total \ha\ flux, which is a significant fraction. If we assume that the primary component is the one that traces the underlying star formation in the galaxy, the calculated rate would decrease to SFR=$2.2$~\msun~yr\me, which should be taken as a lower limit. The secondary over primary (broad over narrow) flux ratio is therefore 0.45 \citep[cf.][]{arribas2014}.
In the case of NGC~5394, if one assumes a unique kinematic component, this single component fit would miss up to $\sim10$\% of the total emitted \ha\ flux, that would remain {\em hidden}  in the emission line wings due to the presence of the broad secondary kinematic component.
\begin{figure}
   \centering
   \includegraphics[angle=0,width=8cm, clip=true]{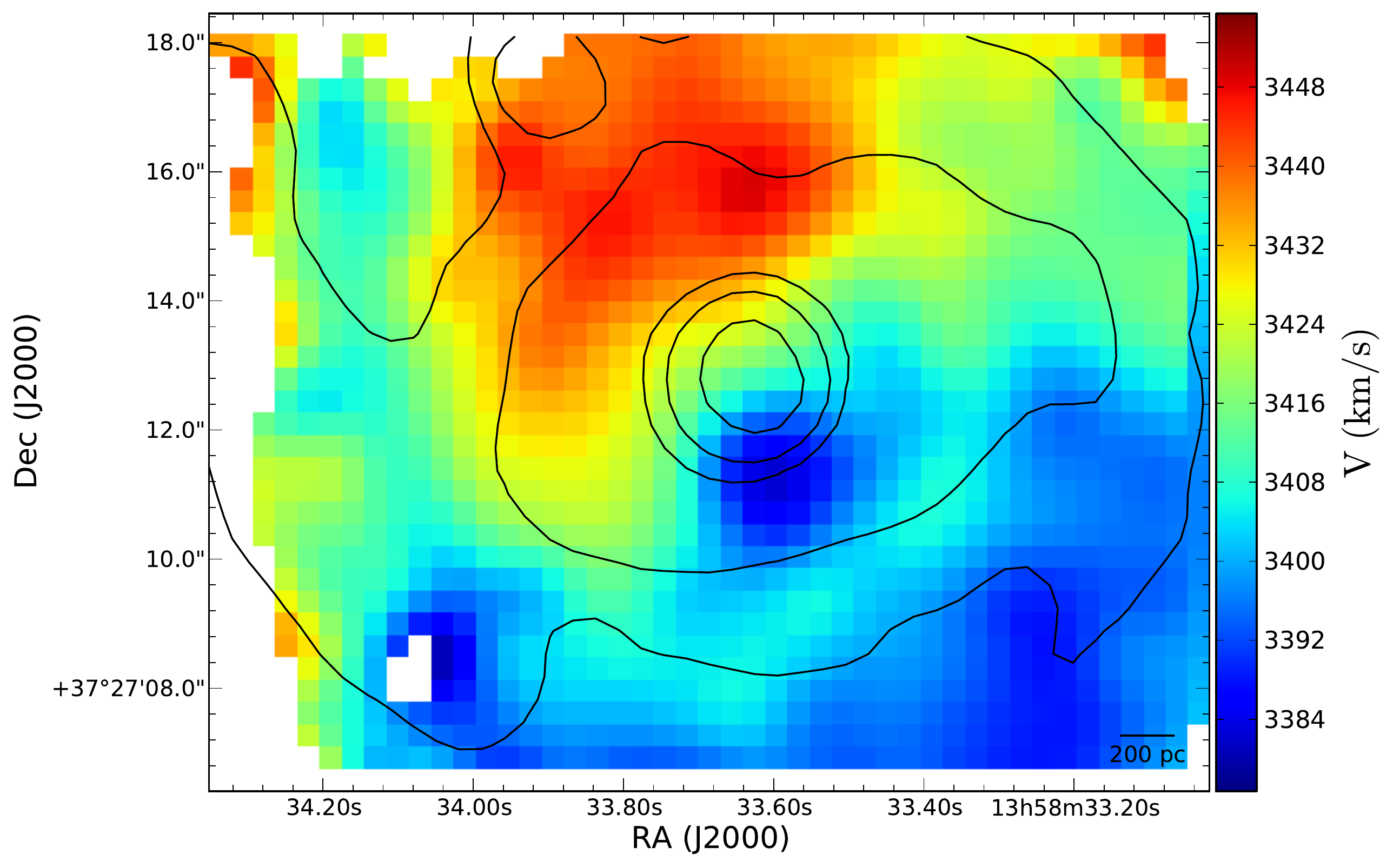}
   \includegraphics[angle=0,width=8cm, clip=true]{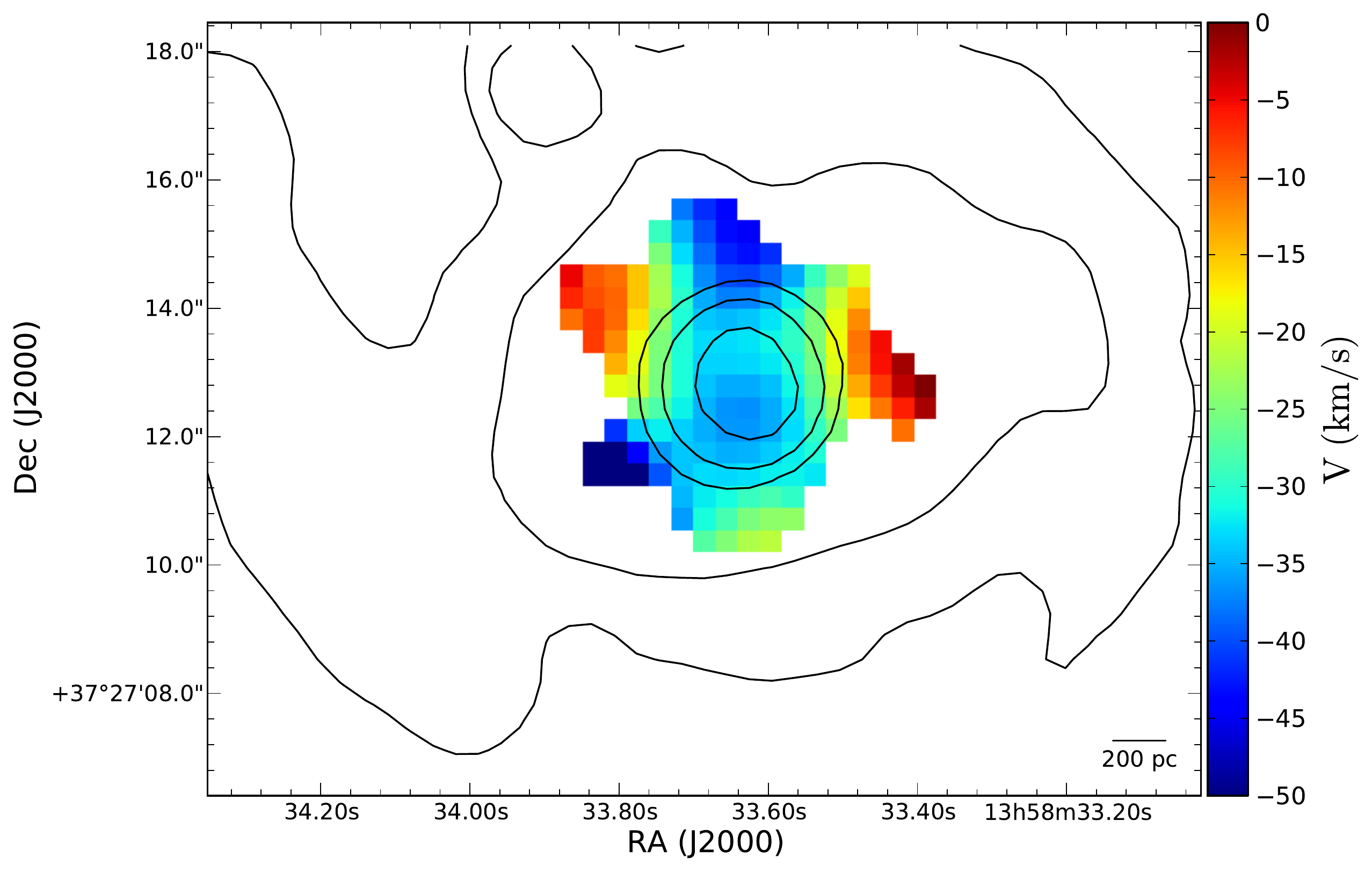}
   \caption{Observed velocity map from a single component fit to the emission lines (top), and residual velocity map for the secondary component (bottom). Overplotted contours show the red continuum emission.}\label{fig:havel}
\end{figure}
\begin{figure}
   \centering
   \includegraphics[angle=0,width=8cm, clip=true]{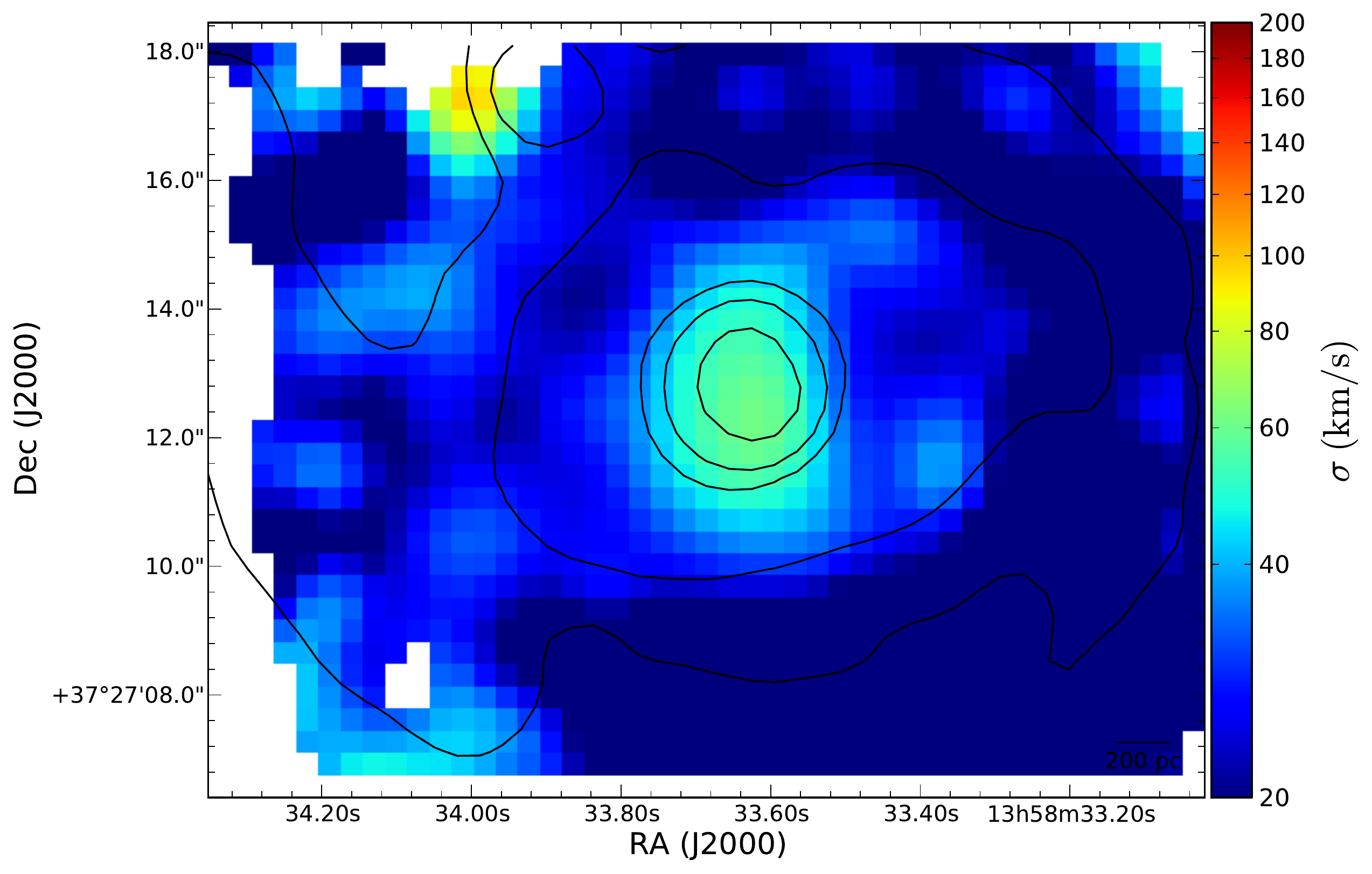}
   \includegraphics[angle=0,width=8cm, clip=true]{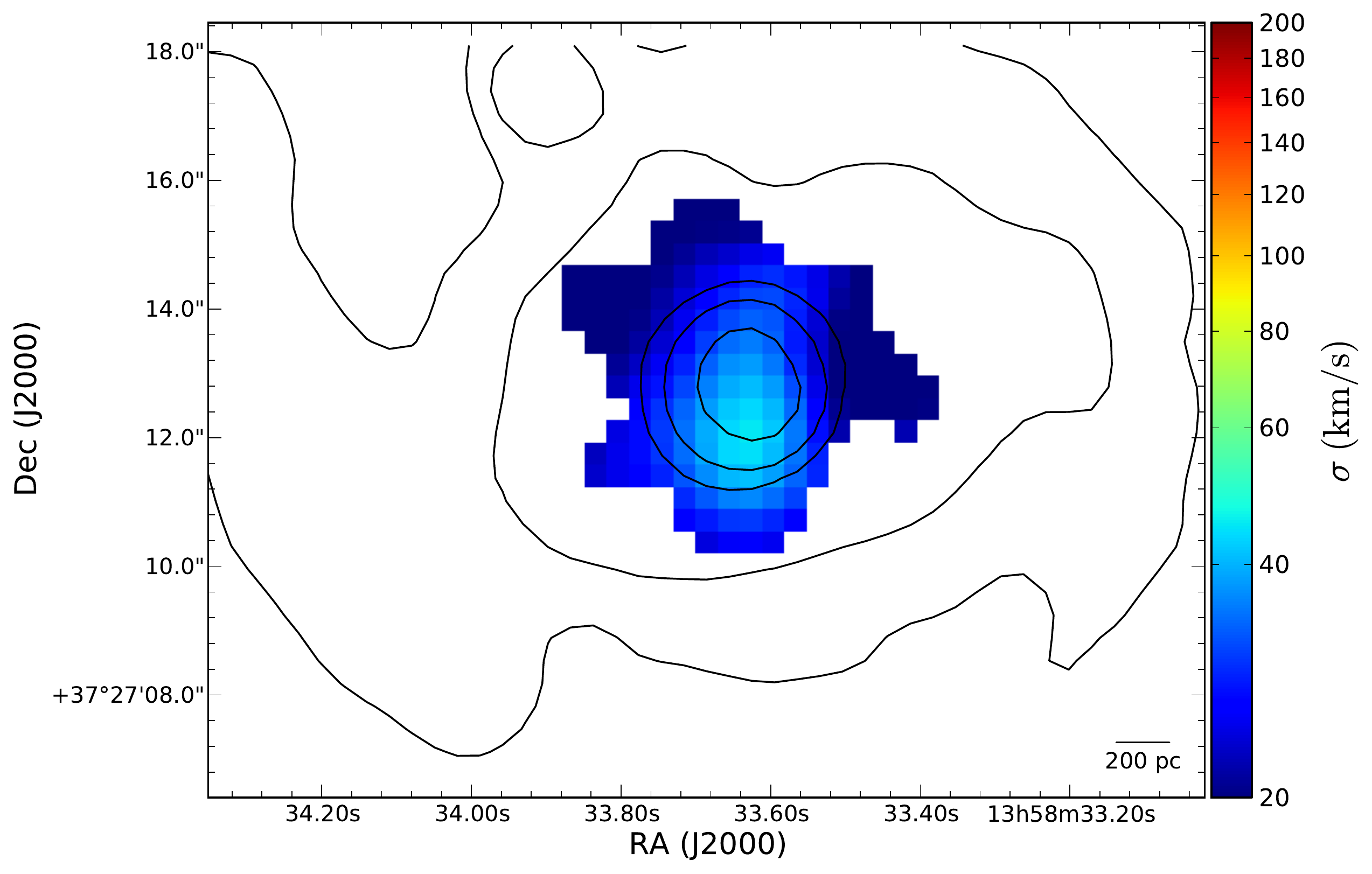}
   \includegraphics[angle=0,width=8cm, clip=true]{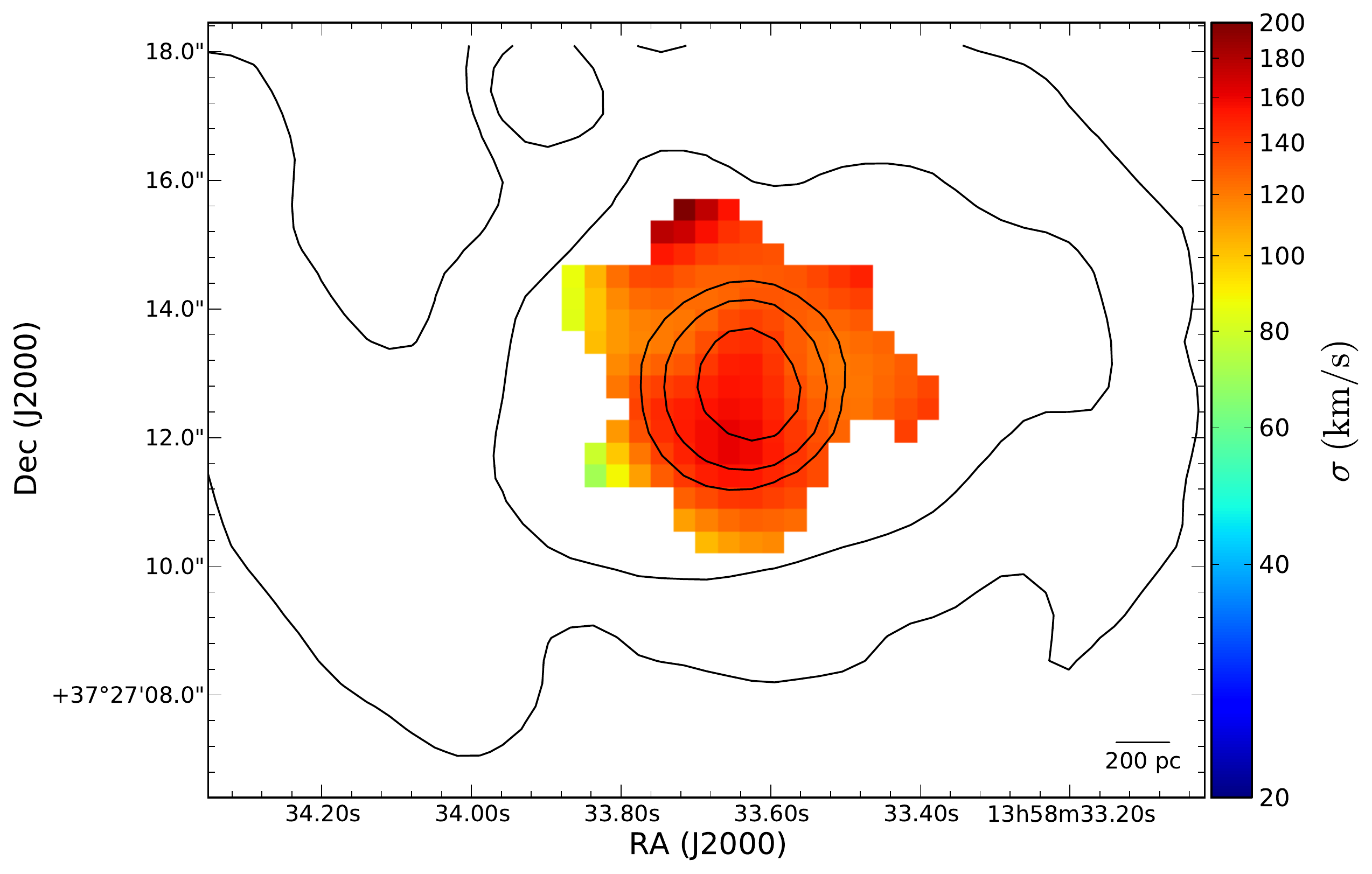}
   \caption{Observed velocity dispersion map for the single component fit to the emission lines (top), and for the primary (middle) and secondary (bottom) kinematic components. Overplotted contours show the red continuum emission.}\label{fig:havel2}
\end{figure}

\subsubsection{Ionised gas kinematics}
The velocity field  obtained from the single-component fit to  the emission lines is shown in Fig.~\ref{fig:havel}. It shows a rotational pattern, albeit rather disturbed in some locations. In particular, there is a rather prominent  redshifted  band towards the W-SW of the nucleus. This band is, as we will see below, spatially coincident with other structural features. Some streaming motions along the spiral arms are also apparent. 

In order to detect gas motions with respect to the expected regular rotation of the galaxy disc, we need to subtract the rotational velocity component along the line-of-sight, from the observed velocity field. Our small FOV and the presence of important non-circular motions prevent us from obtaining a reliable model velocity field for the disc rotation. We will use instead the velocity map obtained from the single-kinematic-component fit (Fig.~\ref{fig:havel}), as a bona fide tracer of the underlying disc gas kinematics with respect to which we will measure residual velocities. 
We are well aware that this map contains non-axisymmetric motions in the galaxy disc, but it will only be used to calculate residuals in the innermost 5\arcsec\ where it is reasonably regular.

\begin{figure}
   \centering
   \includegraphics[angle=0,width=8.5cm, clip=true]{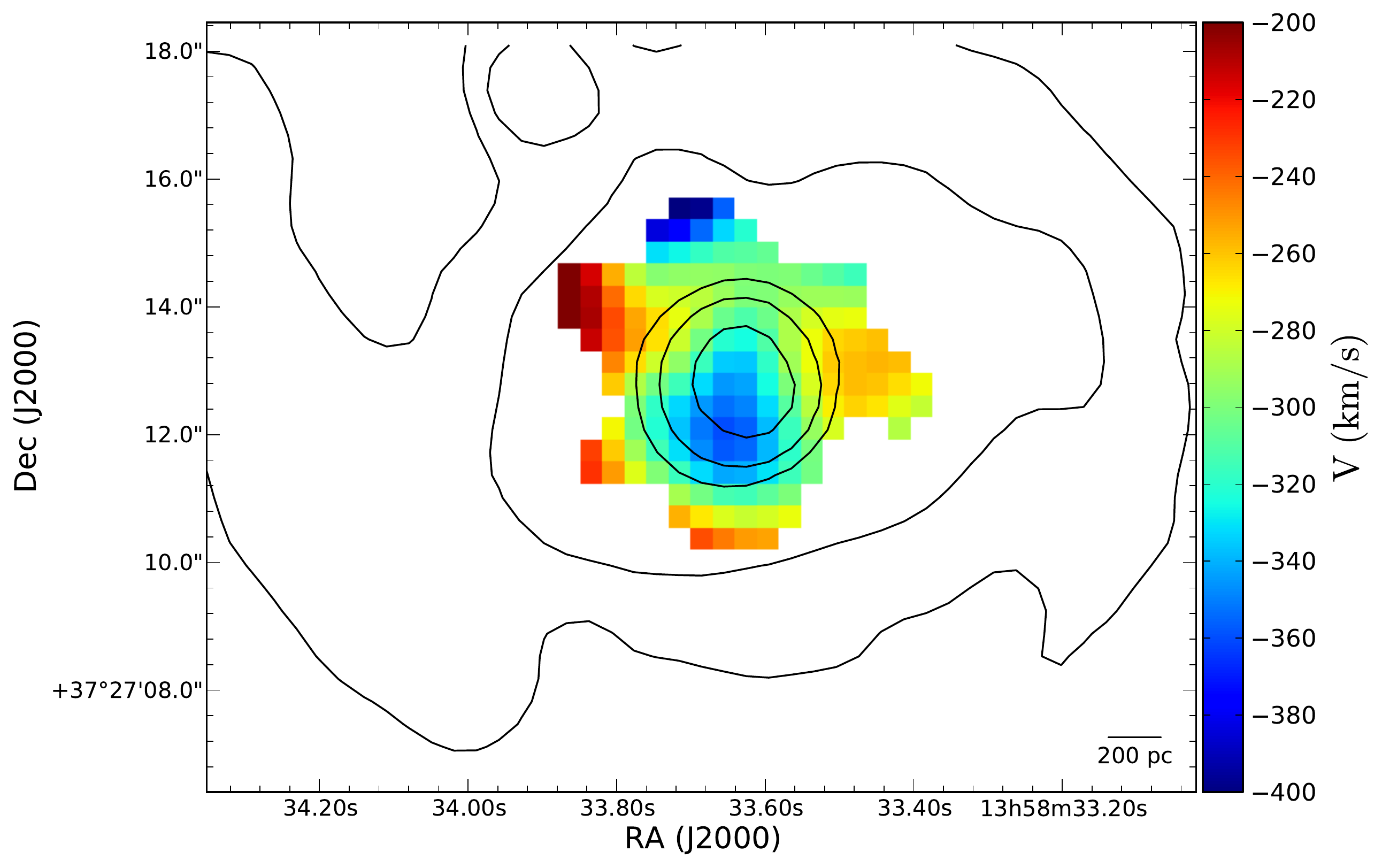}
   \caption{Extreme velocity (v$_{98}$) map of the emission lines for the secondary kinematic component. Overplotted contours show the red continuum emission.}\label{fig:v90ha}
\end{figure}

The velocity field of the primary component is practically indistinguishable from that of the single component, with a root-mean-square (rms) in the difference of order 5~km~s\me. The residual velocity for the secondary component, $v _{sec}$ (i.e. the underlying primary component velocity map has been subtracted from the  secondary component velocity field) is shown in Fig.~\ref{fig:havel} (bottom). It shows blueshifted velocities everywhere, with maximum values around $-80$~km~s\me\ and an average value of $\Delta v _{sec}=-30$~km~s\me. This value is similar to the values reported by \cite{arribas2014} for objects with comparable SFR, and it is most likely indicative of an outflow in this region of the galaxy. The region with largest residual velocities (<-30~km~s\me) resembles an arc-like structure towards the N and SE of the nucleus. 

The presence of a blue wing in the nuclear \ha\ emission line profile for NGC~5394  was already noticed by \cite{kaufman}, who interpreted it as a sign of an outflow in the galaxy nucleus, but it was not detected by \cite{roche}.  These residual velocities may seem, at first sight, too low for an important outflow. This will be discussed in Section~\ref{sec:Disc}.

In Fig.~\ref{fig:havel2} we show the velocity dispersion of the ionised gas for the single component fit. The average velocity dispersion is $\sim40$~km~s\me, and reaches $\sim60$~km~s\me\ in the nuclear region. These values of the velocity dispersion are typical of the most luminous extragalactic star forming regions \citep[e.g.][]{relano,ho2014,arribas2014}.  %LIRGs...casi que son tipicas de regiones de formacion estelar masiva en general...por que LIRGs
The region with maximum velocity dispersion is elongated along roughly  PA$\sim0^{\circ}$ and continues towards the W side of the nucleus forming a C-like structure. This  structure will appear again in the maps of some line ratios below. 

The velocity dispersion of the primary kinematic component follows closely the one obtained for the single component fit, but with somewhat lower values (average of $\sim30$~km~s\me\ and a maximum of $\sim40$~km~s\me). Thus, the dispersions measured using a fit to a single kinematic component are ``contaminated'' by the secondary component, producing higher dispersions. 
The secondary kinematic component presents much higher velocity dispersion (Fig.~\ref{fig:havel2}, bottom), with average values of $\sim130$~km~s\me.
These values cannot be interpreted in terms of thermal broadening, and are most likely indicative of (unresolved) disordered motions along the line of sight. The highest values (up to 190~km~s\me) are located around the nucleus but extending towards the N and SE regions, where the highest residual velocities are also found (Fig.~\ref{fig:havel}, bottom).

The velocity dispersions can be combined with the blueshifted residual velocities to produce a map of the most extreme blueshifted velocities of the ionised gas. This is shown in Fig.~\ref{fig:v90ha}, where we show $v_{98}=\Delta v_{sec}-2\sigma$ \citep[cf.][]{2013ApJ...768...75R}. We see that warm gas with velocities of $\sim-300$~km~s\me\ is present in the central region of this galaxy, with peaks of $\sim350$-400~km~s\me\ towards the N and SE of the nucleus.

\subsubsection{Emission line ratios}
\label{line_ratios}
The secondary-to-primary flux ratio for the [\nii]$\lambda$6583 and [\sii]$\lambda\lambda$6716, 6731 emission lines is larger (38\% and 41\% respectively) than for \ha\ ($\sim30$\%, Section~\ref{ha_emission}). Therefore, the [\nii]/H$\alpha$ and [\sii]/H$\alpha$ line ratios are considerably higher for the secondary component than for the primary. This implies that fluxes obtained from a single kinematic component fit, would underestimate the total [\nii]$\lambda$6583 and [\sii]$\lambda\lambda$6716, 6731 emission by 12\% and 15\%, respectively.
\begin{figure}
   \centering
   \includegraphics[angle=0,width=8cm, clip=true]{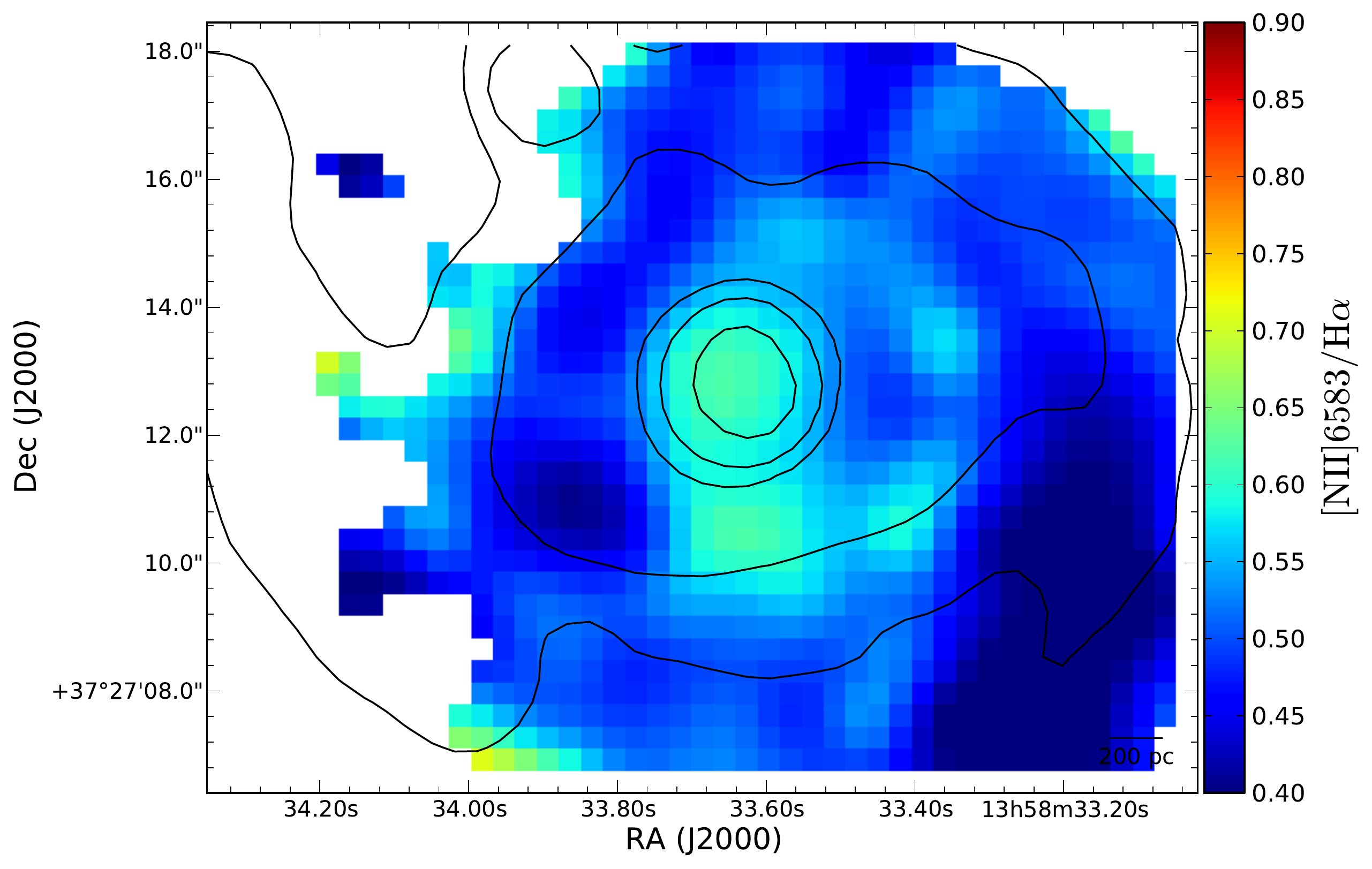}
   \includegraphics[angle=0,width=8cm, clip=true]{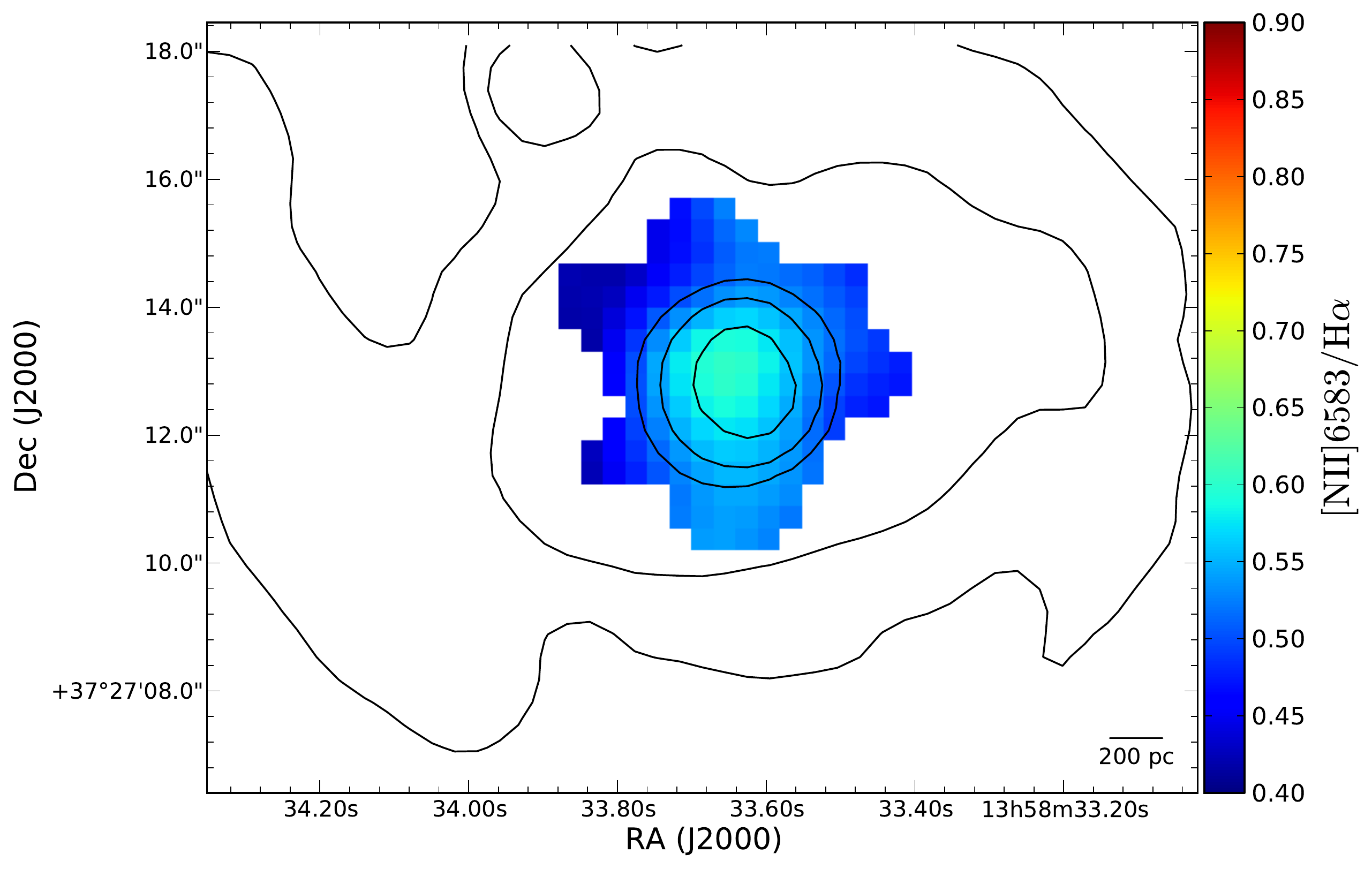}
   \includegraphics[angle=0,width=8cm, clip=true]{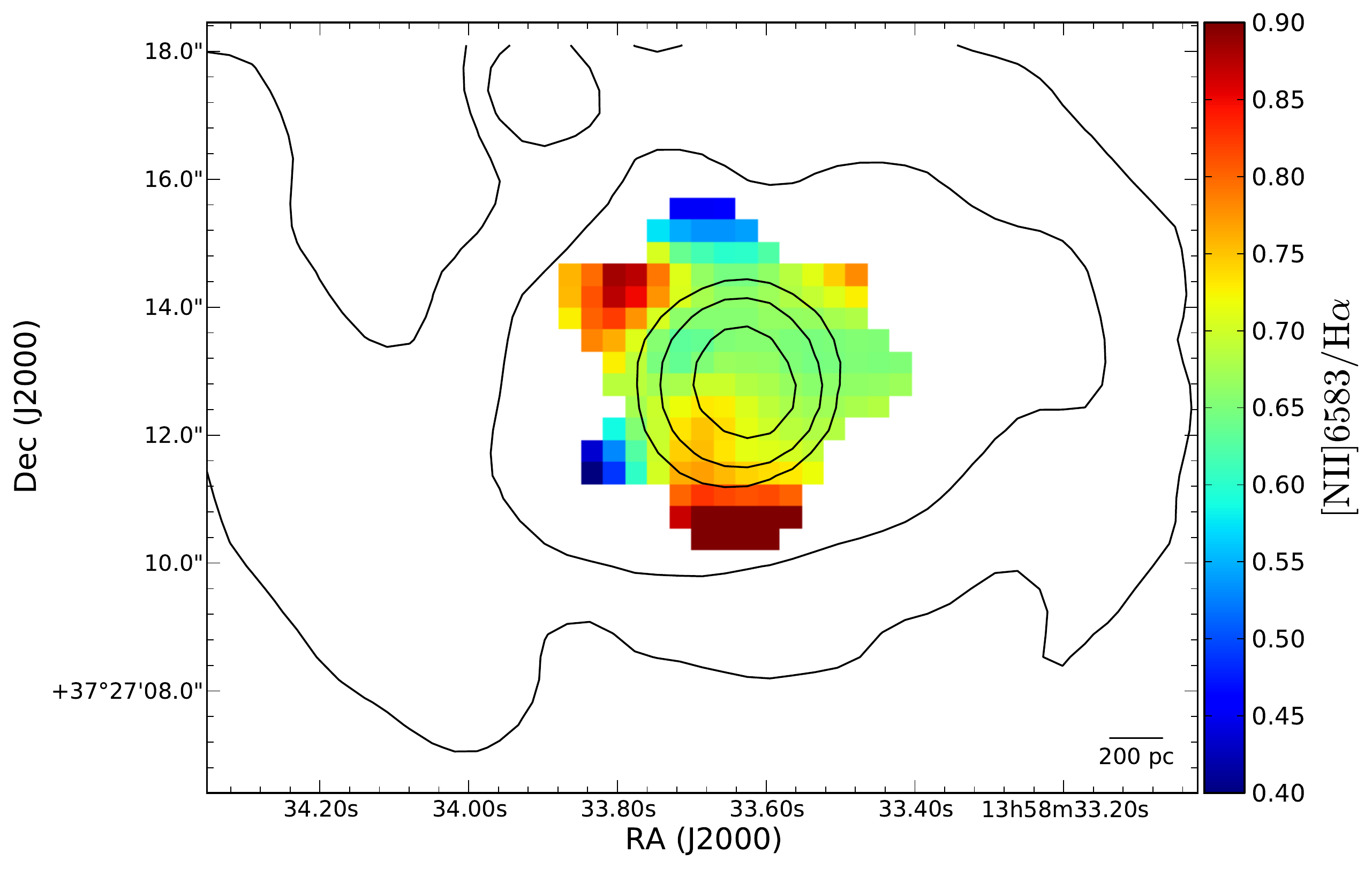}
   \caption{Map of the [\nii]$\lambda6583$/H$\alpha$ ratio for the single (top), primary (middle) and secondary (bottom) component. Overplotted contours show the red continuum emission.}\label{fig:NH1}
\end{figure}

\begin{figure}
   \centering
   \includegraphics[angle=0,width=8cm, clip=true]{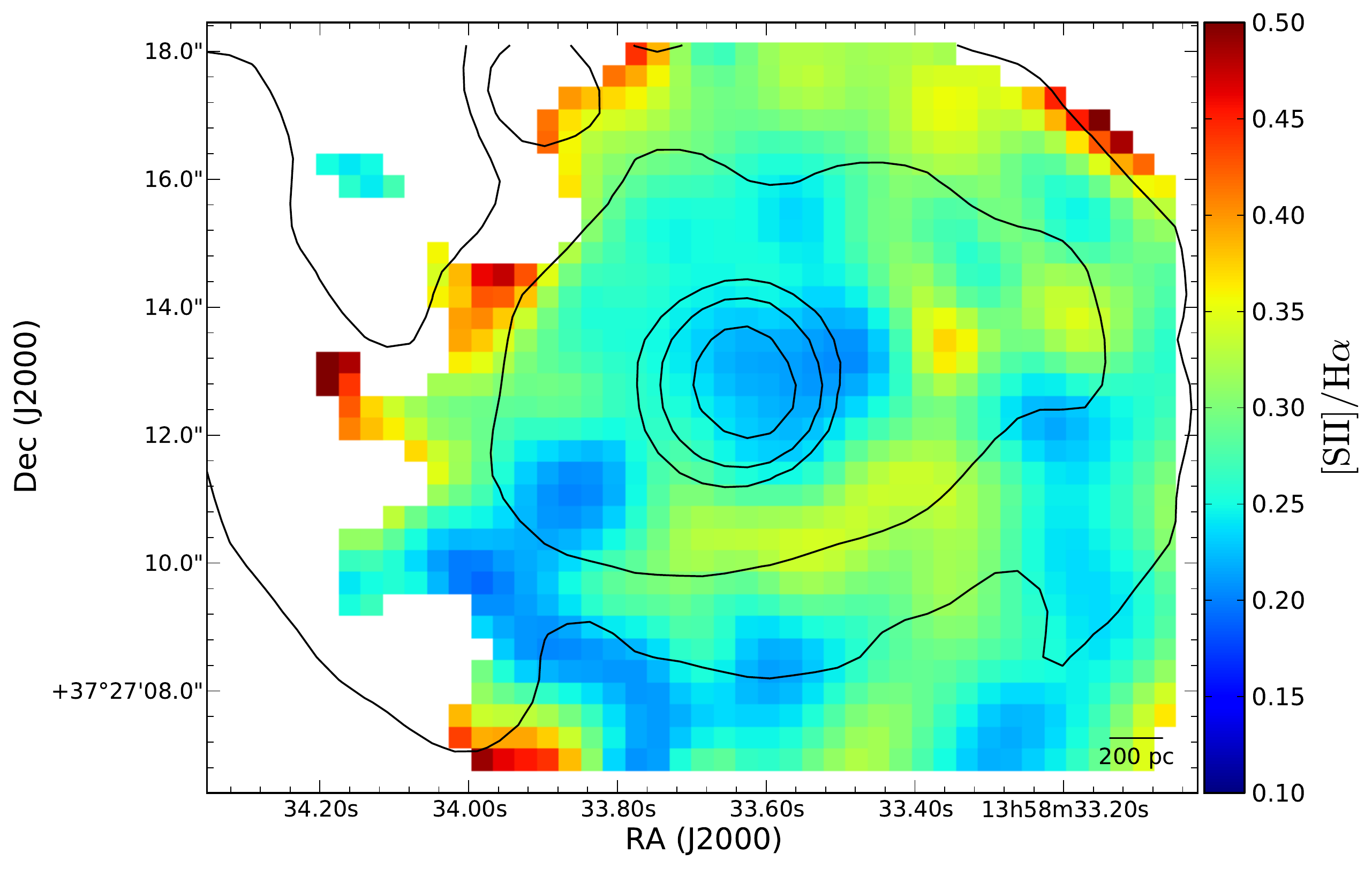}
   \includegraphics[angle=0,width=8cm, clip=true]{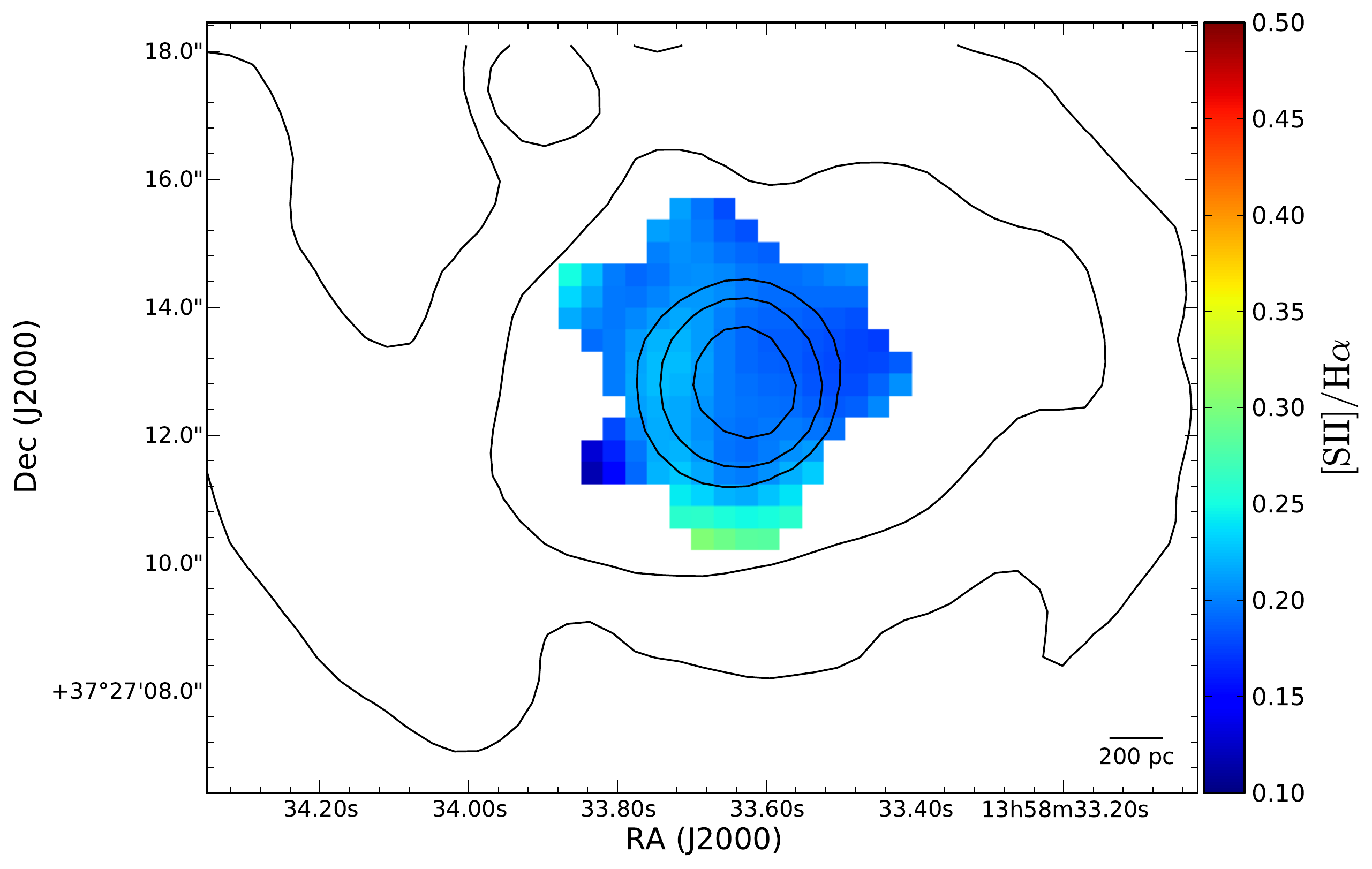}
   \includegraphics[angle=0,width=8cm, clip=true]{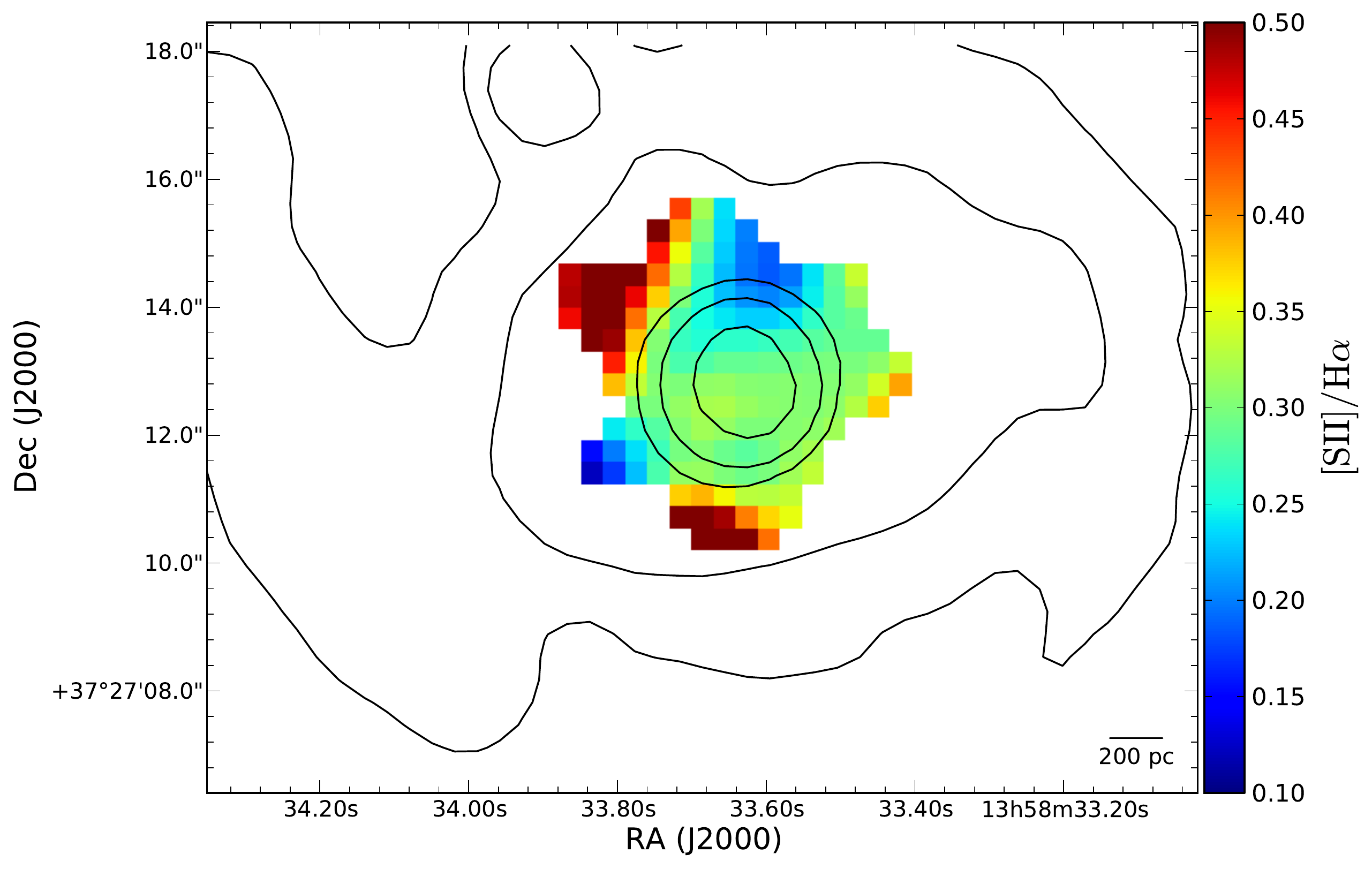}
   \caption{Map of the [\sii]/H$\alpha$ ratio for the single component fit (top), and for the primary (middle) and secondary (bottom) kinematic components. Overplotted contours show the red continuum emission.}\label{fig:SH1}
\end{figure}

\begin{figure*}
   \centering
   \includegraphics[angle=0,width=15cm, clip=true]{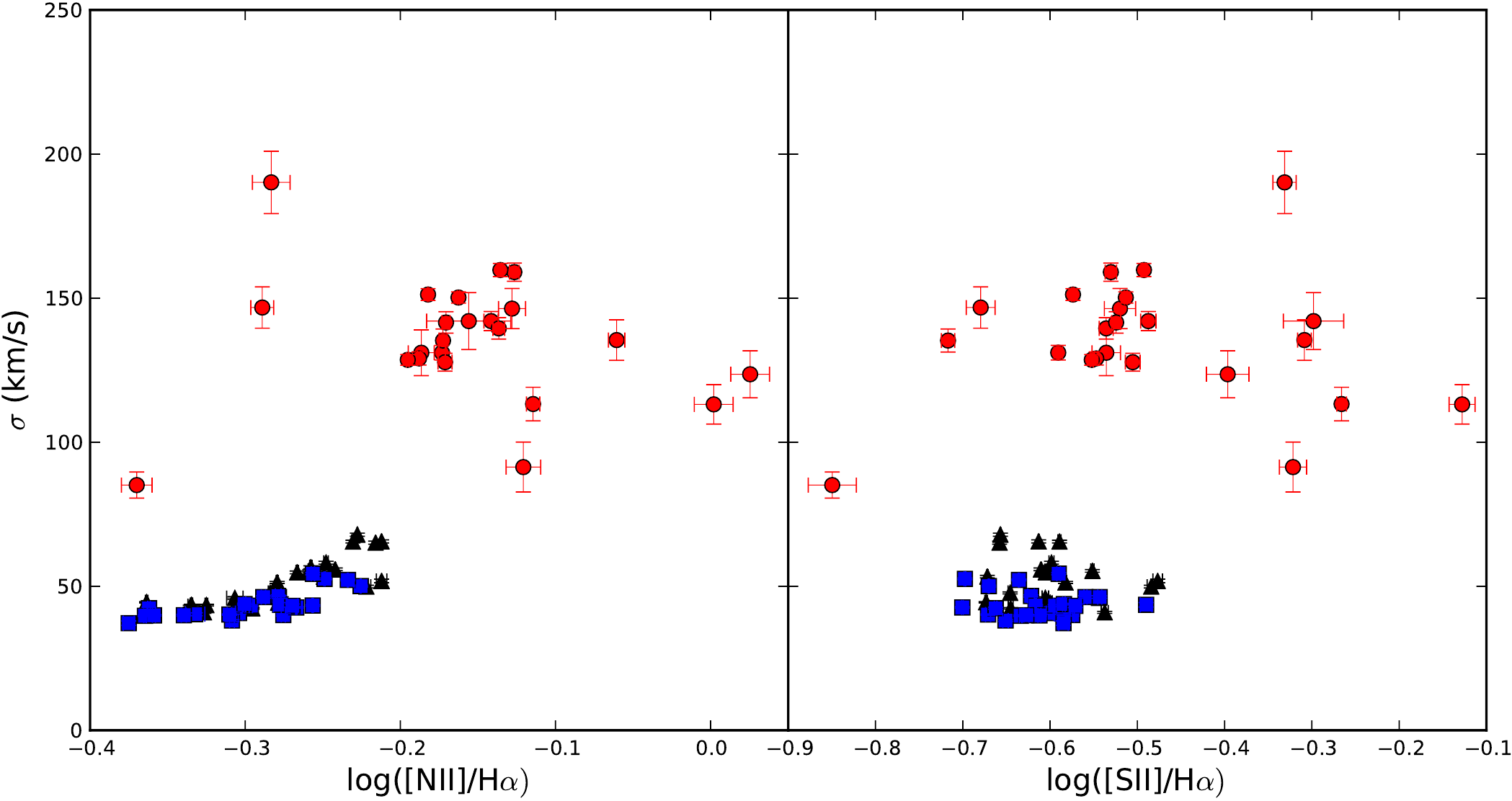}
   \caption{[\nii]/H$\alpha$ (left) and [\sii]/H$\alpha$ (right) ratios vs the velocity dispersion for the three kinematic components. Black triangles, blue squares and red circles represent the single, primary and secondary component respectively.}\label{fig:NIISIIvssig}
\end{figure*}

We have produced maps for the [\nii]/H$\alpha$ and [\sii]/H$\alpha$ line ratios for the the single, primary and secondary components (Figs.~\ref{fig:NH1} and \ref{fig:SH1}). 
The  [\nii]/H$\alpha$  map for the single component fit has an average value of $0.54\pm0.05$, matching well the result by \cite{roche}. The higher spatial re\-so\-lu\-tion of our data, allows us to detect the presence of some structure in the inner region. Enlarged ratios of up to $\sim0.65$ can be seen around the galaxy centre, extending towards the western side of the nucleus into a C-like structure  which resembles the one found in the velocity dispersion map for this primary component, and which is also spatially coincident with the redshifted band in the velocity field of the ionised gas. 

Thus, we observe a spatial correlation  between the regions with large velocity dispersion and enhanced [\nii]/H$\alpha$ for the single component fit. This correlation is even more e\-vi\-dent  in the secondary component. While the primary component has values very similar to (although slightly lower than) the single one (with an average ratio of 0.51), the secondary, broad component, has a much higher ratio which averages to 0.71, with values as high as 0.8. 
High  [\nii]/H$\alpha$ ratios are indicative of ionisation by shocks. This fact, combined with the blueshifted residual velocities and the higher velocity dispersion in the secondary velocity component, reinforces the idea that it traces an outflow from the nuclear region of NGC~5394.

Fig.~\ref{fig:NIISIIvssig} shows the  velocity dispersion (in km~s\me) vs. the [\nii]/H$\alpha$ ratio  for the three kinematic components, for the spaxels where the kinematic decomposition was possible. This plot shows the above correlation very clearly for each kinematic component, but specially among different components. The highest ratios are present in the secondary broad component. This correlation is well known in (U)LIRGS \citep[cf.][]{arribas_colina02, monreal06, rich2011}, but also in normal galaxies \citep{ho2014} and it is a clear indication of shock ionisation.

The [\sii]/H$\alpha$ map for the single component fit (Fig.~\ref{fig:SH1}) also shows enhanced ratios in an arc-like structure towards the SW of the nucleus. This region coincides in part with the regions of enhanced [\nii]/H$\alpha$ and higher velocity dispersion (the C-like structure mentioned above).  The average value of [\sii]/H$\alpha$ in the single component fit map is 0.26, very similar to the mean value of the primary component. On the contrary, the secondary broad component has higher ratios, averaging to 0.35 (but with a higher scatter, reaching values over 0.5). As for the [\nii]/H$\alpha$ ratio, Fig.~\ref{fig:NIISIIvssig} (right) shows that the  kinematic decomposition is able to discriminate between the primary (narrow) component, with ratios compatible with photoionisation by stars, and the  secondary (broad), indicative of a harder radiation field strongly contaminated by shocks.  In contrast with the [\nii]/H$\alpha$ ratio and the velocity dispersion map, the [\sii]/H$\alpha$ ratio has a minimum value at the nucleus. This fact makes the correlation of [\sii]/H$\alpha$ and velocity dispersion very weak and noisy for the primary and single components  although it is still clear among components (cf. Fig.~\ref{fig:NIISIIvssig}, right).

\begin{figure}
   \centering
   \includegraphics[angle=0,width=8cm, clip=true]{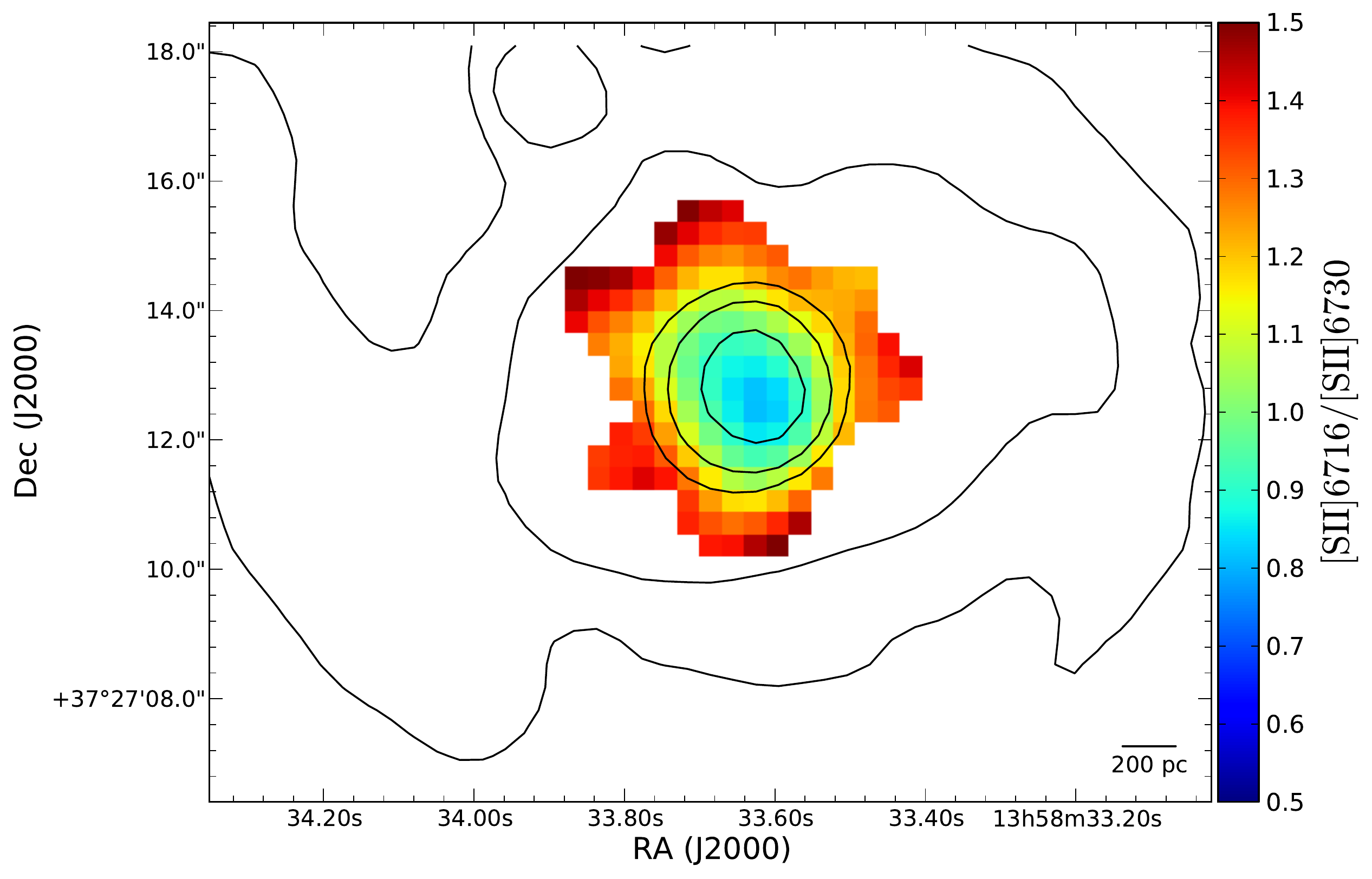}
   \includegraphics[angle=0,width=8cm, clip=true]{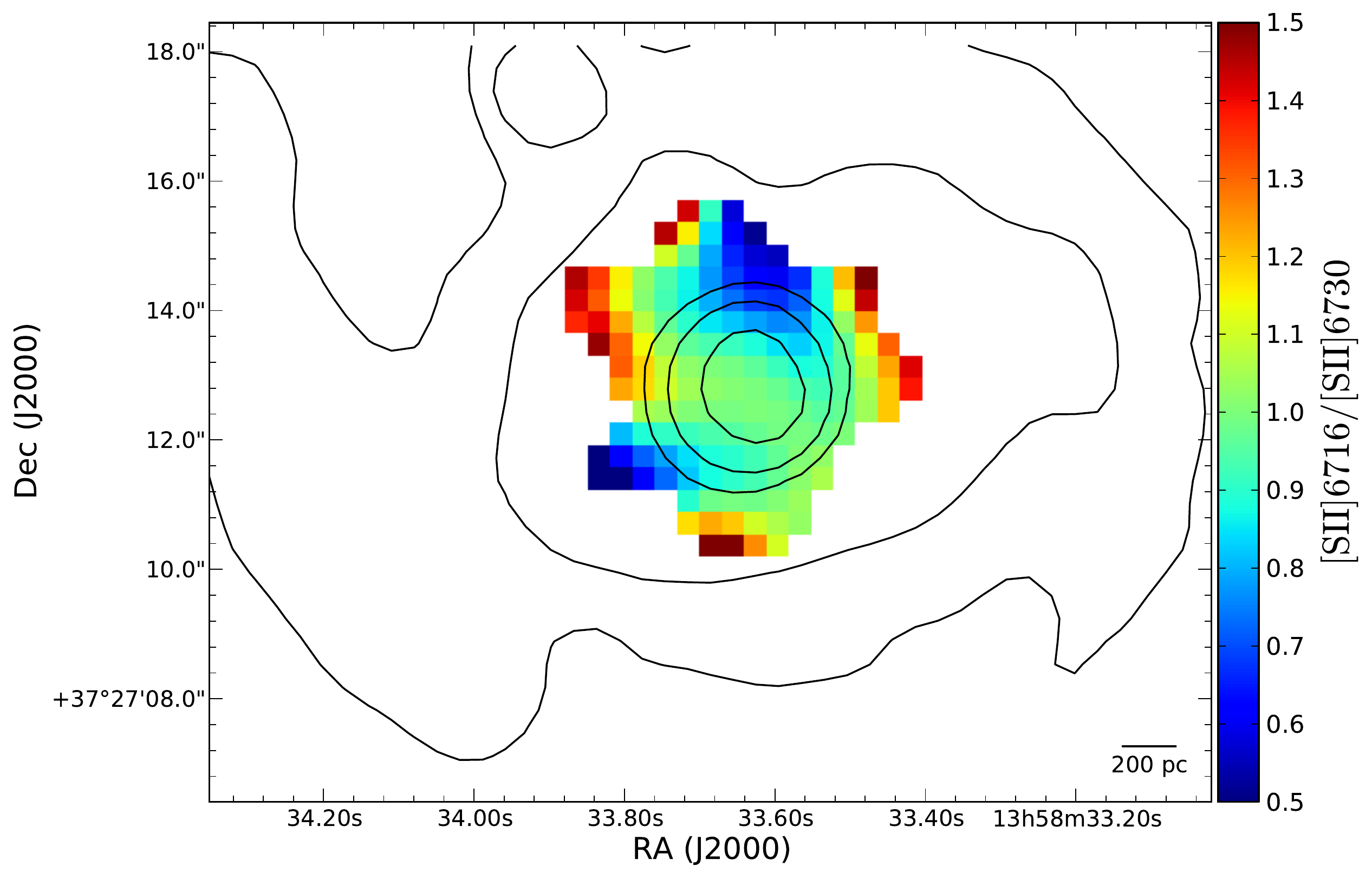}
   \caption{Map of the [\sii]$\lambda$6716/[\sii]$\lambda$6730 ratio for the  primary (top) and secondary (bottom) kinematic components. Overplotted contours show the red continuum emission.}\label{fig:S1S2}
\end{figure}

We have also generated maps (Fig.~\ref{fig:S1S2}) of the [\sii]$\lambda$6716/[\sii]$\lambda$6731 emission-line ratio,  that traces the electron density in the ionised gas. As for the previous emission-line ratio maps, the one obtained from the single and primary component fits are very similar. However, the map for the secondary component presents both different values and a different spatial distribution to the primary component fit, confirming again that these are two physically distinct components. In particular, the [\sii]$\lambda$6716/[\sii]$\lambda$6731 ratio has an average value of only 1.0 in the secondary component map, corresponding to electron densities of $\sim500$~cm$^{-3}$  (assuming a temperature of $10^4$~K), much higher than the average value derived from line ratio of the primary (and single) component fit ($\sim175-200$~cm$^{-3}$ for the same temperature). In addition, the secondary component map presents a different structure. The primary component shows a steep radial gradient, with a peak value of $\sim900$~cm$^{-3}$ at the galaxy centre, while the secondary componet map shows a more uniform distribution.

The [\oi]$\lambda$6300 line is also detected mainly in the region where the secondary component is detected.  Unfortunately, the S/N for this line does not permit to perform the kinematic decomposition.

Table~\ref{table:ionised} summarises all measured properties and relevant emission line ratios for the different kinematic components.

\begin{table*}
\begin{minipage}{140mm}
\centering
\caption{Measured physical properties for the ionised gas forthe different kinematic components of the emission lines.}
\label{table:ionised}
\begin{tabular}{l c c c  c}     % 7 columns
\hline
Parameter                                     & Single  & Primary  & Secondary   & {\em Notes}       \\[1mm]
                                              &         &          &             &                  \\[1.5mm]
\hline
$F(H\alpha)_{obs}$ (erg~s\me~cm$^{-2}$)               &    $3.5\times10^{-13}$ & - & -     & (a)   \\
$L(H\alpha)_{corr}$ (erg~s\me)                        &    $5.7\times10^{41}$ & - & -     & (b)  \\
SFR ($M_\odot$~yr\me)                                 &    2.95       &  2.2  &  -            & (c)  \\
$\Sigma_{SFR}$ ($M_\odot$~yr\me~kpc$^{-2}$)           &         &  2.8  &               &      \\
$A(\ha)$                                              &    1.7  &  -    & -             &  (d) \\
$(F_{comp}$/$F_{total})_{H\alpha}$                    &    0.91  &  0.69   & 0.31          &  (e) \\ 
$(F_{comp}$/$F_{total})_{[NII]6583}$                  &    0.88  &  0.62   & 0.38    &  (e) \\ 
$(F_{comp}$/$F_{total})_{[SII]}$                      &    0.85  &  0.59   & 0.41    &  (e) \\ 
$\mbox{[\sii]}\lambda\lambda6716,6731$/H$\alpha$      &  $0.26\pm0.03$   &    $0.25\pm0.03$       &  $0.35\pm0.13$  &  (f) \\ 
$\mbox{[\nii]}\lambda6583$/H$\alpha$                  &  $0.54\pm0.05$   &    $0.51\pm0.05$       &  $0.71\pm0.14$  &  (f) \\ 
$\mbox{[\sii]}\lambda6716/\mbox{[\sii]}\lambda6731$   &  $1.23\pm0.19$   &    $1.25\pm0.19$       &  $1.02\pm0.31$  &  (f) \\ 
$\sigma$ (km~s\me)                              &     $40\pm10$    &     $30\pm10$          &  $130\pm20$   &  (f)  \\
$\sigma_{max}$ (km~s\me)                        &     $\sim60$     &       $\sim40$         &  $\sim190$    &  (g)  \\
\hline
\end{tabular}
 {\footnotesize \begin{tabular}{l}
$^{a}$  Observed \ha\ flux within the INTEGRAL FOV.\\
$^{b}$  Extinction-corrected \ha\ luminosity within the INTEGRAL FOV, assuming a distance for NGC~5394 \\ 
 of 48.73~kpc \citep{catalan-torrecilla}.\\
$^{c}$  SFR using the Eq.~4 in \cite{catalan-torrecilla}, and considering only the \ha\ emission from the primary \\   
kinematic component.\\
$^{d}$  Average extinction at \ha\ across the INTEGRAL FOV.\\
$^{e}$  Ratio of flux contained in the single, primary or primary component (as indicated) with respect to the total  \\ 
flux emitted in \ha, [\nii]$\lambda$6583 and [\sii]$\lambda\lambda$6716, 6731.\\
$^{f}$  Average value for each kinematic component and each parameter within the INTEGRAL FOV.\\
$^{g}$  Maximum measured value for each kinematic component.\\
\end{tabular}}
\end{minipage}
\end{table*}

\begin{figure*}
   \centering
   \includegraphics[angle=0,width=15cm, clip=true]{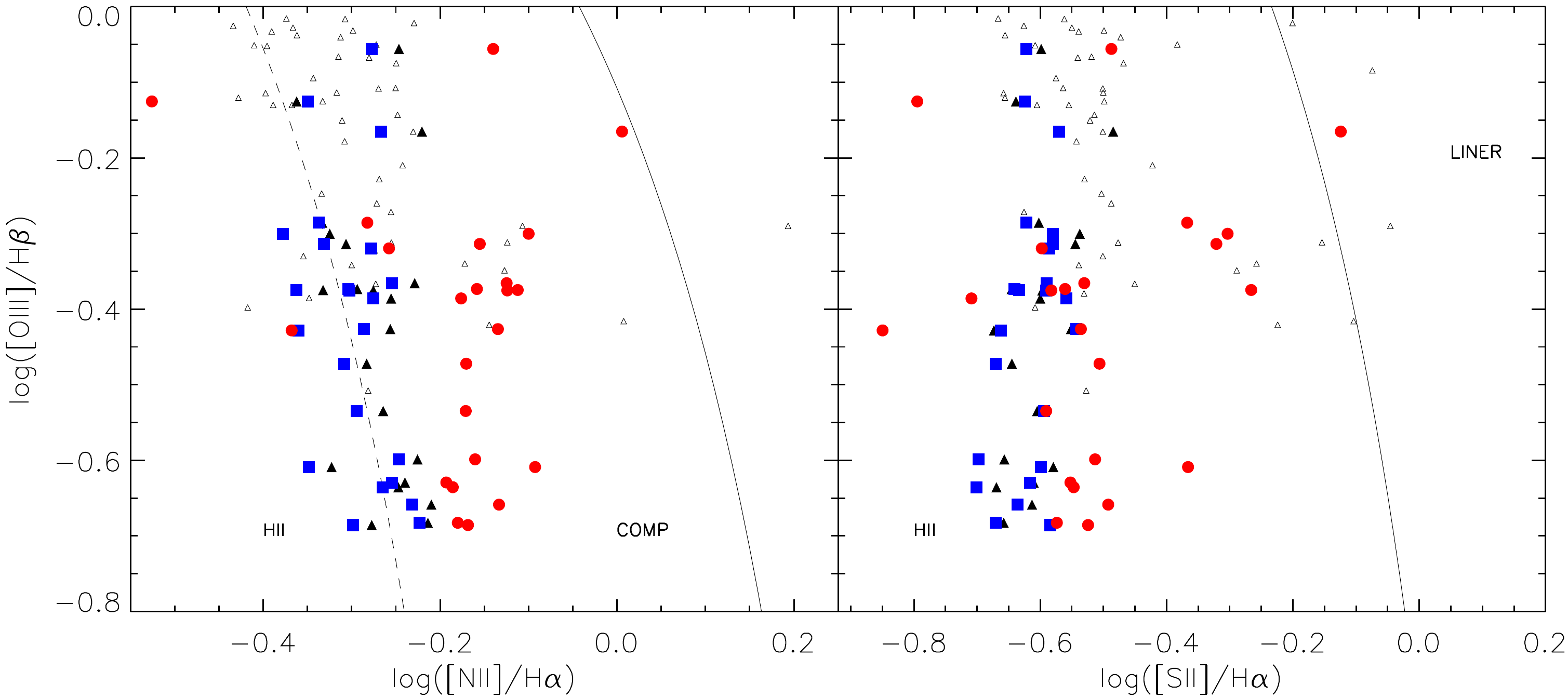}
   \caption{BPT diagrams of the $\log$([\oiii]$\lambda$5007/\hb) versus $\log$([\nii]$\lambda$5383/\ha) (left) and  $\log$([\oiii]$\lambda$5007/\hb) versus $\log$([\sii]$\lambda\lambda$6716, 6731/\ha) for all spaxels. The line ratios are color-coded for the spaxels in which a careful kinematic decomposition (as described in Section~\ref{sec:kinem}) was performed: blue squares correspond to the primary component, red circles for the secondary and black solid triangles for the single component fit.  Open small triangles correspond the single component fit in spaxels in which the kinematic decomposition has not been performed. The [\oiii]$\lambda$5007/\hb\ ratio has been obtained from lower spectral resolution data and correspond in all cases to a single component fit (see text for details).}\label{fig:BPT}
\end{figure*}

\citet[][ BPT]{BPT}  diagrams are powerful tools to diagnose ionisation processes \citep[e.g.][]{2006MNRAS.372..961K}. BPT diagrams for the three kinematic components in different spaxels in NGC~5394 are shown in Fig.~\ref{fig:BPT}. As our wavelength range does not cover neither H$\beta$ nor [\oiii]$\lambda5007$, we have used CALIFA data to measure these emission-line fluxes.  However, the lower spectral resolution of the CALIFA data does not allow to make the kinematic decomposition. Therefore, the y-axis in these diagrams should be taken with caution, as they all belong to the flux ratio for a single-component fit, and vertical shifts could likely occur considering two kinematic components. In this plot, therefore, the three kinematic components at each spaxel share the same value for the [\oiii]/H$\beta$ ratio. Despite this hindrance, the split between the single/primary components and the secondary one is very clear, in particular for the [\nii]$\lambda$5383/\ha\ ratio. The ratios of the secondary component are clearly indicative of strong influence of shocks. This plot also clearly shows slight horizontal shifts between the primary and single components, showing how the latter is a ``weighted average'' between the primary component (most likely tracing the underlying gas ionised by recently formed stars) and the secondary component (which is, in this case, tracing the gas in the galaxy wind). Although this shift is not too extreme for most spaxels, it is important in some of them, and it is indicative of how fluxes hidden in low intensity broad components can alter line ratios when assuming a single kinematic component.

Fig.~\ref{fig:BPT} highlights  the  importance of good spectral resolution and kinematic decomposition in interpreting line ratios. High values from a single kinematic component fit can come from a hidden, kinematically distinct component, which does not trace the underlying gas in the star forming regions, but gas with a different origin and even at different locations along the line-of-sight. 

 The above comparison between the single and double kinematic component fit to the emission lines stresses the importance of higher resolution spectra and the kinematic decomposition for a proper interpretation of the measured kinematics and line ratios, as the single component fit can lead to quite different conclusions. In this particular object, the lower resolution of the data analysed by \cite{roche} did not allow them to perform a kinematic discrimination, which prevented them to detect the shocked (and slightly blueshifted component). This lead them to find no signs of the outflow from their emission line analysis.

\subsection{Neutral gas phase}

\begin{figure}
   \centering
   \includegraphics[angle=0,width=8cm, clip=true]{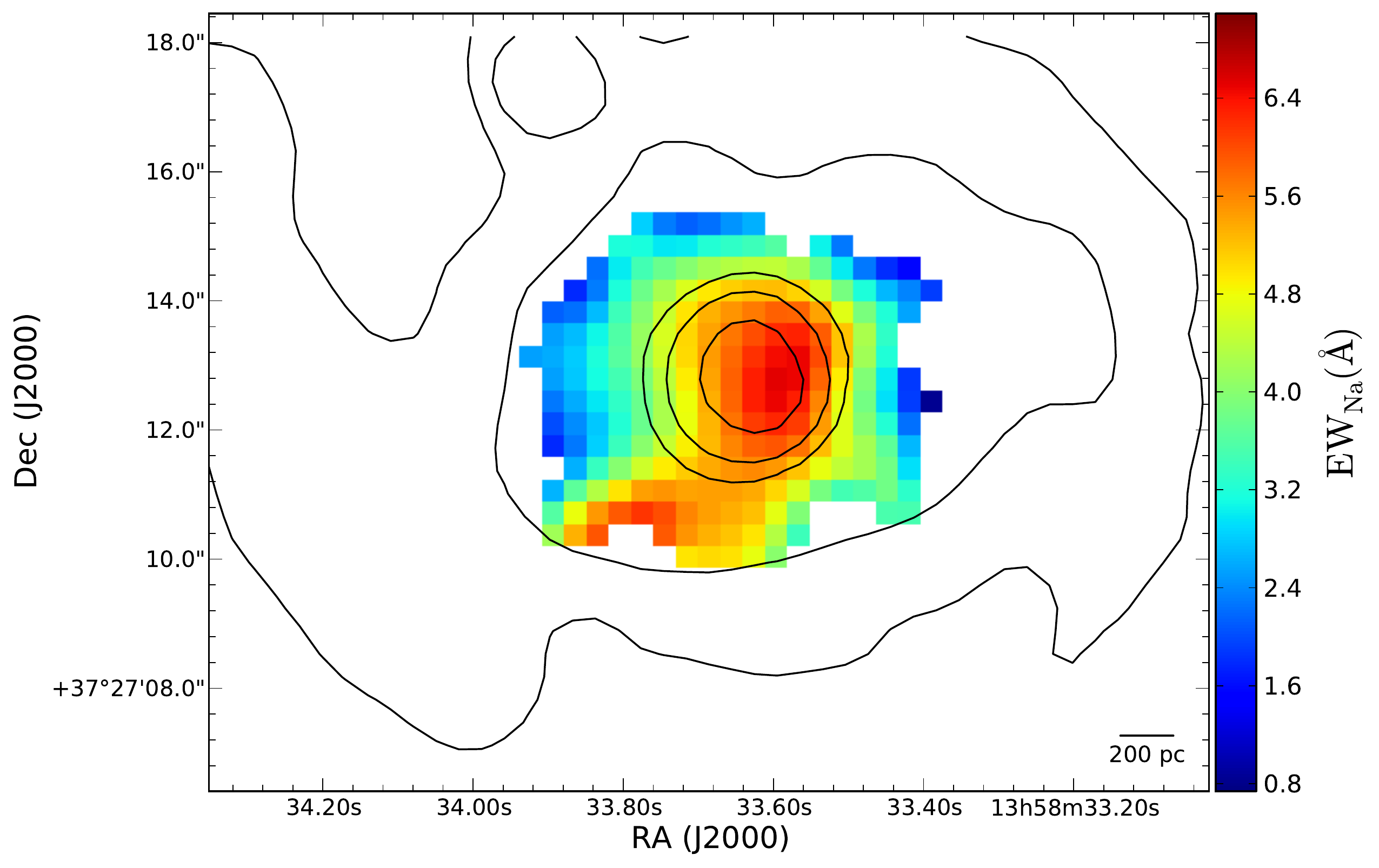}
   \includegraphics[angle=0,width=8cm, clip=true]{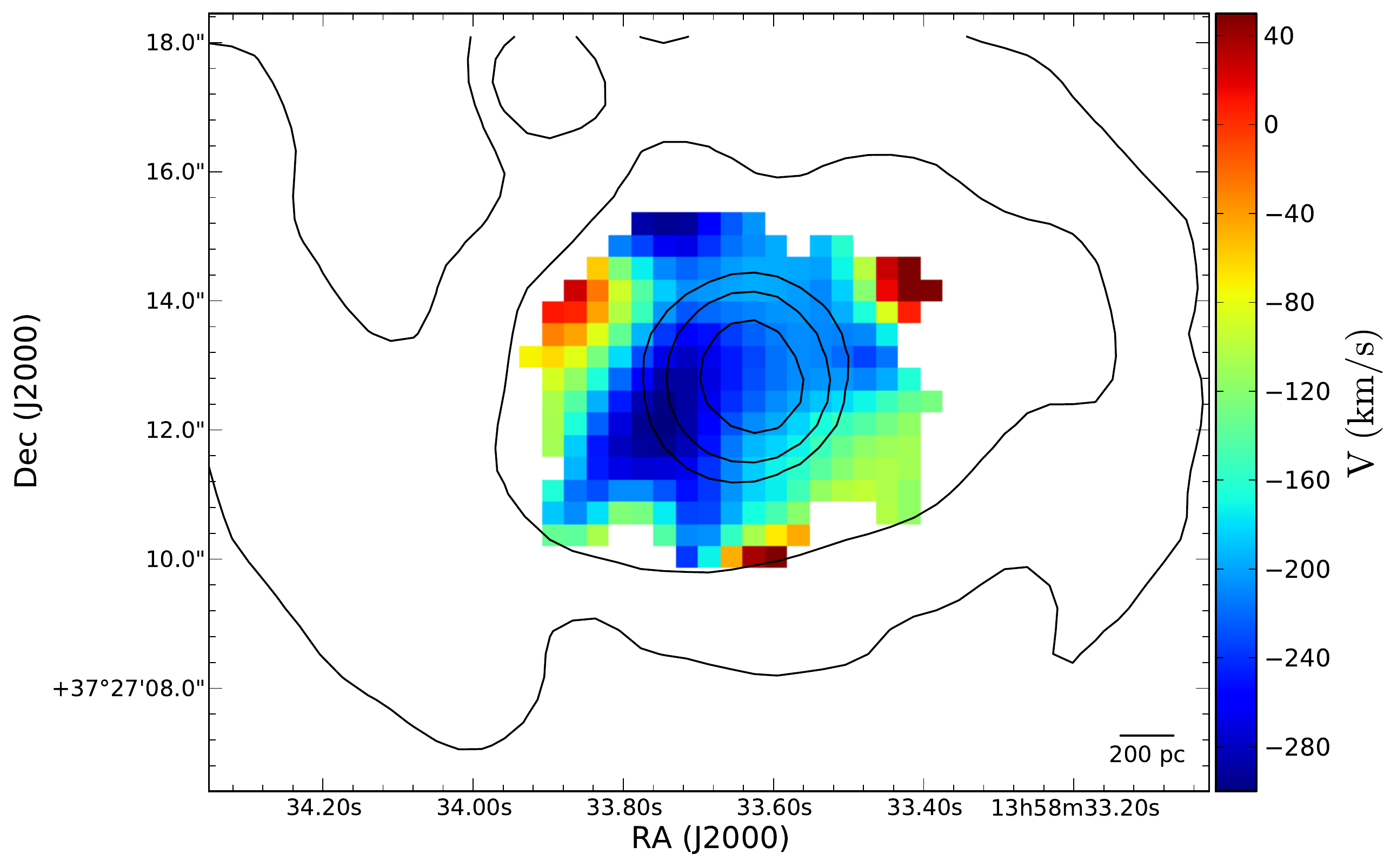}
   \includegraphics[angle=0,width=8cm, clip=true]{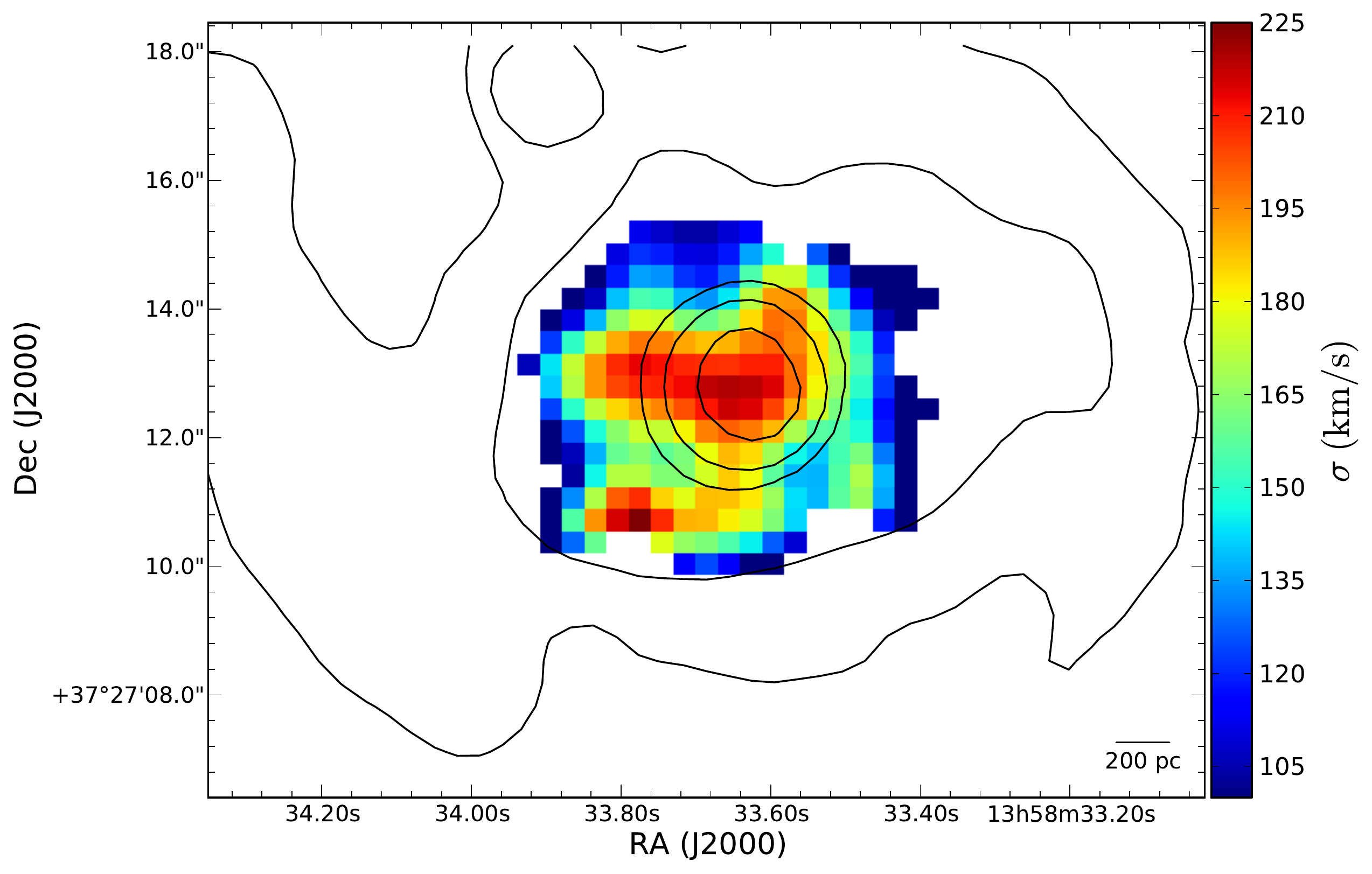}
      \includegraphics[angle=0,width=8cm, clip=true]{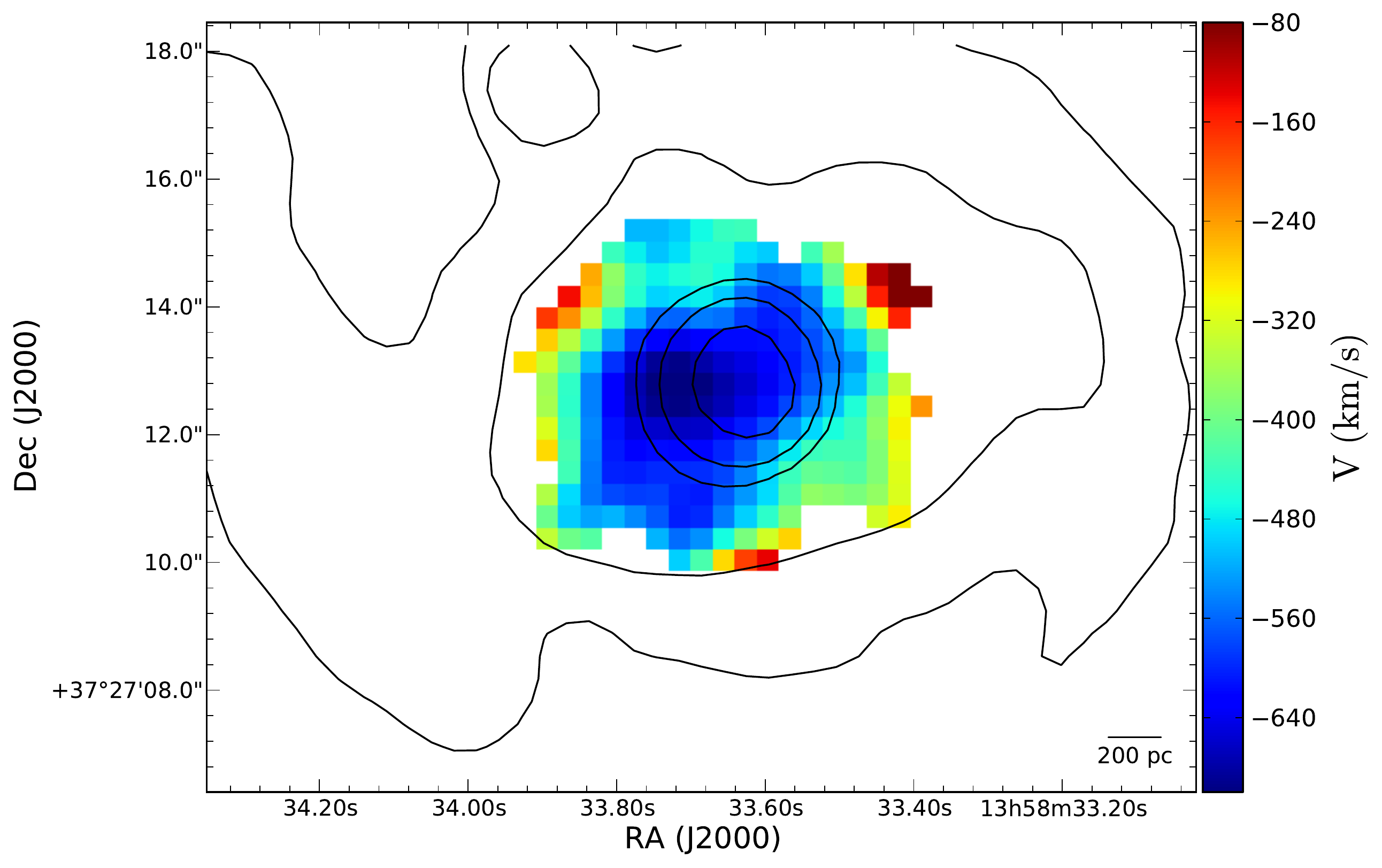}
   \caption{Map of the equivalent width (top), residual velocity (upper middle), velocity dispersion (lower middle) and extreme velocity (v$_{98}$) (bottom) of the interstellar \nai~D doublet. Overplotted contours show the red continuum emission.}\label{fig:NaD}
\end{figure}

With our stellar subtracted spectra we are able to produce maps for the distribution and kinematics of the interstellar cold gas by using the \nai~D doublet.
The top panel of Fig.~\ref{fig:NaD} shows the map of the \nai~D equivalent width. \nai~D absorption is clearly detected with EW$_{Na I}\gtrsim3$~\AA\ in the inner region of the galaxy, basically in the same region where a double kinematic component was found in the emission lines. This excess of interstellar \nai~D was already detected by \cite{roche}, although our better spatial resolution allows to  study the structure of the outflowing gas on smaller spatial scales ($\sim250$~pc). The highest values of EW$_{Na I}$ (larger than 6~\AA) are found around the nucleus of the galaxy, but slightly offset, $\sim1$\arcsec\ towards the E with respect to the maximum in the continuum  and H$\alpha$ intensity maps. This region extends towards the SE, along a region which coincides with the region of highest residual velocities of the secondary kinematic component of the ionised gas, and features in some line ratios as described in the previous section.  
It is worth remarking here that these EW$_{Na I}$  of the interstellar lines  have been measured in spectra which have been previously corrected for stellar absorption, and still produce values much higher than those that could be explained by any stellar origin or contamination. The absorption is very cleanly detected, and it allows the estimate of the velocity and dispersion of the neutral gas traced by the \nai~D. These maps are shown in the middle panels of Fig.~\ref{fig:NaD}.

The residual velocity field of the neutral gas shows blueshifted velocities nearly everywhere it has been detected, which are unambiguously indicative of an outflow in this galaxy. The average velocity is $-165$~km~s\me, although maximum approaching velocities reach  $-280$~km~s\me.
 Our residual velocity map shows slightly higher velocities than \cite[][their Fig.~27b]{roche}, probably due to our better spatial resolution, although our results are consistent with theirs.
Fig.~\ref{fig:NaD} shows that the largest blueshifts trace a band shaped region resembling the structure of the maximum EW$_{Na I}$, although slightly offset $\sim2$~\arcsec\ towards the E. These velocities are significantly larger than the blueshifts found for the warm gas in the emission lines. This is not uncommon in galactic winds \citep[e.g.][]{martin06,2013ApJ...768...75R}. A suitable explanation for this fact in the case of NGC~5394 will be provided in Section~\ref{sec:Disc}. 

The velocity dispersions of the neutral gas have values comparable to that of the warm ionised gas. The residual velocity and velocity dispersion of the neutral gas can be combined, as in the case of the emission lines, to produce a map of the most extreme velocities ($v_{98}=\Delta v-2\sigma$), which is shown in the bottom panel of Fig. \ref{fig:NaD}. This map has an average value of -470~km~s\me, and 
a maximum velocity of -690~km~s\me\ located $\sim2$~\arcsec\ E from the nucleus. These measured extreme blueshifted velocities show that the outflow in NGC~5394  has very high  velocities.

\subsection{Mass and outflow rate of the wind}

We can calculate the amount of neutral gas in the wind from the equivalent width of the \nai\ D doublet. Using the Eqs. 5 and 6 in \cite{hamann97} \citep[see also][]{2000ApJS..129..493H} on the integrated spectrum (for the spaxels with detected interstellar \nai~D), we have determined average values for the covering factor ($<C_f>\sim0.4$) and optical depth of the 5895~\AA\ line ($<\tau_{5985}>\sim0.85$). We have checked the consistency of these values by direct fitting of the absorption doublet to Eq.~2 in \cite{rupke02} to the same spectrum, and to individual spectra of spaxels with the highest S/N, obtaining very similar values.

These values are compatible with the findings of previous studies \citep{martin05,rupke05a,rupke05b,martin06,chen2010,krug2010}.
We then use the curve of growth technique and the equivalent width of \nai~D  to calculate the column density of Na atoms for each spaxel \citep[as in][]{schwartz}.

There are several prescriptions to convert Na column densities into H column densities.  This conversion is affected by rather large uncertainties in the abundance of Na, the degree of ionization of the gas, and the depletion into dust, which can make the final figure vary by a factor of a few. Here, we have used two different prescriptions. On one side, we use the approach of \cite{rupke05a} and \cite{2013ApJ...768...75R}, with the same  ionization, abundance and dust depletion values. On the other hand, we have also used an empirical correction factor as in  \cite{rupke02}. The obtained colum density for the integrated spectrum taking into account these different prescriptions is in the range  $N(H)=1.5-2.5\times10^{21}$cm$^{-2}$ \citep[cf.][]{heckman2015}.

Finally, in order to calculate the mass and outflow rate in neutral gas,  we use the thin shell wind model of  \cite{rupke05b}. They used an opening angle of $\sim65^\circ$, typical of winds in local starbursts \citep[according to][]{veilleux05}. Here we are somewhat more conservative, and we asume a slightly lower opening angle of $50^\circ$ for the wind. We have also used a smaller value of the shell radius of 1.75~kpc (taking into account that the region where we detect the neutral gas has a radius of 0.75~kpc in the plane of the sky). This results in a total mass in the wind of $M(H^0)=1.0-1.8\times 10^7\,M_\odot$. This mass in neutral gas is 2 to 10 times smaller than the estimates for ULIRGS \citep{martin06}, but it is in reasonable agreement with measurements for less powerful starbursts \citep{schwartz}.

In order to calculate the outflow rate, we must take into account the velocity of the wind, which we take, as usual for the neutral gas, the velocity at the line centre. We obtain values of $\dot{M}(H^0)=1.4-2.4 \,M_\odot$~yr\me. We have repeated the calculations taking benefit or our spatial resolution on a spaxel by spaxel basis. The final results are very similar although slightly lower, with masses of $M(H^0)=0.7-1.3\times 10^7 M_\odot$ and outflow rates of  $\dot{M}(H^0)=0.8-1.6 \,M_\odot$~yr\me. Taking into account all these uncertainties, we will assume the average nominal values of $M(H^0)=(1.2\pm0.5)\times 10^7 \,M_\odot$ and  $\dot{M}(H^0)=1.6\pm0.8\,M_\odot$~yr\me\ for the neutral gas in the wind.

Considering the above flow rate and the estimated star formation rate of SFR$=2.95\, M_\odot$~yr\me, we can quantify the importance of the wind by means of the mass entrainment efficiency or loading factor $\eta=\dot{M}/SFR=0.6\pm0.3$. Despite this object has a moderate SFR, this value is in good agreement with previous estimates for ULIRGS, with a muchhigher SFRs \citep{rupke02,rupke05a,2013ApJ...768...75R}. This fact may be explained by the high SFR density, $\Sigma_{SFR}\sim2.8$~\msun~yr\me~kpc$^{2}$ in the nucleus of NGC~5394, which is rather common among (U)LIRGS presenting galactic winds \cite[see e.g. Fig.~4 in][]{arribas2014}.

We can also estimate these quantities for the ionised gas in the outflow of NGC~5394. 
The mass is obtained by integration, over the volume occupied by the wind, of the product of the proton mass, the electron density ($N_e$, as traced by the [\sii] doublet, Section~\ref{line_ratios}) and the volume filling factor \cite[$\epsilon$,][p. 166]{osterbrock}.
$N_e$ for the secondary  broad component is in average $\sim500$~cm$^{-3}$ (Fig.~\ref{fig:S1S2}, bottom).  The filling factor is defined as $\epsilon=(<N_e>_{rms})^2/N_e^2$, where  $<N_e>_{rms}=\sqrt{<N_e^{2}>}$, the root-mean-squared electron density,  can be estimated from the \ha\ emission and the total volume of the wind \citep[see e.g.][]{leonel}. 
Thus, as the volume filling factor is inversely proportional to the wind volume, the ionised gas mass can be obtained from only two measurable  parameters\footnote{The total ionised gas mass in the wind can be obtained from the following expression: $M(H^+)=\frac{2.2 L_{H\alpha}\, m_p}{h\nu_{H\alpha}\, \alpha_B(H^0)\,N_e}$, where $m_p$ is the proton mass, $h\nu_{H\alpha}$ the energy of an \ha\ photon,  $\alpha_B(H^0)$ is the case B recombination coefficient for H, and $L_{H\alpha}$ and $N_e$ are the \ha\ luminosity and the electron density for the secondary component respectively.}: the \ha\ luminosity ($L_{H\alpha}$) and $N_e$ for the secondary (broad) component. In this way, we estimate a total ionized gas mass of $M(H^+)\sim7.9\times10^{5}\,M_\odot$, which is about 10 to 20 times lower than the estimated mass of neutral gas in the wind.
This calculation is in reasonable agreement with the one obtained using \cite{genzel2011} formula, but yields approximately half the mass obtained using the expression in \cite{colina}.

Considering an opening angle of 50$^\circ$ for the wind and a radius $R= 1.75$~kpc, as for the neutral phase calculations, this mass implies a volume filling factor in the wind $\epsilon\sim10^{-5}$ and $<N_e>_{rms}\approx1.6$~cm$^{-3}$.

The ionised mass outflow rate, $\dot{M}(H^+)$, can be estimated by dividing  $M(H^+)$ by the outflow dynamical time $ t = R / v_{out}$. In order to allow for comparisons with previous authors, we have followed the approach by \cite{wood}, who obtained two  different estimates for $\dot{M}(H^+)$, from two different adopted values of $v_{out}$. The first one  assumes that  $v_{out1} = \Delta v_{sec} - \mbox{fwhm}_{sec}/2$, and the second one  $v_{out2} =  \Delta v_{sec} - 2\sigma_{sec}$, where $\Delta v_{sec}$, $\mbox{fwhm}_{sec}$ and  $\sigma_{sec}$ are the velocity blueshift of the secondary velocity component (with respect to the primary), the FWHM of the secondary component, and the corresponding Gaussian sigma value, respectively.
The obtained mass outflow rate is 0.08 and 0.13~$\,M_\odot$~yr\me\ using $v_{out1}$ and $v_{out2}$, respectively. These values,  imply loading factors for the ionised component alone of $\sim0.04$ - 0.06.

These values are considerably lower than the one obtained for the neutral phase of this galaxy, indicating that the outflow material is dominated by neutral gas in the wind in NGC 5394. The derived value for the loading factor for the ionised gas is also  considerably lower than in non-AGN LIRGs and ULIRGS (cf. $\sim0.3$-0.5, in average) studied by \cite{arribas2014,wood}, but are in good agreement with the corresponding estimates for some of the ULIRGS analysed by \cite{2013ApJ...768...75R} (their Tables~7 and 8), and \cite{soto}.

\begin{table*}
\begin{minipage}{160mm}
%\centering
\caption{Measured properties of the galactic NGC~5394 wind for its neutral and ionised phases, from the emission lines  and \nai~D absorption doublet analysis, respectively.}
\label{table:wind}
\begin{tabular}{l c c c c c c c }     % 7 columns
\hline
Phase       & $<v_{out}>$$^{(a)}$ & $<\sigma>$   & $(v_{out})_{max}$$^{(b)}$ & $v_{98}$$^{(c)}$   & $M$                 & $\dot{M}$  & $\eta$            \\[1mm]
            & (km~s\me)        &  (km~s\me)    & (km~s\me)       & (km~s\me)        & ($M_{\odot}$)         & ($M_{\odot}$~yr\me)   & (=$\dot{M}/SFR$)\\[1.5mm]
\hline 
Neutral     & $-165\pm100$    &  $150\pm90$      &   $-280$       &  $-470\pm160$   & $(1.2\pm0.5)\times10^{7}$ & $(1.6\pm0.8)$   &  $0.6\pm0.3$    \\  
Ionised     & $-30\pm15$      &  $130\pm20$      &   $-80$        &  $-290\pm50$    & $\sim7.9\times10^{5}$     & 0.08-0.13       &   0.04 - 0.06  \\
\hline
\end{tabular}
 {\footnotesize \begin{tabular}{l}
$^{a}$  Average outflow velocity (measured with respect to the velocity of the primary component in the ionised \\
gas) in all fibres.\\
$^{b}$  Maximum measured outflow velocity.\\
$^{c}$  $v_{98} = v_{out} - 2\sigma$\\
\end{tabular}}
\end{minipage}
\end{table*}

\begin{figure}
   \centering
   \includegraphics[angle=0,width=8cm, clip=true]{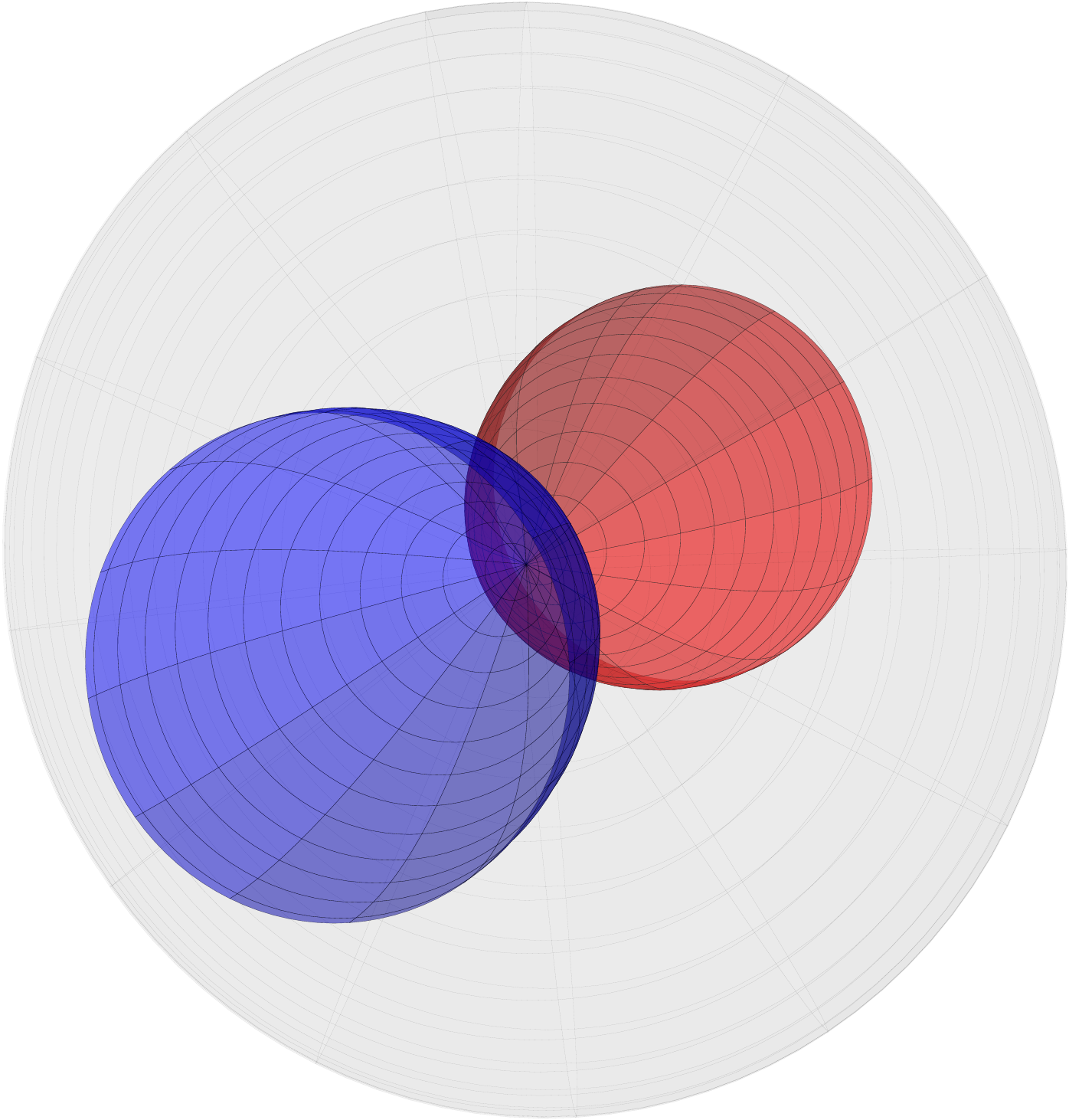}
   \caption{Sketch showing the bipolar outflow in NGC 5394. The approaching and receding sides are shown in blue and red respectively. The underlying galaxy disc is shown in light grey.}
\label{fig:bipolar}
\end{figure}

\section{Discussion}
\label{sec:Disc}
The findings described in the previous section unambiguously support the existence  of an important galactic wind in the nuclear region of NGC 5394 in both, the neutral and ionized gas phases. The kinematical discrimination of the emission lines has allowed us to separate the wind component from the underlying gas disc, with significantly different physical conditions (i.e. velocity dispersion, line ratios and local electron densities). On the other hand, the neutral phase traced by the interstellar \nai~D absorption lines, has made possible to estimate  the total mass  and mass flow of neutral gas in the wind. 

It is very often found that galactic winds follow a bipolar structure \citep[cf.][]{veilleux05}, with the gas flowing mainly perpendicular to the plane of the galaxy disc. We find that in the case of NGC~5394, the most important observational features of the wind can also be easily understood in terms of  a bipolar wind model as the one sketched in Fig.~\ref{fig:bipolar}. The blue and red cone-like structures represent the blueshifted and redshifted parts of the wind, while the light grey disc represents the underlying galaxy disc.

According to simulations \citep[cf.][]{2000ApJS..129..493H,fujita09} the cold gas in a wind, detected via the \nai\ D lines, is expected to trace dense ambient gas entrained into the wind flow, or fragments of the blowing shell swept up by the wind. In any case, it is most likely to be found at the base and walls of the (approaching side of) this bipolar structure. This scenario seems to be compatible with the map of the EW of \nai\ D shown in Fig.~\ref{fig:NaD} (top). This map shows a narrow structure in the EW$_{Na I}$ which is slightly offset with respect to the nuclear starburst, and which could be tracing the (approaching side of the) wall of the biconical structure as shown in Fig.~\ref{fig:bipolar}. In the same way, the maximum velocities of this neutral gas component are expected to be found also close to the walls of the biconical structure, with the largest values found at the locations where the outflowing gas moves along the line of sight, towards the observer. Again, the observed velocity map of the \nai\ D shown in the middle top panel of Fig.~\ref{fig:NaD} is perfectly compatible with this scenario, showing the maximum velocity of the \nai\ D in NGC~5394 close to, but at the inner side of the wall of the approaching cone. The velocity dispersion of the neutral gas is maximum close to the centre of the structure, probably due to integration along the line of sight of clumps with different line of sight velocities.

The observations for the warmer gas, traced by the emission lines, can also be easily understood within this framework. Unfortunately, the low inclination of NGC~5394 makes the emission from the wind  outshined by the underlying gas in the galaxy. The kinematical discrimination can help us to disentangle both components, but only at the locations with highest gas densities (more easily detectable).
Although the secondary component undoubtedly traces the outflowing gas, signatures of the wind can also be hinted in the single  component  fit.

While for the absorption lines we can be sure that they are generated in front of the galaxy disc, emission can be generated, in principle, anywhere  along the line of sight, throughout the whole biconical structure. As emission lines are less sensitive to low density gas, it is not strange that emission from the ionised gas is only detected at the very centre of the  structure, where integration along the line of sight contributes to increase the detected flux. \cite{wood} have shown that the broad component
of the ionized gas is most likely generated near the galaxy plane and not far from
the starburst, where densities are higher and the gas still shares much of the kinematics
of the undelying disc. 
This scenario not only explains the fact that we only detect this broad component in a rather compact nuclear region but also the lower velocity with respect to the bulk of
the neutral gas, as this ionized gas close to the wind base has not yet been accelerated to the high velocities detected in the neutral components.
There could still be a more extended and higher velocity ionized gas component of the
wind, but which is most likely too faint (due to its lower density) to be detected by this means.

%There is also the issue of the high velocity
%dispersion of this gas. \cite{wood} also discuss several 
%If emission from both ends of the structure is indeed detected at the same projected location, the (possibly more extincted) redshifted emission from the far side, can combine with the blueshifted emission in the near side to lower the average velocity of the line, but producing a higher velocity dispersion \citep[cf.][]{wood}. 
%The secondary component does indeed present higher velocity dispersions, up to 190~km~s\me, in the region where the wind is found. 

But not only the secondary component does show some structure in velocity dispersion related with the wind. The C-like structure towards the W of the nucleus found in the single component maps, can tentatively be interpreted within this scenario as tracing (part of) the receding side of the biconical structure. Indeed, as commented in the previous section, this structure can also be seen in the [\nii]/\ha\ map of the single component, and even some hints of it are present in the [\sii]/\ha\ and in the velocity field of this same component. This is not unexpected, as shocks and turbulence at the walls of the wind structure can give rise to higher line ratios and velocity dispersions, as well as to velocity perturbations.

The measured residual velocities of the wind gas are significantly higher in the neutral (averaging to $\sim-165$~km~s\me) than in the ionized gas (averaging to $\sim-30$~km~s\me). On the other hand, velocity dispersions are much more alike, as are the most extreme velocities for both phases ($\sim-300$~km~s\me\ for the ionized gas and $\sim-450$~km~s\me\ for the neutral gas).

Our estimates for the  mass of neutral gas in the wind ($M(H^0) \sim 1.2\times10^{7}\, M_\odot$) and the outflow rates ($dM(H^0)/dt \sim 1.6 M_\odot$~yr\me) agree with previous estimates for this \citep[][]{roche} and other similar objects \citep{schwartz}, and are compatible with the wind being originated by the mechanical energy input from the central starburst. The estimated loading factor ($\eta\sim0.6$) is similar to the values found in other similar studies \citep{rupke02,rupke05a,2013ApJ...768...75R}.

We have estimated that the mass of ionised gas in the wind is about 10 to 16 times lower than the neutral gas mass ($M(H^+) \sim 8\times 10^5 M_\odot$) with an outflow rate also considerably smaller  ($dM(H^+)/dt \sim 0.08 - 0.13 M_\odot$ yr\me). The NGC 5394 wind is therefore dominated by the neutral component. This is also the case for some of the few  galaxies so far, for which both the ionised and the neutral phases of the wind have been studied simultaneously  \citep{2013ApJ...768...75R}. However, in other galaxies (mostly ULIRGS) the ionised gas seems to represent a considerable amount or even dominates the outflowing material mass \citep{arribas2014}. 

%{\bf Comment on mass and mass outflow rate in ionized gas and compare with neutral phase.}

Finally, in order  to answer the question of whether a significant fraction of gas and/or metals are able to escape the potential well of the galaxy
in this wind,  we can make a rough estimate of the escape velocity by using a singular isothermal sphere potential truncated at $r_{max}$ as in \cite[][]{rupke02}.
We have $v_{esc}(r)=\sqrt{2}v_c[1+\ln(r_{max}/r)]^{-1/2}$. We will assume $r_{max}/r=10$, although the final result is rather insensitive to this value \citep[cf.][]{rupke02}.  
We take a nominal value for the circular velocity of $v_c=156$ km s\me\ \citep[][]{roche}. \cite{kaufman} estimated $v_c=163$ km s\me\ using \hi\  observations. With these values, we obtain $v_{esc}\sim400$ km s\me. We can compare this value with the maps of the extreme velocities ($v_{98}$) measured for NGC 5394. The escape velocity is comparable to the maximum velocities for the ionised component in the gas, so a very tiny fraction of the ionised gas should be able to escape the potential well of the galaxy, if at all. Moreover, as shown above, the ionised gas represents only $\sim$10\% of the total mass in the wind, and therefore, a very small amount of the ionised gas is expected to be able to escape. 

For the neutral gas, the situation is very different, as the average value for the $v_{98}$ map ($\sim470$~km~s\me) is above the escape velocity (and the maximum values comfortably exceed that value). Taking into account that the mass in neutral gas in the wind is $\sim1.2\times10^7 M_\odot$ and that the calculated escape velocity of 400~km~s\me\ corresponds to gas (blueshifted) $1.56\times \sigma$ above the average velocity, we can calculate that the escaping gas is 6.5\% of the EW in the doublet, from which we can estimate that a mass of {\bf $\sim0.8\times10^6 M_\odot$} may be able to escape the galaxy.
\cite{kaufman} estimated a total mass of $7.3\times10^8 M_\odot$ in \hi, and four times this amount in molecular gas in this galaxy.
The calculated amount of gas escaping the galaxy represents a tiny fraction of the available mass in the central region of the galaxy, and is therefore not expected to affect the star formation processes.
It can nevertheless have an impact in contributing to metal enrichment of the IGM, which is, unfortunately, difficult to quantify with the available information.

The gas in the wind which is not able to escape may still have an important effect on the host galaxy evolution, as it falls back and redistributes metals and dust within the disc. This effect can contribute to the flattening of the 
metallicity gradient in NGC~5394 found by \cite{roche}.

\section{Conclusions}
\label{sec:Conclusions}
In this paper, we have confirmed the presence of an important galactic wind powered by the nuclear starburst in the face-on galaxy NGC~5394. We have
been able to study the cold and warm phases of the gas in the wind. The neutral gas has been traced by means of the interstellar \nai\ $\lambda\lambda 5890, 5896\,$\AA\ (\nai\ D) absorption lines, while the ionized gas has been traced by several emission lines (H$\alpha$, [\nii]$\lambda\lambda$6548, 6583, [\sii]$\lambda\lambda$6716, 6731). From the analysis of the observations, the main conclusions of this work can be summarised as follows:

\begin{itemize}

\item Kinematical discrimination of the emission lines is critical to disentangle emission from the wind and the underlying disc. Both components present
clearly distinctive kinematical (velocity and velocity dispersion) and ionization (line ratios) properties, indicating that they likely trace gas with different origins. 

\item The secondary component of the ionised gas has significantly larger
velocity dispersion and enhanced [\nii]/\ha\ and [\sii]/\ha\ line ratios compared to the single or primary components. These are clear 
signatures of shock ionization. This component is also blueshifted with respect 
to the underlying disc by $\sim30$~km s\me\ (reaching $\sim80$~km s\me\ at
some locations). All these facts support that this
component traces the ionised phase of the galactic outflow in NGC~5394.

\item After careful correction of the observed spectra for 
ste\-llar absorption, we find 
a significant interstellar component of the \nai\ D doublet. This interstellar absorption is detected in a region
of $\sim 1$~kpc around the galaxy nucleus, roughly coincident with the region
where two kinematic conponents can be found in the emission lines. The 
neutral gas traced by this absorption is blueshifted by an average of $\sim165$~km~s\me\ with respect to the underlying disc (reaching a maximum of  $\sim280$ km s\me). 

\item The mass of the ionized gas in the galactic wind ($7.9\times 10^5 M_\odot$) represents a small but significant fraction ($\sim7\%$) of the mass in form of neutral gas ($1.2\times 10^7 M_\odot$). Although these estimates are subject 
to large uncertainties (particularly in the neutral phase, for which the ionization fraction, Na abundance and dust depletion are not well known), our 
results show that the majority of the wind material is in form of neutral gas.

\item The measured loading factors for the neutral ($\sim0.6$) and ionized ($\sim0.1$) gas phases are also in a similar ratio 1/6.  The mass flow rate in ionized gas is therefore nearly an order of magnitude lower than for the neutral gas. This result, as well as the obtained values for the loading factors in both phases agree well with results from previous studies including both gas phases \citep[cf.][]{2013ApJ...768...75R}.

\item We estimate that the vast majority of the detected ionised gas in the wind will 
not be able to escape the potential well of the galaxy.
On the contrary, a fraction of roughly $5\%-10\%$ of the neutral gas, amounting to 
$\sim10^6 M_\odot$, could be able to escape to the IGM contributing to its metal 
enrichment and dust content. The rest of wind material,
not fast enough to escape, will eventually fall back to the galaxy, potentially 
redistributing its metal and dust content.

\item The distribution and kinematics of the neutral and ionized gas, as well as the 
ionization structure of the galactic wind in NGC~5394 can be easily understood 
in the scenario of a bipolar wind geometry. 

\end{itemize}

The present study shows that massive galaxies with starbursts of moderate SFR can host important winds if the star formation is concentrated enough. These winds in not so extreme
starburst galaxies may be rather frequent at low redshift, and can have a non negligible 
impact in the metal enrichment of the IGM  and the secular evolution of their hosts.

\section*{Acknowledgements}
We are grateful to the anonymous referee for useful suggestions that improved the presentation of this work. P.M.F., J.J.V. and A.Z. acknowledge support from the project AYA2014-53506-P financed by the Spanish ``Ministerio de Econom\'{\i}a y Competividad'' and by FEDER (Fondo Europeo de Desarrollo Regional), and from the  ``Junta de Andaluc\'\i a'' local government through the FQM-108 project. EMG is supported by the Spanish ``Ministerio de Econom\'{\i}a y Competividad'' through project AYA2013-47744-C3-1. ACM also thanks the support from the ``Plan Nacional de Investigaci\'on y Desarrollo'' funding program AYA2013-46724-P.\par

Based on observations made with the William Herschel Telescope operated on the island of La Palma by the Isaac Newton Group in the Spanish Observatorio del Roque de los Muchachos of the Instituto de Astrof\'\i sica de Canarias.\par

Funding for SDSS-III has been provided by the Alfred P. Sloan Foundation, the Participating Institutions, the National Science Foundation, and the U.S. Department of Energy Office of Science. The SDSS-III web site is http://www.sdss3.org/.\par

%SDSS-III is managed by the Astrophysical Research Consortium for the Participating Institutions of the SDSS-III Collaboration including the University of Arizona, the Brazilian Participation Group, Brookhaven National Laboratory, Carnegie Mellon University, University of Florida, the French Participation Group, the German Participation Group, Harvard University, the Instituto de Astrofisica de Canarias, the Michigan State/Notre Dame/JINA Participation Group, Johns Hopkins University, Lawrence Berkeley National Laboratory, Max Planck Institute for Astrophysics, Max Planck Institute for Extraterrestrial Physics, New Mexico State University, New York University, Ohio State University, Pennsylvania State University, University of Portsmouth, Princeton University, the Spanish Participation Group, University of Tokyo, University of Utah, Vanderbilt University, University of Virginia, University of Washington, and Yale University. \par

This study makes uses of the data provided by the Calar Alto Legacy Integral Field Area (CALIFA) survey (http://califa.caha.es/).
Based on observations collected at the Centro Astron\'omico Hispano Alem\'an (CAHA) at Calar Alto, operated jointly by the Max-Planck-Institut fur Astronomie and the Instituto de Astrofisica de Andalucia (CSIC).

%%%%%%%%%%%%%%%%%%%%%%%%%%%%%%%%%%%%%%%%%%%%%%%%%%

%%%%%%%%%%%%%%%%%%%% REFERENCES %%%%%%%%%%%%%%%%%%

% The best way to enter references is to use BibTeX:

\bibliographystyle{mnras}
\bibliography{./bib} % if your bibtex file is called example.bib

\begin{thebibliography}{}
\makeatletter
\relax
\def\mn@urlcharsother{\let\do\@makeother \do\$\do\&\do\#\do\^\do\_\do\%\do\~}
\def\mn@doi{\begingroup\mn@urlcharsother \@ifnextchar [ {\mn@doi@}
  {\mn@doi@[]}}
\def\mn@doi@[#1]#2{\def\@tempa{#1}\ifx\@tempa\@empty \href
  {http://dx.doi.org/#2} {doi:#2}\else \href {http://dx.doi.org/#2} {#1}\fi
  \endgroup}
\def\mn@eprint#1#2{\mn@eprint@#1:#2::\@nil}
\def\mn@eprint@arXiv#1{\href {http://arxiv.org/abs/#1} {{\tt arXiv:#1}}}
\def\mn@eprint@dblp#1{\href {http://dblp.uni-trier.de/rec/bibtex/#1.xml}
  {dblp:#1}}
\def\mn@eprint@#1:#2:#3:#4\@nil{\def\@tempa {#1}\def\@tempb {#2}\def\@tempc
  {#3}\ifx \@tempc \@empty \let \@tempc \@tempb \let \@tempb \@tempa \fi \ifx
  \@tempb \@empty \def\@tempb {arXiv}\fi \@ifundefined
  {mn@eprint@\@tempb}{\@tempb:\@tempc}{\expandafter \expandafter \csname
  mn@eprint@\@tempb\endcsname \expandafter{\@tempc}}}

\bibitem[\protect\citeauthoryear{{Arp}}{{Arp}}{1966}]{arp66}
{Arp} H.,  1966, \mn@doi [\apjs] {10.1086/190147}, \href
  {http://cdsads.u-strasbg.fr/abs/1966ApJS...14....1A} {14, 1}

\bibitem[\protect\citeauthoryear{{Arribas} \& {Colina}}{{Arribas} \&
  {Colina}}{2002}]{arribas_colina02}
{Arribas} S.,  {Colina} L.,  2002, \mn@doi [\apj] {10.1086/340755}, \href
  {http://cdsads.u-strasbg.fr/abs/2002ApJ...573..576A} {573, 576}

\bibitem[\protect\citeauthoryear{{Arribas} et~al.,}{{Arribas}
  et~al.}{1998}]{1998ASPC..152..149A}
{Arribas} S.,  et~al., 1998, in {Arribas} S.,  {Mediavilla} E.,   {Watson} F.,
  eds,  Astronomical Society of the Pacific Conference Series Vol. 152, Fiber
  Optics in Astronomy III. p.~149

\bibitem[\protect\citeauthoryear{{Arribas}, {Colina}, {Bellocchi}, {Maiolino}
  \& {Villar-Mart{\'{\i}}n}}{{Arribas} et~al.}{2014}]{arribas2014}
{Arribas} S.,  {Colina} L.,  {Bellocchi} E.,  {Maiolino} R.,
  {Villar-Mart{\'{\i}}n} M.,  2014, \mn@doi [\aap]
  {10.1051/0004-6361/201323324}, \href
  {http://cdsads.u-strasbg.fr/abs/2014A\%26A...568A..14A} {568, A14}

\bibitem[\protect\citeauthoryear{{Baldwin}, {Phillips}  \&
  {Terlevich}}{{Baldwin} et~al.}{1981}]{BPT}
{Baldwin} J.~A.,  {Phillips} M.~M.,   {Terlevich} R.,  1981, \mn@doi [\pasp]
  {10.1086/130766}, \href {http://cdsads.u-strasbg.fr/abs/1981PASP...93....5B}
  {93, 5}

\bibitem[\protect\citeauthoryear{{Becker}, {White}  \& {Helfand}}{{Becker}
  et~al.}{1995}]{becker}
{Becker} R.~H.,  {White} R.~L.,   {Helfand} D.~J.,  1995, \mn@doi [\apj]
  {10.1086/176166}, \href {http://cdsads.u-strasbg.fr/abs/1995ApJ...450..559B}
  {450, 559}

\bibitem[\protect\citeauthoryear{{Belfiore}, {Maiolino}  \&
  {Bothwell}}{{Belfiore} et~al.}{2016}]{belfiore}
{Belfiore} F.,  {Maiolino} R.,   {Bothwell} M.,  2016, \mn@doi [\mnras]
  {10.1093/mnras/stv2332}, \href
  {http://cdsads.u-strasbg.fr/abs/2016MNRAS.455.1218B} {455, 1218}

\bibitem[\protect\citeauthoryear{{Bellocchi}, {Arribas}, {Colina}  \&
  {Miralles-Caballero}}{{Bellocchi} et~al.}{2013}]{bellocchi}
{Bellocchi} E.,  {Arribas} S.,  {Colina} L.,   {Miralles-Caballero} D.,  2013,
  \mn@doi [\aap] {10.1051/0004-6361/201221019}, \href
  {http://cdsads.u-strasbg.fr/abs/2013A%26A...557A..59B} {557, A59}

\bibitem[\protect\citeauthoryear{{Bingham}, {Gellatly}, {Jenkins}  \&
  {Worswick}}{{Bingham} et~al.}{1994}]{1994SPIE.2198...56B}
{Bingham} R.~G.,  {Gellatly} D.~W.,  {Jenkins} C.~R.,   {Worswick} S.~P.,
  1994, in {Crawford} D.~L.,  {Craine} E.~R.,  eds,  Society of Photo-Optical
  Instrumentation Engineers (SPIE) Conference Series Vol. 2198, Instrumentation
  in Astronomy VIII. pp 56--64

\bibitem[\protect\citeauthoryear{{Bregman}}{{Bregman}}{1980}]{bregman}
{Bregman} J.~N.,  1980, \mn@doi [\apj] {10.1086/157776}, \href
  {http://adsabs.harvard.edu/abs/1980ApJ...236..577B} {236, 577}

\bibitem[\protect\citeauthoryear{{Castillo-Morales}, {Jim{\'e}nez-Vicente},
  {Mediavilla}  \& {Battaner}}{{Castillo-Morales} et~al.}{2007}]{castillo07}
{Castillo-Morales} A.,  {Jim{\'e}nez-Vicente} J.,  {Mediavilla} E.,
  {Battaner} E.,  2007, \mn@doi [\mnras] {10.1111/j.1365-2966.2007.12104.x},
  \href {http://cdsads.u-strasbg.fr/abs/2007MNRAS.380..489C} {380, 489}

\bibitem[\protect\citeauthoryear{{Catal{\'a}n-Torrecilla}
  et~al.,}{{Catal{\'a}n-Torrecilla} et~al.}{2015}]{catalan-torrecilla}
{Catal{\'a}n-Torrecilla} C.,  et~al., 2015, \mn@doi [\aap]
  {10.1051/0004-6361/201526023}, \href
  {http://cdsads.u-strasbg.fr/abs/2015A\%26A...584A..87C} {584, A87}

\bibitem[\protect\citeauthoryear{{Cazzoli}, {Arribas}, {Colina},
  {Piqueras-L{\'o}pez}, {Bellocchi}, {Emonts}  \& {Maiolino}}{{Cazzoli}
  et~al.}{2014}]{cazzoli}
{Cazzoli} S.,  {Arribas} S.,  {Colina} L.,  {Piqueras-L{\'o}pez} J.,
  {Bellocchi} E.,  {Emonts} B.,   {Maiolino} R.,  2014, \mn@doi [\aap]
  {10.1051/0004-6361/201323296}, \href
  {http://cdsads.u-strasbg.fr/abs/2014A%26A...569A..14C} {569, A14}

\bibitem[\protect\citeauthoryear{{Chen}, {Tremonti}, {Heckman}, {Kauffmann},
  {Weiner}, {Brinchmann}  \& {Wang}}{{Chen} et~al.}{2010}]{chen2010}
{Chen} Y.-M.,  {Tremonti} C.~A.,  {Heckman} T.~M.,  {Kauffmann} G.,  {Weiner}
  B.~J.,  {Brinchmann} J.,   {Wang} J.,  2010, \mn@doi [\aj]
  {10.1088/0004-6256/140/2/445}, \href
  {http://cdsads.u-strasbg.fr/abs/2010AJ....140..445C} {140, 445}

\bibitem[\protect\citeauthoryear{{Colina}, {Lipari}  \& {Macchetto}}{{Colina}
  et~al.}{1991}]{colina}
{Colina} L.,  {Lipari} S.,   {Macchetto} F.,  1991, \mn@doi [\apj]
  {10.1086/170489}, \href {http://cdsads.u-strasbg.fr/abs/1991ApJ...379..113C}
  {379, 113}

\bibitem[\protect\citeauthoryear{{Dalcanton}}{{Dalcanton}}{2007}]{dalcanton}
{Dalcanton} J.~J.,  2007, \mn@doi [\apj] {10.1086/508913}, \href
  {http://adsabs.harvard.edu/abs/2007ApJ...658..941D} {658, 941}

\bibitem[\protect\citeauthoryear{{Fujita}, {Martin}, {Mac Low}, {New}  \&
  {Weaver}}{{Fujita} et~al.}{2009}]{fujita09}
{Fujita} A.,  {Martin} C.~L.,  {Mac Low} M.-M.,  {New} K.~C.~B.,   {Weaver} R.,
   2009, \mn@doi [\apj] {10.1088/0004-637X/698/1/693}, \href
  {http://adsabs.harvard.edu/abs/2009ApJ...698..693F} {698, 693}

\bibitem[\protect\citeauthoryear{{Garnett}}{{Garnett}}{2002}]{garnett}
{Garnett} D.~R.,  2002, \mn@doi [\apj] {10.1086/344301}, \href
  {http://adsabs.harvard.edu/abs/2002ApJ...581.1019G} {581, 1019}

\bibitem[\protect\citeauthoryear{{Genzel} et~al.,}{{Genzel}
  et~al.}{2011}]{genzel2011}
{Genzel} R.,  et~al., 2011, \mn@doi [\apj] {10.1088/0004-637X/733/2/101}, \href
  {http://cdsads.u-strasbg.fr/abs/2011ApJ...733..101G} {733, 101}

\bibitem[\protect\citeauthoryear{{Guti{\'e}rrez} \& {Beckman}}{{Guti{\'e}rrez}
  \& {Beckman}}{2010}]{leonel}
{Guti{\'e}rrez} L.,  {Beckman} J.~E.,  2010, \mn@doi [\apjl]
  {10.1088/2041-8205/710/1/L44}, \href
  {http://cdsads.u-strasbg.fr/abs/2010ApJ...710L..44G} {710, L44}

\bibitem[\protect\citeauthoryear{{Hamann}, {Barlow}, {Junkkarinen}  \&
  {Burbidge}}{{Hamann} et~al.}{1997}]{hamann97}
{Hamann} F.,  {Barlow} T.~A.,  {Junkkarinen} V.,   {Burbidge} E.~M.,  1997,
  \apj, \href {http://cdsads.u-strasbg.fr/abs/1997ApJ...478...80H} {478, 80}

\bibitem[\protect\citeauthoryear{{Heckman}, {Lehnert}, {Strickland}  \&
  {Armus}}{{Heckman} et~al.}{2000}]{2000ApJS..129..493H}
{Heckman} T.~M.,  {Lehnert} M.~D.,  {Strickland} D.~K.,   {Armus} L.,  2000,
  \mn@doi [\apjs] {10.1086/313421}, \href
  {http://adsabs.harvard.edu/abs/2000ApJS..129..493H} {129, 493}

\bibitem[\protect\citeauthoryear{{Heckman}, {Alexandroff}, {Borthakur},
  {Overzier}  \& {Leitherer}}{{Heckman} et~al.}{2015}]{heckman2015}
{Heckman} T.~M.,  {Alexandroff} R.~M.,  {Borthakur} S.,  {Overzier} R.,
  {Leitherer} C.,  2015, \mn@doi [\apj] {10.1088/0004-637X/809/2/147}, \href
  {http://cdsads.u-strasbg.fr/abs/2015ApJ...809..147H} {809, 147}

\bibitem[\protect\citeauthoryear{{Ho} et~al.,}{{Ho} et~al.}{2014}]{ho2014}
{Ho} I.-T.,  et~al., 2014, \mn@doi [\mnras] {10.1093/mnras/stu1653}, \href
  {http://cdsads.u-strasbg.fr/abs/2014MNRAS.444.3894H} {444, 3894}

\bibitem[\protect\citeauthoryear{{Hopkins}, {Somerville}, {Hernquist}, {Cox},
  {Robertson}  \& {Li}}{{Hopkins} et~al.}{2006}]{hopkins}
{Hopkins} P.~F.,  {Somerville} R.~S.,  {Hernquist} L.,  {Cox} T.~J.,
  {Robertson} B.,   {Li} Y.,  2006, \mn@doi [\apj] {10.1086/508503}, \href
  {http://adsabs.harvard.edu/abs/2006ApJ...652..864H} {652, 864}

\bibitem[\protect\citeauthoryear{{Howarth}}{{Howarth}}{1983}]{howarth}
{Howarth} I.~D.,  1983, \mnras, \href
  {http://adsabs.harvard.edu/abs/1983MNRAS.203..301H} {203, 301}

\bibitem[\protect\citeauthoryear{{Husemann}, {Jahnke}, {S{\'a}nchez},
  {Barrado}, {Bekerait{\'e}}, {Bomans}, {Castillo-Morales}  et~al.}{{Husemann}
  et~al.}{2013}]{2013A&A...549A..87H}
{Husemann} B.,  {Jahnke} K.,  {S{\'a}nchez} S.~F.,  {Barrado} D.,
  {Bekerait{\'e}} S.,  {Bomans} D.~J.,  {Castillo-Morales} A.,   et~al., 2013,
  \mn@doi [\aap] {10.1051/0004-6361/201220582}, \href
  {http://adsabs.harvard.edu/abs/2013A\%26A...549A..87H} {549, A87}

\bibitem[\protect\citeauthoryear{{Jeong}, {Yi}, {Kyeong}, {Sarzi}, {Sung}  \&
  {Oh}}{{Jeong} et~al.}{2013}]{jeong2013}
{Jeong} H.,  {Yi} S.~K.,  {Kyeong} J.,  {Sarzi} M.,  {Sung} E.-C.,   {Oh} K.,
  2013, \mn@doi [\apjs] {10.1088/0067-0049/208/1/7}, \href
  {http://cdsads.u-strasbg.fr/abs/2013ApJS..208....7J} {208, 7}

\bibitem[\protect\citeauthoryear{{Jim{\'e}nez-Vicente}, {Castillo-Morales},
  {Mediavilla}  \& {Battaner}}{{Jim{\'e}nez-Vicente} et~al.}{2007}]{jjv2007}
{Jim{\'e}nez-Vicente} J.,  {Castillo-Morales} A.,  {Mediavilla} E.,
  {Battaner} E.,  2007, \mn@doi [\mnras] {10.1111/j.1745-3933.2007.00380.x},
  \href {http://cdsads.u-strasbg.fr/abs/2007MNRAS.382L..16J} {382, L16}

\bibitem[\protect\citeauthoryear{{Jim{\'e}nez-Vicente}, {Mediavilla},
  {Castillo-Morales}  \& {Battaner}}{{Jim{\'e}nez-Vicente}
  et~al.}{2010}]{jjv2010}
{Jim{\'e}nez-Vicente} J.,  {Mediavilla} E.,  {Castillo-Morales} A.,
  {Battaner} E.,  2010, \mn@doi [\mnras] {10.1111/j.1365-2966.2010.16692.x},
  \href {http://cdsads.u-strasbg.fr/abs/2010MNRAS.406..181J} {406, 181}

\bibitem[\protect\citeauthoryear{{Kaufman}, {Brinks}, {Elmegreen}, {Elmegreen},
  {Klari{\'c} }, {Struck}, {Thomasson}  \& {Vogel}}{{Kaufman}
  et~al.}{1999a}]{1999AJ....118.1577K}
{Kaufman} M.,  {Brinks} E.,  {Elmegreen} B.~G.,  {Elmegreen} D.~M.,
  {Klari{\'c} } M.,  {Struck} C.,  {Thomasson} M.,   {Vogel} S.,  1999a,
  \mn@doi [\aj] {10.1086/301030}, \href
  {http://adsabs.harvard.edu/abs/1999AJ....118.1577K} {118, 1577}

\bibitem[\protect\citeauthoryear{{Kaufman}, {Wolfire}, {Hollenbach}  \&
  {Luhman}}{{Kaufman} et~al.}{1999b}]{kaufman}
{Kaufman} M.~J.,  {Wolfire} M.~G.,  {Hollenbach} D.~J.,   {Luhman} M.~L.,
  1999b, \mn@doi [\apj] {10.1086/308102}, \href
  {http://cdsads.u-strasbg.fr/abs/1999ApJ...527..795K} {527, 795}

\bibitem[\protect\citeauthoryear{{Kaufman}, {Sheth}, {Struck}, {Elmegreen},
  {Thomasson}, {Elmegreen}  \& {Brinks}}{{Kaufman}
  et~al.}{2002}]{2002AJ....123..702K}
{Kaufman} M.,  {Sheth} K.,  {Struck} C.,  {Elmegreen} B.~G.,  {Thomasson} M.,
  {Elmegreen} D.~M.,   {Brinks} E.,  2002, \mn@doi [\aj] {10.1086/338433},
  \href {http://adsabs.harvard.edu/abs/2002AJ....123..702K} {123, 702}

\bibitem[\protect\citeauthoryear{{Kewley}, {Groves}, {Kauffmann}  \&
  {Heckman}}{{Kewley} et~al.}{2006}]{2006MNRAS.372..961K}
{Kewley} L.~J.,  {Groves} B.,  {Kauffmann} G.,   {Heckman} T.,  2006, \mn@doi
  [\mnras] {10.1111/j.1365-2966.2006.10859.x}, \href
  {http://adsabs.harvard.edu/abs/2006MNRAS.372..961K} {372, 961}

\bibitem[\protect\citeauthoryear{{Krug}, {Rupke}  \& {Veilleux}}{{Krug}
  et~al.}{2010}]{krug2010}
{Krug} H.~B.,  {Rupke} D.~S.~N.,   {Veilleux} S.,  2010, \mn@doi [\apj]
  {10.1088/0004-637X/708/2/1145}, \href
  {http://cdsads.u-strasbg.fr/abs/2010ApJ...708.1145K} {708, 1145}

\bibitem[\protect\citeauthoryear{{Lanz} et~al.,}{{Lanz} et~al.}{2013}]{lanz}
{Lanz} L.,  et~al., 2013, \mn@doi [\apj] {10.1088/0004-637X/768/1/90}, \href
  {http://cdsads.u-strasbg.fr/abs/2013ApJ...768...90L} {768, 90}

\bibitem[\protect\citeauthoryear{{Lehnert} \& {Heckman}}{{Lehnert} \&
  {Heckman}}{1996}]{lehnert96}
{Lehnert} M.~D.,  {Heckman} T.~M.,  1996, \mn@doi [\apj] {10.1086/177180},
  \href {http://cdsads.u-strasbg.fr/abs/1996ApJ...462..651L} {462, 651}

\bibitem[\protect\citeauthoryear{{Lehnert}, {Tasse}, {Nesvadba}, {Best}  \&
  {van Driel}}{{Lehnert} et~al.}{2011}]{lehnert11}
{Lehnert} M.~D.,  {Tasse} C.,  {Nesvadba} N.~P.~H.,  {Best} P.~N.,   {van
  Driel} W.,  2011, \mn@doi [\aap] {10.1051/0004-6361/201117323}, \href
  {http://cdsads.u-strasbg.fr/abs/2011A\%26A...532L...3L} {532, L3}

\bibitem[\protect\citeauthoryear{{Martin}}{{Martin}}{2005}]{martin05}
{Martin} C.~L.,  2005, \mn@doi [\apj] {10.1086/427277}, \href
  {http://cdsads.u-strasbg.fr/abs/2005ApJ...621..227M} {621, 227}

\bibitem[\protect\citeauthoryear{{Martin}}{{Martin}}{2006}]{martin06}
{Martin} C.~L.,  2006, \mn@doi [\apj] {10.1086/504886}, \href
  {http://adsabs.harvard.edu/abs/2006ApJ...647..222M} {647, 222}

\bibitem[\protect\citeauthoryear{{Monreal-Ibero}, {Arribas}  \&
  {Colina}}{{Monreal-Ibero} et~al.}{2006}]{monreal06}
{Monreal-Ibero} A.,  {Arribas} S.,   {Colina} L.,  2006, \mn@doi [\apj]
  {10.1086/498257}, \href {http://cdsads.u-strasbg.fr/abs/2006ApJ...637..138M}
  {637, 138}

\bibitem[\protect\citeauthoryear{{Ocvirk}, {Pichon}, {Lan{\c c}on}  \&
  {Thi{\'e}baut}}{{Ocvirk} et~al.}{2006}]{2006MNRAS.365...74O}
{Ocvirk} P.,  {Pichon} C.,  {Lan{\c c}on} A.,   {Thi{\'e}baut} E.,  2006,
  \mn@doi [\mnras] {10.1111/j.1365-2966.2005.09323.x}, \href
  {http://adsabs.harvard.edu/abs/2006MNRAS.365...74O} {365, 74}

\bibitem[\protect\citeauthoryear{{Oke}}{{Oke}}{1990}]{oke}
{Oke} J.~B.,  1990, \mn@doi [\aj] {10.1086/115444}, \href
  {http://cdsads.u-strasbg.fr/abs/1990AJ.....99.1621O} {99, 1621}

\bibitem[\protect\citeauthoryear{{Osterbrock} \& {Ferland}}{{Osterbrock} \&
  {Ferland}}{2006}]{osterbrock}
{Osterbrock} D.~E.,  {Ferland} G.~J.,  2006, {Astrophysics of gaseous nebulae
  and active galactic nuclei}

\bibitem[\protect\citeauthoryear{{Rela{\~n}o}, {Beckman}, {Zurita}, {Rozas}  \&
  {Giammanco}}{{Rela{\~n}o} et~al.}{2005}]{relano}
{Rela{\~n}o} M.,  {Beckman} J.~E.,  {Zurita} A.,  {Rozas} M.,   {Giammanco} C.,
   2005, \mn@doi [\aap] {10.1051/0004-6361:20040483}, \href
  {http://cdsads.u-strasbg.fr/abs/2005A\%26A...431..235R} {431, 235}

\bibitem[\protect\citeauthoryear{{Rich}, {Kewley}  \& {Dopita}}{{Rich}
  et~al.}{2011}]{rich2011}
{Rich} J.~A.,  {Kewley} L.~J.,   {Dopita} M.~A.,  2011, \mn@doi [\apj]
  {10.1088/0004-637X/734/2/87}, \href
  {http://cdsads.u-strasbg.fr/abs/2011ApJ...734...87R} {734, 87}

\bibitem[\protect\citeauthoryear{{Roche}, {Humphrey}, {Gomes}, {Papaderos},
  {Lagos}  \& {S{\'a}nchez}}{{Roche} et~al.}{2015}]{roche}
{Roche} N.,  {Humphrey} A.,  {Gomes} J.~M.,  {Papaderos} P.,  {Lagos} P.,
  {S{\'a}nchez} S.~F.,  2015, \mn@doi [\mnras] {10.1093/mnras/stv1669}, \href
  {http://cdsads.u-strasbg.fr/abs/2015MNRAS.453.2349R} {453, 2349}

\bibitem[\protect\citeauthoryear{{Rupke} \& {Veilleux}}{{Rupke} \&
  {Veilleux}}{2013}]{2013ApJ...768...75R}
{Rupke} D.~S.~N.,  {Veilleux} S.,  2013, \mn@doi [\apj]
  {10.1088/0004-637X/768/1/75}, \href
  {http://adsabs.harvard.edu/abs/2013ApJ...768...75R} {768, 75}

\bibitem[\protect\citeauthoryear{{Rupke}, {Veilleux}  \& {Sanders}}{{Rupke}
  et~al.}{2002}]{rupke02}
{Rupke} D.~S.,  {Veilleux} S.,   {Sanders} D.~B.,  2002, \mn@doi [\apj]
  {10.1086/339789}, \href {http://adsabs.harvard.edu/abs/2002ApJ...570..588R}
  {570, 588}

\bibitem[\protect\citeauthoryear{{Rupke}, {Veilleux}  \& {Sanders}}{{Rupke}
  et~al.}{2005a}]{rupke05a}
{Rupke} D.~S.,  {Veilleux} S.,   {Sanders} D.~B.,  2005a, \mn@doi [\apjs]
  {10.1086/432886}, \href {http://adsabs.harvard.edu/abs/2005ApJS..160...87R}
  {160, 87}

\bibitem[\protect\citeauthoryear{{Rupke}, {Veilleux}  \& {Sanders}}{{Rupke}
  et~al.}{2005b}]{rupke05b}
{Rupke} D.~S.,  {Veilleux} S.,   {Sanders} D.~B.,  2005b, \mn@doi [\apjs]
  {10.1086/432889}, \href {http://adsabs.harvard.edu/abs/2005ApJS..160..115R}
  {160, 115}

\bibitem[\protect\citeauthoryear{{S{\'a}nchez-Bl{\'a}zquez}
  et~al.,}{{S{\'a}nchez-Bl{\'a}zquez} et~al.}{2006}]{2006MNRAS.371..703S}
{S{\'a}nchez-Bl{\'a}zquez} P.,  et~al., 2006, \mn@doi [\mnras]
  {10.1111/j.1365-2966.2006.10699.x}, \href
  {http://adsabs.harvard.edu/abs/2006MNRAS.371..703S} {371, 703}

\bibitem[\protect\citeauthoryear{{S{\'a}nchez}, {Kennicutt}, {Gil de Paz}, {van
  de Ven}, {V{\'{\i}}lchez}, {Wisotzki}, {Walcher}  et~al.}{{S{\'a}nchez}
  et~al.}{2012}]{2012A&A...538A...8S}
{S{\'a}nchez} S.~F.,  {Kennicutt} R.~C.,  {Gil de Paz} A.,  {van de Ven} G.,
  {V{\'{\i}}lchez} J.~M.,  {Wisotzki} L.,  {Walcher} C.~J.,   et~al., 2012,
  \mn@doi [\aap] {10.1051/0004-6361/201117353}, \href
  {http://adsabs.harvard.edu/abs/2012A\%26A...538A...8S} {538, A8}

\bibitem[\protect\citeauthoryear{{Schwartz} \& {Martin}}{{Schwartz} \&
  {Martin}}{2004}]{schwartz}
{Schwartz} C.~M.,  {Martin} C.~L.,  2004, \mn@doi [\apj] {10.1086/421546},
  \href {http://cdsads.u-strasbg.fr/abs/2004ApJ...610..201S} {610, 201}

\bibitem[\protect\citeauthoryear{{Seaton}}{{Seaton}}{1979}]{seaton}
{Seaton} M.~J.,  1979, \mnras, \href
  {http://cdsads.u-strasbg.fr/abs/1979MNRAS.187P..73S} {187, 73P}

\bibitem[\protect\citeauthoryear{{Shapiro} \& {Field}}{{Shapiro} \&
  {Field}}{1976}]{shapiro}
{Shapiro} P.~R.,  {Field} G.~B.,  1976, \mn@doi [\apj] {10.1086/154332}, \href
  {http://adsabs.harvard.edu/abs/1976ApJ...205..762S} {205, 762}

\bibitem[\protect\citeauthoryear{{Sharp} \& {Bland-Hawthorn}}{{Sharp} \&
  {Bland-Hawthorn}}{2010}]{sharp}
{Sharp} R.~G.,  {Bland-Hawthorn} J.,  2010, \mn@doi [\apj]
  {10.1088/0004-637X/711/2/818}, \href
  {http://cdsads.u-strasbg.fr/abs/2010ApJ...711..818S} {711, 818}

\bibitem[\protect\citeauthoryear{{Shih} \& {Rupke}}{{Shih} \&
  {Rupke}}{2010}]{shih10}
{Shih} H.-Y.,  {Rupke} D.~S.~N.,  2010, \mn@doi [\apj]
  {10.1088/0004-637X/724/2/1430}, \href
  {http://cdsads.u-strasbg.fr/abs/2010ApJ...724.1430S} {724, 1430}

\bibitem[\protect\citeauthoryear{{Soto}, {Martin}, {Prescott}  \&
  {Armus}}{{Soto} et~al.}{2012}]{soto}
{Soto} K.~T.,  {Martin} C.~L.,  {Prescott} M.~K.~M.,   {Armus} L.,  2012,
  \mn@doi [\apj] {10.1088/0004-637X/757/1/86}, \href
  {http://cdsads.u-strasbg.fr/abs/2012ApJ...757...86S} {757, 86}

\bibitem[\protect\citeauthoryear{{Spitzer}}{{Spitzer}}{1978}]{spitzer}
{Spitzer} L.,  1978, {Physical processes in the interstellar medium}

\bibitem[\protect\citeauthoryear{{Tenorio-Tagle} \&
  {Mu{\~n}oz-Tu{\~n}{\'o}n}}{{Tenorio-Tagle} \&
  {Mu{\~n}oz-Tu{\~n}{\'o}n}}{1998}]{TenorioTagle98}
{Tenorio-Tagle} G.,  {Mu{\~n}oz-Tu{\~n}{\'o}n} C.,  1998, \mn@doi [\mnras]
  {10.1046/j.1365-8711.1998.01194.x}, \href
  {http://cdsads.u-strasbg.fr/abs/1998MNRAS.293..299T} {293, 299}

\bibitem[\protect\citeauthoryear{{Tremonti} et~al.,}{{Tremonti}
  et~al.}{2004}]{tremonti}
{Tremonti} C.~A.,  et~al., 2004, \mn@doi [\apj] {10.1086/423264}, \href
  {http://adsabs.harvard.edu/abs/2004ApJ...613..898T} {613, 898}

\bibitem[\protect\citeauthoryear{{Vazdekis}, {S{\'a}nchez-Bl{\'a}zquez},
  {Falc{\'o}n-Barroso}, {Cenarro}, {Beasley}, {Cardiel}, {Gorgas}  \&
  {Peletier}}{{Vazdekis} et~al.}{2010}]{2010MNRAS.404.1639V}
{Vazdekis} A.,  {S{\'a}nchez-Bl{\'a}zquez} P.,  {Falc{\'o}n-Barroso} J.,
  {Cenarro} A.~J.,  {Beasley} M.~A.,  {Cardiel} N.,  {Gorgas} J.,   {Peletier}
  R.~F.,  2010, \mn@doi [\mnras] {10.1111/j.1365-2966.2010.16407.x}, \href
  {http://adsabs.harvard.edu/abs/2010MNRAS.404.1639V} {404, 1639}

\bibitem[\protect\citeauthoryear{{Veilleux}, {Cecil}  \&
  {Bland-Hawthorn}}{{Veilleux} et~al.}{2005}]{veilleux05}
{Veilleux} S.,  {Cecil} G.,   {Bland-Hawthorn} J.,  2005, \mn@doi [\araa]
  {10.1146/annurev.astro.43.072103.150610}, \href
  {http://cdsads.u-strasbg.fr/abs/2005ARA\%26A..43..769V} {43, 769}

\bibitem[\protect\citeauthoryear{{Westmoquette}, {Clements}, {Bendo}  \&
  {Khan}}{{Westmoquette} et~al.}{2012}]{westmoquette2012}
{Westmoquette} M.~S.,  {Clements} D.~L.,  {Bendo} G.~J.,   {Khan} S.~A.,  2012,
  \mn@doi [\mnras] {10.1111/j.1365-2966.2012.21214.x}, \href
  {http://cdsads.u-strasbg.fr/abs/2012MNRAS.424..416W} {424, 416}

\bibitem[\protect\citeauthoryear{{Wood}, {Tremonti}, {Calzetti}, {Leitherer},
  {Chisholm}  \& {Gallagher}}{{Wood} et~al.}{2015}]{wood}
{Wood} C.~M.,  {Tremonti} C.~A.,  {Calzetti} D.,  {Leitherer} C.,  {Chisholm}
  J.,   {Gallagher} J.~S.,  2015, \mn@doi [\mnras] {10.1093/mnras/stv1471},
  \href {http://cdsads.u-strasbg.fr/abs/2015MNRAS.452.2712W} {452, 2712}

\makeatother
\end{thebibliography}

% Alternatively you could enter them by hand, like this:
% This method is tedious and prone to error if you have lots of references
% \begin{thebibliography}{99}
% \bibitem[\protect\citeauthoryear{Author}{2012}]{Author2012}
% Author A.~N., 2013, Journal of Improbable Astronomy, 1, 1
% \bibitem[\protect\citeauthoryear{Others}{2013}]{Others2013}
% Others S., 2012, Journal of Interesting Stuff, 17, 198
% \end{thebibliography}

%%%%%%%%%%%%%%%%%%%%%%%%%%%%%%%%%%%%%%%%%%%%%%%%%%

%%%%%%%%%%%%%%%%% APPENDICES %%%%%%%%%%%%%%%%%%%%%

%\appendix

%\section{Some extra material}

%If you want to present additional material which would interrupt the flow of the main paper,
%it can be placed in an Appendix which appears after the list of references.

%\input{/home/pablo/Trabajo/INTEGRAL/2014_reducidas/NGC5394/mapas/palatex.tex}

%%%%%%%%%%%%%%%%%%%%%%%%%%%%%%%%%%%%%%%%%%%%%%%%%

% Don't change these lines
\bsp	% typesetting comment
\label{lastpage}
\end{document}